\documentclass{jfm}
\usepackage{graphicx}
\usepackage{graphicx,subfigure}
\usepackage{natbib}
\usepackage{multirow}
\usepackage{amsmath}
\usepackage{amsfonts,amssymb}
\usepackage[colorlinks=true,citecolor = blue]{hyperref}

\usepackage[switch, modulo]{lineno}

\ifCUPmtlplainloaded \else
  \checkfont{eurm10}
  \iffontfound
    \IfFileExists{upmath.sty}
      {\typeout{^^JFound AMS Euler Roman fonts on the system,
                  using the 'upmath' package.^^J}%
      \usepackage{upmath}}
      {\typeout{^^JFound AMS Euler Roman fonts on the system, but you
                  don't seem to have the}%
      \typeout{'upmath' package installed. jfm.cls can take advantage
                 of these fonts,^^Jif you use 'upmath' package.^^J}%
      }
  \else
  \fi
\fi

\ifCUPmtlplainloaded \else
  \checkfont{msam10}
  \iffontfound
    \IfFileExists{amssymb.sty}
      {\typeout{^^JFound AMS Symbol fonts on the system, using the
                'amssymb' package.^^J}%
       \usepackage{amssymb}%
         \let\leq=\leqslant
         \let\geq=\geqslant
      }{}
  \fi
\fi

\ifCUPmtlplainloaded \else
  \IfFileExists{amsbsy.sty}
    {\typeout{^^JFound the 'amsbsy' package on the system, using it.^^J}%
     \usepackage{amsbsy}}
    {\providecommand\boldsymbol[1]{\mbox{\boldmath $##1$}}}
\fi

\newcommand\widebar[1]{\mathop{\overline{#1}}}

\title[Mean zonal flows in rotating spheroids]{Mean zonal flows induced by weak mechanical forcings in rotating spheroids}

\author[D. C\'ebron \etal]{David C\'ebron$^1$\footnote{Email address for correspondence: david.cebron@univ-grenoble-alpes.fr}, J\'er\'emie Vidal$^{1}$, Nathana\"el Schaeffer$^{1}$, Antonin Borderies$^{1}$ and Alban Sauret$^2$}
\affiliation{$^1$ ISTerre, Universit\'e Grenoble Alpes, CNRS, 38000 Grenoble, France \\ $^2$ Department of Mechanical Engineering, University of California,
Santa Barbara CA 93106, USA}

\begin{document}

\label{firstpage}
\maketitle

\begin{abstract}
The generation of mean flows is a long-standing issue in rotating fluids.
Motivated by planetary objects, we consider here a rapidly rotating fluid-filled spheroid, which is subject to weak perturbations of either the boundary (e.g. tides) or the rotation vector (e.g. in direction by precession, or in magnitude by longitudinal librations). 
Using boundary-layer theory, we determine the mean zonal flows generated by nonlinear interactions within the viscous Ekman layer.
These flows are of interest because they survive in the relevant planetary regime of both vanishing forcings and viscous effects. 
We extend the theory to take into account (i) the combination of spatial and temporal perturbations, providing new mechanically driven zonal flows (e.g. driven by latitudinal librations), and (ii) the spheroidal geometry relevant for planetary bodies. 
Wherever possible, our analytical predictions are validated with direct numerical simulations. 
The theoretical solutions are in good quantitative agreement with the simulations, with expected discrepancies (zonal jets) in the presence of inertial waves generated at the critical latitudes (as for precession).
Moreover, we find that the mean zonal flows can be strongly affected in spheroids. Guided by planetary applications, we also revisit the scaling laws for the geostrophic shear layers at the critical latitudes, and the influence of a solid inner core.
\end{abstract}


\section{Introduction}
\subsection{Physical context}
Global rotation tends to sustain two-dimensional mean flows that are almost invariant along the rotation axis in rapidly rotating systems. 
These mean flows are indeed often obtained in various models of rotating turbulence \citep[e.g.][]{guervilly2014large,godeferd2015structure} and planetary core flows \citep[e.g.][]{aubert2005steady,schaeffer2017turbulent,monville2019rotating}. 
In the latter context, they are believed to play an important role in the exchange of angular momentum between liquid layers and surrounding solid domains \citep[e.g.][]{roberts2012theory}, which drives the long-term dynamical evolution of planetary bodies. 
Moreover, mean flows could be unstable in the rapidly rotating regime \citep[e.g.][]{sauret2014tide,favier2014non}, which could sustain space-filling turbulence and mixing. 
Therefore, understanding the formation of mean flows is essential to model the fluid dynamics of many rapidly rotating systems. 

A commonly observed feature of geostrophic flows is that they are spontaneously generated by nonlinear effects, for instance involving small-scale eddies \citep[e.g.][]{christensen2002zonal,aubert2002observations} or waves.
Rapidly rotating fluids are indeed characterised by the ubiquitous presence of inertial waves \citep[e.g.][]{zhang2017theory}, whose restoring force is the Coriolis force.
However, \citet{greenspan1969non} demonstrated that inviscid nonlinear interactions of inertial waves do not produce significant geostrophic flows in the rapidly rotating regime.
The combination of some nonlinear interactions and viscous effects is thus essential to generate mean geostrophic flows, and various wave-induced mechanisms have been explored. 
Local wave interactions in the weakly viscous interior could transfer energy from the inertial waves to the geostrophic flows, either through wave-wave interactions \citep[e.g.][]{newell1969rossby,smith1999transfer} or wave-induced secondary instabilities \citep[e.g.][]{kerswell1999secondary,brunet2020shortcut}. 
The aforementioned mechanisms have been explored in Cartesian or cylindrical geometries for computational simplicity. 
In these previous studies, the container depth does not vary in the direction perpendicular to the rotation axis.
However, this so-called beta effect is known to be important for planetary configurations \citep{busse1970thermal}, and also strongly modifies the geostrophic flows \citep{greenspanbook}. 
Thus, although these local mechanisms are certainly generic, the geostrophic flows investigated in these studies are not directly relevant for (large-scale) planetary core flows. 

Another mechanism, which is relevant for planetary applications, has been proposed by \citet{busse1968}. 
Most planetary fluid bodies are subject to mechanical forcings (e.g. librations, precession, or tides) because of the presence of orbital companions. 
Mechanical forcings have received a renewed interest in fluid mechanics, because of their non-negligible contribution in the internal fluid dynamics of planetary bodies \citep[e.g.][]{le2015flows}.
They are indeed responsible for differential motions of the rigid boundary with respect to the fluid. 
These motions can be transmitted to the bulk by viscous coupling, generating inertial waves \citep[e.g.][]{aldridge1969axisymmetric,noir2001experimental,sauret2013spontaneous}, and mean geostrophic flows resulting from nonlinear interactions of the flows within the Ekman boundary layer \citep[as considered in][]{busse1968}.
The latter mechanism has been then confirmed experimentally and numerically for various mechanical forcings \citep[e.g.][]{noir2001,noir2012experimental,lin2020libration}. \\

\begin{figure}
    \centering
    \begin{tabular}{cc}
    \includegraphics[width=0.47\textwidth]{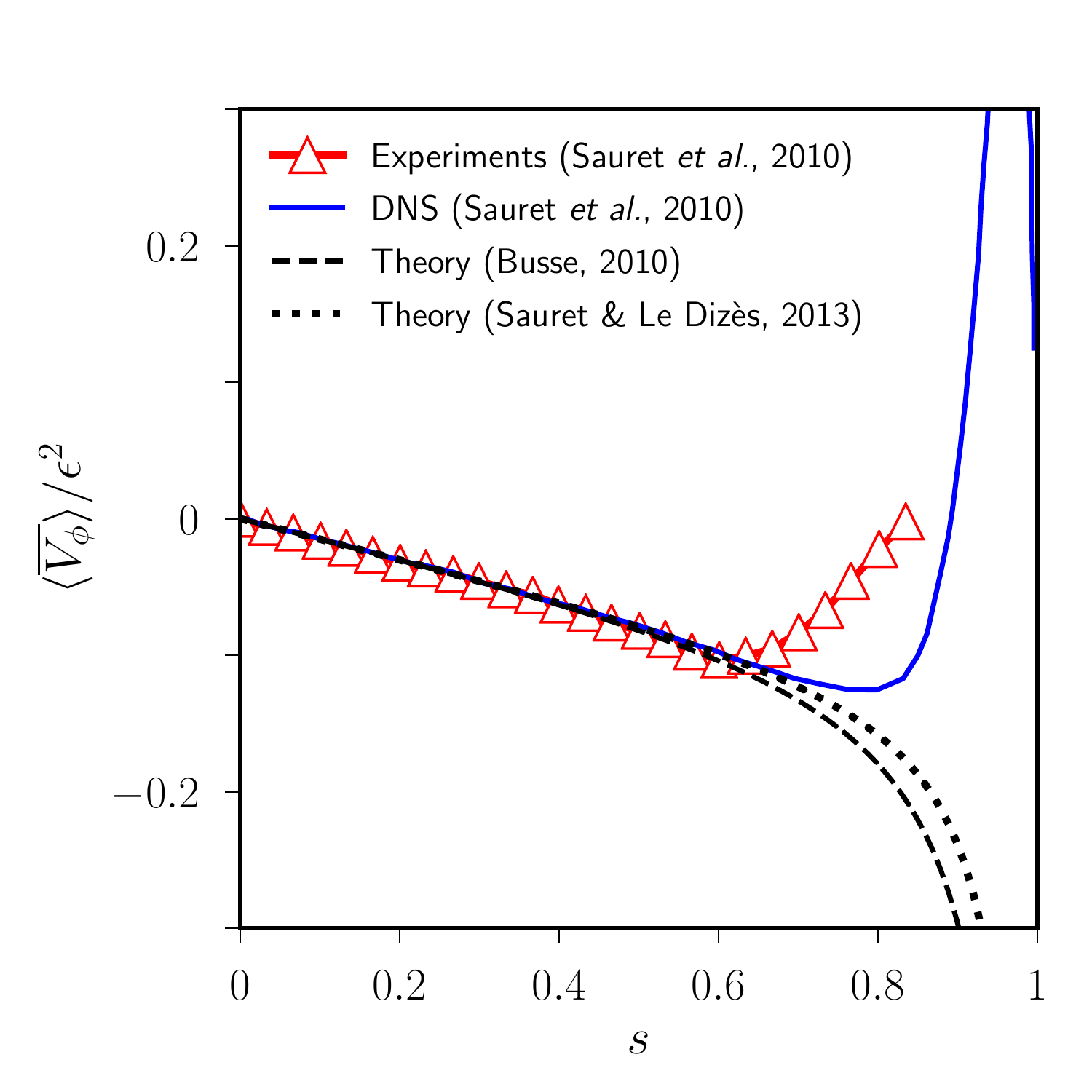} & \includegraphics[width=0.47\textwidth]{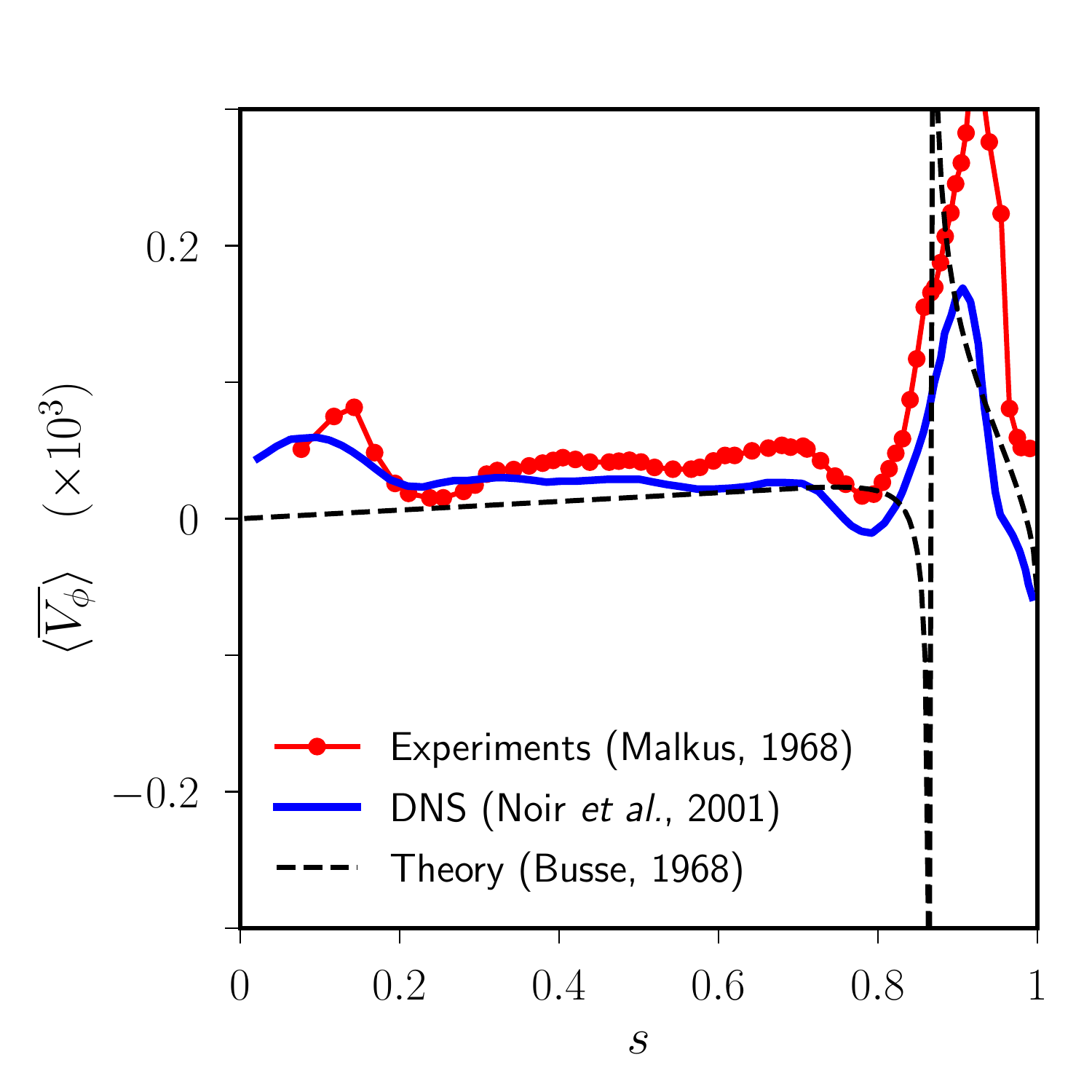} \\
    (a) & (b) \\
    \end{tabular}
    \caption{Comparison of the azimuthal component of the time-averaged zonal flow $\langle \overline{V}_\phi \rangle$, as a function of cylindrical radius $s$ in the equatorial plane $z_0=0$, between theory (dashed curves) and experiments/numerics (solid curves). (a) Libration-driven zonal flows with dimensionless angular frequency $\omega=0.1$ in a full sphere, extracted from figure 3b and figure 4 in \citet{sauret2010}. Experiment with $E = 1.15 \times 10^{-5}$, and $\epsilon=0.08$. DNS with $E = 5 \times 10^{-5}$, and $\epsilon=0.2$.  
    The vertical axis has been normalised by $\epsilon^2$, where $\epsilon$ is the dimensionless forcing amplitude. 
    (b) Precession-driven zonal flows extracted from figure 8 in \citet{noir2001}.}
    \label{fig:intro}
\end{figure}

\subsection{Motivations}
Rotating flows are usually characterised by the Ekman number $E$, which compares viscous to rotational effects.
As outlined in \citet{busse1968,busse2010}, mechanical forcings of typical amplitude $\mathcal{O}(\epsilon)$ can induce a mean zonal flow in the bulk of typical amplitude $\mathcal{O}(\epsilon^2)$, which is independent of $E$ in the regime $E \to 0$ (as shown below). 
Hence, this mechanism gives a non-zero bulk flow driven by viscous effects that survives in the planetary regime of vanishing viscosity $E \ll 1$. 
However, most studies about mean zonal flows in spherical-like domains have employed laboratory experiments or direct numerical simulations (DNS) with moderate values $E \geq 10^{-6}$, whereas rapidly rotating planetary flows are characterised by much smaller values (typically $E = 10^{-15}-10^{-12}$).
Thus, since viscous effects are overestimated in experimental or numerical works, an analytical study is directly relevant to obtain rigorous results about zonal flows in the planetary regime.

Only a few theoretical studies have hitherto investigated mean zonal flows driven by mechanical forcings. 
The case of a rotating cylindrical tank subject to longitudinal librations has been recently revisited analytically \citep{sauret2015mean}, and the theory has been convincingly compared with experiments \citep{wang1970cylindrical} and simulations \citep{sauret2012fluid}. 
However, a successful validation of mechanically driven zonal flows is generally missing in spherical and ellipsoidal geometries when $E$ is vanishingly small. 
For instance, considering longitudinal librations in spheres, two different results have been obtained for low libration frequencies  \citep{busse2010,sauret2013libration}. 
As shown in figure \ref{fig:intro}(a), they can both explain the experimental results of \citet{sauret2010}. 
Concerning precession, experiments \citep{malkus1968precession} or numerical simulations \citep{noir2001} have never properly validated the theory of \citet{busse1968}, as illustrated in figure \ref{fig:intro}(b). 
Similarly, the theoretical zonal flows driven by tides \citep{suess1971} do not agree with experimental findings (as we will show below). 
Consequently, theoretical predictions remain to be thoroughly validated before they can be extrapolated for planets.

Finally, singularities have been found in the boundary-layer calculation due to the presence of the critical latitudes,
where the flows should be smoothed out by additional viscous effects \citep[e.g.][]{kerswell1995,kida2011steady} not taken into account in the theory \citep[as in][]{busse1968,sauret2013spontaneous}.
Around these locations, the mean flows are known to take the form of narrow geostrophic shear layers aligned with the axis of rotation \citep[e.g.][]{calkins2010axisymmetric}.
The variations of the geostrophic shear amplitude with the Ekman number are however still disputed \citep{noir2001,lin2020libration}, such that planetary extrapolations remain speculative. 
Thus, targeted DNS in the regime $E\ll1$ are also worth performing to explore the behaviour of the geostrophic shear layers.

Solving the full mathematical problem of mechanically driven flows is complex, but analytical progress can be made for planetary parameters \citep[as undertaken in][]{busse1968,busse2010}. 
Since planetary interiors are characterised by small forcing amplitudes $\epsilon \ll 1$ and small viscous effects $E \ll 1$, we will employ asymptotic theory in $\epsilon$ and $E$. 
Moreover, our zonal flow calculation will also assume that the spin-up time scale of the fluid \citep{greenspanbook} is much longer than the characteristic time scale of the mechanical forcing (in the fluid rotating frame), such that no global spin-up of the fluid will occur during the dynamics. 
Finally, following previous works on mean zonal flows \citep[e.g.][]{busse2010,sauret2013libration}, we neglect in the theory viscous effects at the critical latitudes (associated with internal shear layers) and our theoretical bulk basic flow is taken as a solid-body rotation (for its spatial dependency). Thus, we assume that no inertial mode is excited by the forcing on top of this basic flow \cite[which is exact if the forcing frequency is larger than twice the mean fluid rotation rate, e.g.][]{greenspanbook}. 
Using DNS, where these effects are fully taken into account, we will revisit the proposed associated scaling laws for planetary extrapolations \citep[e.g.][]{noir2001,lin2020libration}.
The paper is organised as follows. 
We introduce the problem and the methods in \S\ref{sec:pb}.
We describe the asymptotic weakly nonlinear analysis in \S\ref{sec:asymp}, and present the theoretical and numerical results in \S\ref{sec:results}. 
We discuss the results in \S\ref{sec:discussion}, and we finally conclude the paper in \S\ref{sec:ccl}.

\section{Description of the problem and methods}
\label{sec:pb}
\subsection{Mathematical description}
\begin{figure}
    \centering
    \includegraphics[width=0.85\textwidth]{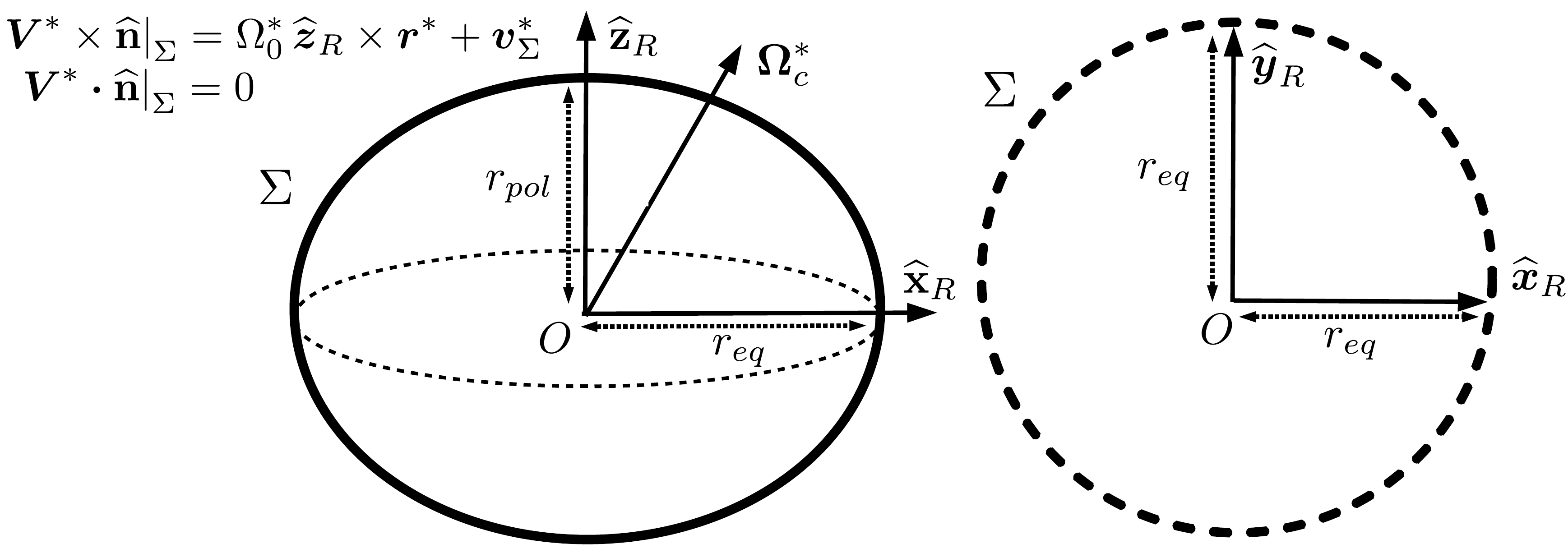}
    \caption{\emph{Left}: Spheroidal geometry of the forced problem. \emph{Right}: equatorial (circular) section.}
    \label{fig:geometry}
\end{figure}

We consider an incompressible and homogeneous Newtonian fluid of kinematic viscosity $\nu$ and density $\rho$, enclosed in a spheroidal container of semi-axis length $r_{pol}$ along the revolution axis, while the other one is noted $r_{eq}$ (the spheroid is oblate when $r_{eq}>r_{pol}$, and prolate when $r_{eq}<r_{pol}$). 
We introduce the Cartesian basis vector $(\widehat{\boldsymbol{x}}_I,\widehat{\boldsymbol{y}}_I,\widehat{\boldsymbol{z}}_I)$ of the inertial frame, whose origin $O$ is the centre of the spheroidal container. 
In the following, we work in a frame of reference where the spheroidal shape of the container boundary is stationary. 
We use a Cartesian basis vectors $(\widehat{\boldsymbol{x}}_R,\widehat{\boldsymbol{y}}_R,\widehat{\boldsymbol{z}}_R)$ where $\widehat{\boldsymbol{z}}_R$ is aligned with the spheroid revolution axis, as illustrated in figure \ref{fig:geometry}. 
The rotation vector of this reference frame, denoted  $\boldsymbol{\Omega}_c^* (t)$ in the following, is along $\widehat{\boldsymbol{z}}_I$ in the absence of perturbation. 
In this reference frame, the velocity $\boldsymbol{V}^*$ satisfies the no-slip boundary conditions (BC) on $\Sigma$
\begin{subequations} \label{eq:condav}
\begin{equation}
\left . \boldsymbol{V}^* \boldsymbol{\cdot} \widehat{\boldsymbol{n}} \right |_{\Sigma} = 0, \quad 
    \left . \boldsymbol{V}^* \times \widehat{\boldsymbol{n}} \right |_{\Sigma} = \boldsymbol{V}_\Sigma^*,
    \tag{\theequation \emph{a,b}}
\end{equation}
\end{subequations}
where ${\widehat{\boldsymbol{n}}}$ is the unit vector normal to the boundary,
and $\boldsymbol{V}_\Sigma^* $ is mainly a solid-body rotation at $\Omega_0^*$ around $ \widehat{\boldsymbol{z}}_I$, possibly perturbed by a small flow $\boldsymbol{v}_\Sigma^*$. 
We thus consider
\begin{equation}
\boldsymbol{V}_\Sigma^* = 
\Omega_0^* \, \widehat{\boldsymbol{z}}_R \times \boldsymbol{r}^* + \boldsymbol{v}_\Sigma^*,
\label{eq:noslip}
\end{equation}
where $\boldsymbol{r}^*$ is the position vector, and $\boldsymbol{v}_\Sigma^*$ is an imposed tangential velocity related to the considered mechanical forcing. 
One can choose a frame of reference where the spheroidal boundary is steady, that is with $\boldsymbol{V}_\Sigma^*=\boldsymbol{0}$.
This frame is referred as the wall frame, or the mantle frame in planetary sciences.
We denote the associated Cartesian basis vectors $(\widehat{\boldsymbol{x}}_M, \widehat{\boldsymbol{y}}_M, \widehat{\boldsymbol{z}}_M)$ centred on $O$, where $\widehat{\boldsymbol{z}}_M$ is along the spheroid revolution axis.

We work below using dimensionless units, denoting the dimensionless variables without the superscript $^*$ for the sake of clarity. 
We choose $r_{eq}$ as the length scale, and $|\Omega_s^*|^{-1}=(|\Omega_0^*|+|\overline{{\Omega}}_c^*|)^{-1}$ as the time scale, where $\overline{{\Omega}}_c^*$ is the time average of ${{\Omega}}_c^*$.
In the reference frame rotating at $\boldsymbol{\Omega}_c$, the dimensionless fluid velocity $ \boldsymbol{V} $ is governed by
\begin{subequations}
\label{syst:eqn_sl}
\begin{equation}
\partial_t  \boldsymbol{V} +  (\boldsymbol{V} \boldsymbol{\boldsymbol{\cdot}} \boldsymbol{\nabla}) \, \boldsymbol{V} +2 \,  \boldsymbol{\Omega}_c \times \boldsymbol{V}  + \dot{\boldsymbol{\Omega}}_c \times \boldsymbol{r} = -\nabla \Pi + E \, \boldsymbol{\nabla}^2 \boldsymbol{V}, \quad \boldsymbol{\nabla} \boldsymbol{\cdot} \boldsymbol{V} = 0,
\tag{\theequation \emph{a,b}}
\end{equation}
\end{subequations}
where $\Pi$ is the reduced pressure (taking into account the centrifugal effects), $\dot{\boldsymbol{\Omega}}_c=\mathrm{d}_t {\boldsymbol{\Omega}}_c$ is the time derivative of  ${\boldsymbol{\Omega}}_c$, $\partial_t  \boldsymbol{V}$ is the partial time derivative of $ \boldsymbol{V}$, and $E=\nu/( \Omega_s^* r_{eq}^2)$ is the (dimensionless) Ekman number. 

\subsection{Mechanical forcings}
\label{sec:mecha}
We note $\boldsymbol{U}$ the inviscid bulk flow driven by the forcing. 
In the absence of any mechanical forcings, that is with $\boldsymbol{V}_\Sigma=\boldsymbol{0}$ and $\dot{\boldsymbol{\Omega}}_c=\boldsymbol{0}$, $\boldsymbol{U}$ reduces to a solid-body rotation with the angular velocity $ \Omega_0 \, \widehat{\boldsymbol{z}}_R$.
However, the latter flow is perturbed by weak harmonic perturbations generated by mechanical forcings.
The general framework described in this study allows us to consider various mechanical forcings that are described below.

\begin{enumerate}
    \item Multipolar tidal-like forcing corresponds to $\boldsymbol{\Omega}_c={\Omega}_c \,  \widehat{\boldsymbol{z}}_R$ (with here $\widehat{\boldsymbol{z}}_R=\widehat{\boldsymbol{z}}_I$), and 
\begin{equation}
\boldsymbol{v}_\Sigma^{} = \epsilon s^{q} \cos(m {\phi}) \cos(\omega t) \, \widehat{\boldsymbol{z}}_R \times \boldsymbol{r},
\label{cond_limite567}
\end{equation}
with the azimuthal angle $\phi$ with respect to $\widehat{\boldsymbol{z}}_R$ (see figure \ref{fig:geometry}), the forcing amplitude $\epsilon$, and the azimuthal wavenumber $m$ of the spatial deformation \cite[this boundary velocity has also been considered by][see e.g. his equation 2.14.2]{greenspanbook}. 
In our calculation, $q$ is taken as an independent parameter but, for regularity along the rotation axis \citep{lewis1990physical}, and to consider multipolar flows \citep[e.g.][]{cebron2014libration,sauret2015mean}, we must consider $q=|m-1|$. In expression (\ref{cond_limite567}), the case $m=2$, $\Omega_c=\omega=0$ has been considered in
\citet{suess1971}, which is extended here to account for both multipolar deformations and oscillations at the frequency $\omega$ \citep[e.g.][]{sauret2013libration}.

    \item Longitudinal librations are investigated with ${\Omega}_s=1$ and $\widehat{\boldsymbol{z}}_R=\widehat{\boldsymbol{z}}_I$. 
    Introducing the forcing amplitude $\epsilon$, rotating spheroids can be studied, in an equivalent way, either (i) in the mantle frame of reference \citep[e.g.][]{favier2015generation} with $\Omega_0=0$, $\boldsymbol{\Omega}_c=[1+\epsilon \cos(\omega t)] \, \widehat{\boldsymbol{z}}_R$ and $\boldsymbol{V}_\Sigma = \boldsymbol{0}$, (ii) in the mean rotating frame of reference \citep[e.g.][]{busse2010} with $\Omega_0=0$, $\boldsymbol{\Omega}_c=\widehat{\boldsymbol{z}}_R$ and $\boldsymbol{V}_\Sigma = \epsilon \cos(\omega t) \, \widehat{\boldsymbol{z}}_R \times \boldsymbol{r}$, or (iii) in the inertial frame of reference with $\Omega_0=1$, $\boldsymbol{\Omega}_c=\boldsymbol{0}$ and $\boldsymbol{V}_\Sigma = [1+\epsilon \cos(\omega t)] \, \widehat{\boldsymbol{z}}_R \times \boldsymbol{r}$ . Note that the case (ii) can actually be recovered with the particular case $m=\Omega_0=0$ of the multipolar tidal-like forcing.
    
    \item Latitudinal librations are modelled with $\Omega_0=0$, and the general case of a rigid spheroidal container can only be studied in the mantle frame, where the $\Sigma$ is stationary and $\boldsymbol{V}_\Sigma = \boldsymbol{0}$. 
    The corresponding forcing in this frame is \citep[see in][]{vantieghem2015latitudinal}
    \begin{subequations}
    \begin{align}
        \boldsymbol{\Omega}_c &= \dot{\Theta} \, \widehat{\boldsymbol{x}}_M + \sin(\Theta) \, \widehat{\boldsymbol{y}}_M + \cos(\Theta) \,  \widehat{\boldsymbol{z}}_M, \\ \dot{\boldsymbol{\Omega}}_c &= \ddot{\Theta} \, \widehat{\boldsymbol{x}}_M + \dot{\Theta} \cos(\Theta) \, \widehat{\boldsymbol{y}}_M - \dot{\Theta} \sin(\Theta) \, \widehat{\boldsymbol{z}}_M,
    \end{align}
    \end{subequations}
    where $\Theta=(\epsilon/\omega) \sin(\omega t)$ is the instantaneous libration angle, and where $\epsilon$ is the forcing amplitude. In the limit $\epsilon \ll 1$ considered for the analytical calculations performed in this work, these expressions read (at first order in $\epsilon$)
    \begin{subequations}
    \begin{align}
        \boldsymbol{\Omega}_c&= \epsilon [ \cos(\omega t) \, \widehat{\boldsymbol{x}}_M + \omega^{-1} \, \sin(\omega t) \, \widehat{\boldsymbol{y}}_M ] +    \widehat{\boldsymbol{z}}_M , \\ \dot{\boldsymbol{\Omega}}_c &=  \epsilon [ - \omega \sin(\omega t) \, \widehat{\boldsymbol{x}}_M +  \cos(\omega t) \, \widehat{\boldsymbol{y}}_M ],
    \end{align}
    \end{subequations}
    Note that the particular case of the sphere can also be studied analytically and numerically in the mean rotating frame with $\boldsymbol{\Omega}_c = \widehat{\boldsymbol{z}}_R= \widehat{\boldsymbol{z}}_I$ and $\boldsymbol{V}_\Sigma = \epsilon \cos(\omega t)\, \widehat{\boldsymbol{x}}_R \times \boldsymbol{r}$.
    
    \item Precession can be considered in the precession frame \citep[e.g.][]{cebron2019precessing} by using  $\boldsymbol{V}_\Sigma = 
\Omega_0 \, \widehat{\boldsymbol{z}}_R \times \boldsymbol{r}$ with $\Omega_0=(1+P_o)^{-1}$, and 
    \begin{equation} \label{eq:skd9xx}
    \boldsymbol{\Omega}_c=\Omega_0\, P_o \,  [ \sin (\alpha) \,  \widehat{\boldsymbol{x}}_R + \cos (\alpha) \, \widehat{\boldsymbol{z}}_R ],
    \end{equation}
    where $\alpha$ is the precession angle, and $P_o$ is the Poincar\'e number (ratio of the precession and the boundary rotation rates). 
    The associated bulk flow $\boldsymbol{U}$ is then mainly a tilted (stationary) solid-body rotation $\boldsymbol{U}=\boldsymbol{\omega}_f \times \boldsymbol{r}$. 
    For weak precession forcing, $\boldsymbol{\omega}_f \approx (1+P_o)^{-1} \widehat{\boldsymbol{z}}_R$ at the order $\epsilon$, where $\epsilon$ characterises the small misalignment of $\boldsymbol{\omega}_f$ and $\widehat{\boldsymbol{z}}_R$ \citep[i.e. the forcing amplitude, see further details in the seminal work of][]{busse1968}.
\end{enumerate}

\subsection{Numerical modelling}
We integrate equations (\ref{syst:eqn_sl}) using two open-source codes. 
Equations in spherical geometries are solved using the parallel pseudo-spectral code \textsc{xshells} \citep[e.g.][]{schaeffer2017turbulent}, based on a poloidal-toroidal decomposition of the velocity field onto spherical harmonics of degree $l \leq l_{\max}$ and azimuthal wavenumber $m \leq m_{\max}$ using the \textsc{shtns} library \citep[][]{schaeffer2013efficient}, and second-order finite differences with $N_r$ points are used in the radial direction.
The code has been validated for full-sphere computations, including flows crossing the origin \citep{marti2014full}, and details about the implementation at the centre are given in appendix \ref{sec:jvreg}.
To solve the dynamical equations, the code can use several semi-implicit time-stepping schemes, which treat the diffusive terms implicitly and the other ones explicitly.
Most of the simulations have been performed using the accurate semi-implicit backward difference formula of order 3 \citep[SBDF3, see][]{ascher1995implicit}.
The typical spatial resolution at $E=10^{-7}$ is $N_r=576$, $l_{\max}=159$, $m_{\max}=5$. 

In spheroidal geometries, we solve the nonlinear equations in their weak variational form using the spectral-element code \textsc{Nek5000} \citep[e.g.][]{fischer2007simulation}, which combines the geometrical flexibility of finite element methods with the accuracy of spectral methods.
The computational domain is made of $\mathcal{E} = 3584$ non-overlapping hexahedral elements in coreless geometries (or $\mathcal{E} = 3840$ in spheroidal shells, see below).
Moreover, the velocity (and pressure) is represented within each element as Lagrange polynomials of order $\mathcal{N}=13$ (respectively, $\mathcal{N}-2$) on the Gauss-Lobatto-Legendre (Gauss-Legendre) points. 
Temporal discretisation is accomplished by a third-order method, based on an adaptive and semi-implicit scheme in which the nonlinear and Coriolis terms are treated explicitly, and the remaining linear terms are treated implicitly. 
We have checked the numerical accuracy in targeted simulations by varying the polynomial order from $\mathcal{N}=13$ to $\mathcal{N}=15$, and found that the resolution of the Ekman boundary layers is appropriate with at least ten grid points within the layer.

\subsection{Extracting the mean zonal flows from DNS}
In the planetary limit $\epsilon \ll 1$ considered in this work, the forced flow is mainly a solid-body rotation in the bulk for all the aforementioned forcings.  
Consequently, when $E \to 0$, the mean zonal flows tend to geostrophic flows, which are invariant along the fluid rotation axis and are established on the dimensionless spin-up time scale $E^{-1/2}$ \citep{greenspanbook}.
Thus, for every DNS, we have simulated the dynamics over several spin-up time scale, ensuring that the mean zonal flows are well established.
We have also used typical time steps $\mathrm{d}t = 10^{-3}-10^{-2}$, which were sufficient to integrate the dynamics.

\begin{figure}
    \centering
    \begin{tabular}{cc}
    \includegraphics[width=0.49\textwidth]{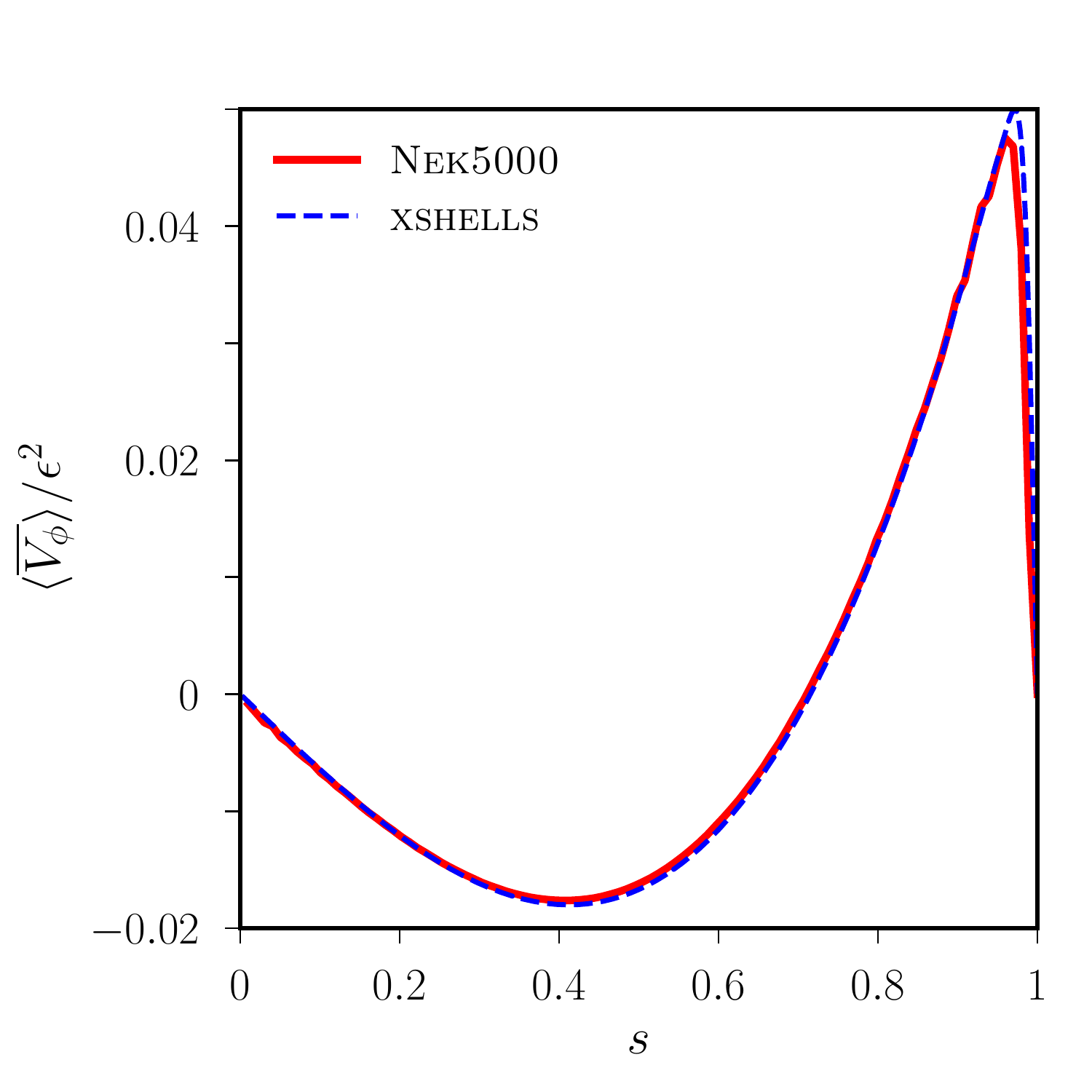} & 
    \includegraphics[width=0.49\textwidth]{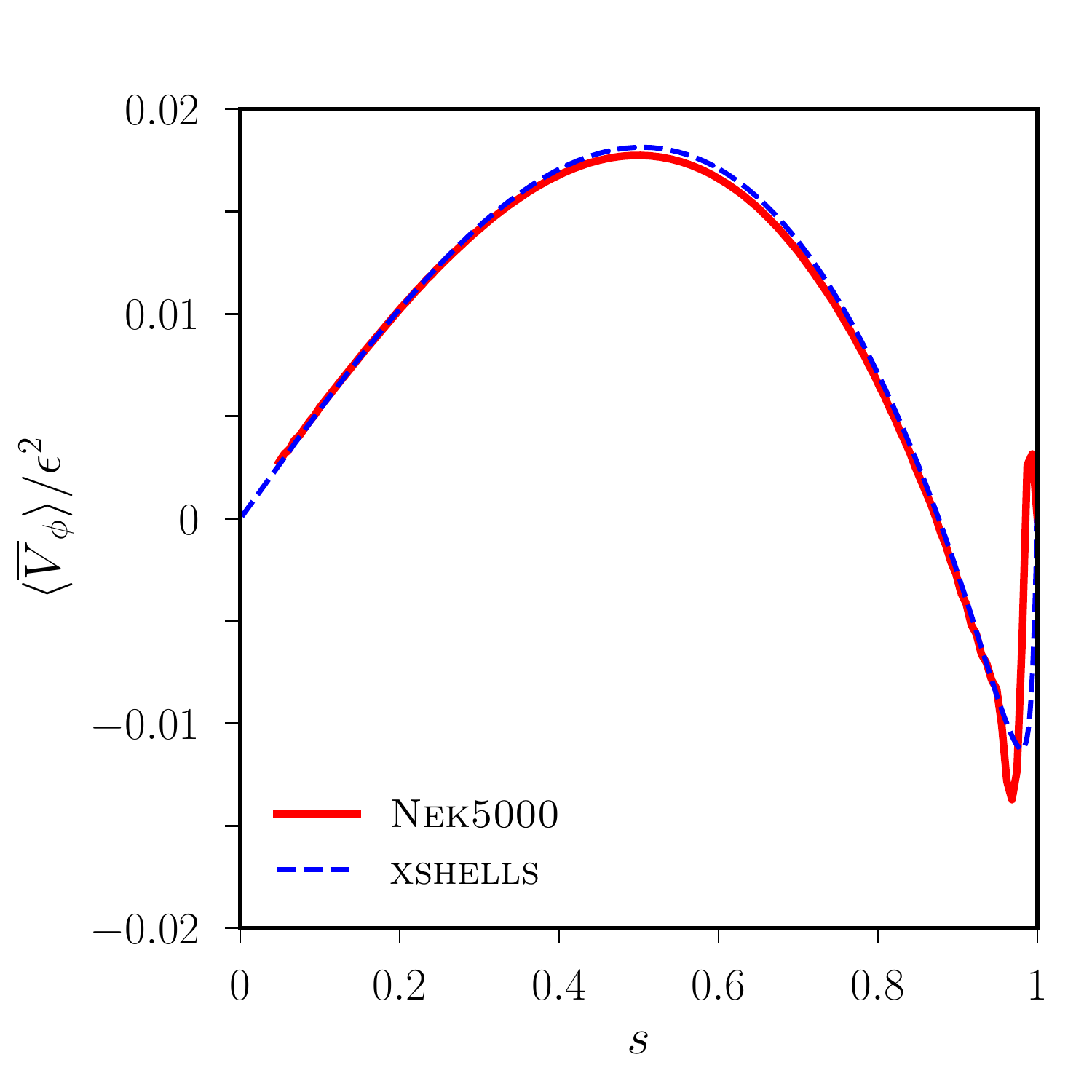} \\
    (a) & (b)\\
    \end{tabular}
    \caption{Azimuthal component of the time-averaged zonal flow $\langle \overline{V}_\phi \rangle/\epsilon^2$ as a function of cylindrical radius $s$ in the equatorial plane $z_0=0$. DNS in a sphere with $E=10^{-4}$, $\epsilon=10^{-2}$ and forcing frequency $\omega=\pi$, using \textsc{Nek5000} (red solid curves) and \textsc{xshells} (blue dashed curves). (a) Longitudinal librations. (b) Latitudinal librations.}
    \label{fig:benchDNS}
\end{figure}

The mean zonal flow is computed from the three-dimensional flow $\boldsymbol{V}$ by considering the cylindrical radial variation of $\langle \widebar{V}_\phi \rangle \, (s, z=z_0)$ in the horizontal plane $z=z_0$, where $\widebar{X}$ and $\langle X \rangle$ are the time and azimuthal averages of the quantity $X$ (respectively). 
With \textsc{xshells}, the mean zonal flow is computed from the time-averaged $m=0$ component of the toroidal scalar in the plane $z=z_0$. 
In spheroids, the azimuthal component  $\langle \overline{V}_\phi \rangle$ of the mean zonal flow is estimated as \citep{favier2015generation}
\begin{equation}
    \langle \overline{V}_\phi \rangle \, (s, z=z_0) \simeq \frac{1}{N}  \sum_{|z-z_0| \leq z_{\max}} \quad \sum_{0 \leq \phi \leq 2\pi} \quad \sum_{s-\mathrm{d}s < s < s+\mathrm{d}s} \overline{\boldsymbol{V}} \boldsymbol{\cdot} \widehat{\boldsymbol{\phi}},
    \label{eq:zonalNek5000}
\end{equation}
where $N$ is the total number of grid points used to evaluate expression (\ref{eq:zonalNek5000}). 
The \textsc{Nek5000} DNS have been performed at $E \geq 10^{-6}$, contrary to the \textsc{xshells} DNS performed at $E\geq 10^{-7}$. 
Hence, the \textsc{Nek5000} DNS are more influenced by Ekman pumping when approaching the boundary. 
To properly estimate the geostrophic components, we $z$-average the flows over the vertical positions $|z-z_0|\leq z_{\max}$. 
In the \textsc{xshells} DNS, the mean zonal flow is defined as the value of the $m=0$ azimuthal velocity in the plane $z=z_0$. 
To be consistent, we have here considered $z_{\max} = 0.1$.
We have also checked that the mean flow computations are unchanged when using $z_{\max}/r_{pol}\leq 0.4-0.5$. 
Moreover, the approximate number of points in each direction in the \textsc{Nek5000} DNS is here $\mathcal{E}^{1/3} \mathcal{N} \approx 200$. 
We have thus averaged the azimuthal component over one hundred different shells along the cylindrical radius $s$, and over the vertical positions $|z-z_0| \leq 0.1$. We show in figure \ref{fig:benchDNS} the mean zonal flows computed from DNS in spheres with $E=10^{-4}$ and $\epsilon=10^{-2}$, with \textsc{Nek5000} in the mantle frame of reference and with \textsc{xshells} in the frame rotating at $\widehat{\boldsymbol{z}}_R$.
We find a very good agreement between the two codes, which validates our procedure to compute the mean zonal flows. 

\subsection{Perturbation approach}
Current DNS cannot be performed at the extremely small values of $E$ reached in planetary liquid cores. 
Thus, we solve analytically equations (\ref{syst:eqn_sl}) to gain physical insights into the asymptotic regime $E \ll 1$.
We assume that there is no significant shear in the interior and use viscous boundary-layer theory \citep[BLT, e.g.][]{greenspanbook} to write
$[\boldsymbol{V}, \Pi] = [\boldsymbol{U}, P] + [\boldsymbol{u}, p]$, where $\boldsymbol{U}$ describes the interior flow for which viscous effects can be neglected, and a boundary-layer flow $\boldsymbol{u}$. 
The latter contribution takes into account the viscous effects near the outer boundary, and decays exponentially towards the interior of the container. 
The governing equations, obtained from (\ref{syst:eqn_sl}), are in the limit $E \ll 1$
\begin{subequations}
\label{eq:BLT}
\begin{align}
\partial_t \boldsymbol{U} + (\boldsymbol{U}\boldsymbol{\cdot}\boldsymbol{\nabla}) \, \boldsymbol{U} + 2\, \boldsymbol{\Omega}_c  \times \boldsymbol{U}  + \dot{\boldsymbol{\Omega}}_c  \times \boldsymbol{r}  & = -\nabla P, \label{eq:NS_volume} \\   
\partial_t \boldsymbol{u} + (\boldsymbol{u}\boldsymbol{\cdot}\boldsymbol{\nabla}) \, \boldsymbol{U} + (\boldsymbol{u}+\boldsymbol{U})\boldsymbol{\cdot}\boldsymbol{\nabla}\boldsymbol{u}+2\, \boldsymbol{\Omega}_c  \times \boldsymbol{u} & = -\nabla p + E \, \boldsymbol{\nabla}^2\boldsymbol{u}, 
\label{eq:NS_CL}
\end{align}
\end{subequations}
Within the boundary layer, we introduce the stretched coordinate  $\zeta=(\boldsymbol{r}_{|\Sigma}-\boldsymbol{r})\boldsymbol{\cdot} \widehat{\boldsymbol{n}}/E^{1/2}$, where $\boldsymbol{r}_{|\Sigma}$ is the position vector on the boundary $\Sigma$.
We also assume that the field gradients along the boundary are negligible compared to the gradients normal to the boundary, that is $\widehat{\boldsymbol{n}} \boldsymbol{\cdot} \boldsymbol{\nabla} \simeq - E^{1/2} \partial_\zeta$ \citep[e.g.][]{greenspanbook}.
Then, the mass conservation equation reduces to its usual boundary layer approximation \citep[see p. 25 in][]{greenspanbook}
\begin{equation}
\boldsymbol{\nabla}\boldsymbol{\cdot}\boldsymbol{u} \simeq - E^{1/2} \partial_\zeta (\boldsymbol{u}
\boldsymbol{\cdot} \widehat{\boldsymbol{n}}) +\widehat{\boldsymbol{n}}
\boldsymbol{\cdot}\boldsymbol{\nabla}\times(\widehat{\boldsymbol{n}}\times\boldsymbol{u})=0.
\label{eq:consmass}
\end{equation}
To solve the BLT equations, we use asymptotic theory with the small forcing amplitude $\epsilon \ll 1$. Noting $\boldsymbol{V}_{\Sigma}^0=\Omega_0 \, \widehat{\boldsymbol{z}}_R \times \boldsymbol{r} $, we write
\begin{subequations}
\begin{equation}
\boldsymbol{V}_\Sigma= \boldsymbol{V}_\Sigma^0+\epsilon \, \boldsymbol{V}_\Sigma^1+\epsilon^2 \, \boldsymbol{V}_\Sigma^2+..., \quad \boldsymbol{\Omega}_c =  \Omega_c^0 \, \widehat{\boldsymbol{z}}_R+\epsilon \, \boldsymbol{\Omega}^1_c+\epsilon^2 \, \boldsymbol{\Omega}^2_c+... ,
\tag{\theequation \emph{a,b}}
\end{equation}
\end{subequations}
where the time average of $\boldsymbol{\Omega}_c$ is $\overline{\boldsymbol{\Omega}}_c= \Omega_c^0\, \widehat{\boldsymbol{z}}_R$ (since the perturbations are harmonic).
The BC (\ref{eq:condav}) imposes then $\boldsymbol{V}_\Sigma^k \boldsymbol{\cdot}  \widehat{\boldsymbol{n}} = \boldsymbol{0} $ at every order $\epsilon^k$. 
To perform the boundary-layer and perturbation calculations, we also expand $[\boldsymbol{V}, \Pi]$ in double power series involving the asymptotic parameters $ \epsilon \ll 1$ and $E \ll 1$. 
Note that we formally neglect the possible critical latitudes \citep[e.g.][]{kerswell1995}, although they can modify the mean zonal flows \citep[as previously found in cylinders by][see also below]{sauret2012fluid}. 
Since the Ekman layer scales as $E^{1/2}$ outside the critical latitudes, we use the double power series expansions for all our unknowns
\begin{subequations}
\label{eq:dev_velocity}
\begin{equation}
\left [ \boldsymbol{U}, \boldsymbol{u} \right ] = \sum_{i,j=0}^{+\infty}E^{\frac{1}{2}i}\,\epsilon^j\, \left [\boldsymbol{U}^{j}_i, \boldsymbol{u}^{j}_i \right ], \quad 
\left [ P, p \right ] = \sum_{i,j=0}^{+\infty} E^{\frac{1}{2}i}\,\epsilon^j\, \left [ P^{j}_i, p^{j}_i \right ].
\tag{\theequation \emph{a,b}}
\end{equation}
\end{subequations}
The substitution of equations (\ref{eq:dev_velocity}) in the governing equations (\ref{eq:BLT})-(\ref{eq:consmass}) leads to a sequence of equations for the interior and boundary-layer flows. We anticipate that our flows may also a priori vary slowly on the time scale $\tau = \mathcal{O}(E^{-1/2})$ and, thus, we also expand below the time in powers of $E^{-1/2}$ \cite[as e.g. done when calculating the Ekman layer damping of inertial modes, see][]{greenspanbook}.

\subsection{Governing equations}
At the leading order $\epsilon^0 E^0$, a natural solution of the interior zeroth-order equation is $ \boldsymbol{U}^0_0 =\Omega_0 \, \widehat{\boldsymbol{z}}_R \times \boldsymbol{r}$, which satisfies the BC since ${\boldsymbol{\Omega}}_c^0$ is constant and $\boldsymbol{v}_\Sigma = \boldsymbol{0} $ at this order. Since $\boldsymbol{U}^0_0 $ verifies the BC, we obtain that   $\boldsymbol{u}^0_0 = \boldsymbol{0}$.

Noting that, for an arbitrary velocity field $\boldsymbol{v}$, we have
\begin{equation}
(\boldsymbol{U}^0_0 \boldsymbol{\cdot} \boldsymbol{\nabla}) \, \boldsymbol{v} + (\boldsymbol{v} \boldsymbol{\cdot} \boldsymbol{\nabla}) \, \boldsymbol{U}^0_0 = 2\, \Omega_0\, \widehat{\boldsymbol{z}}_R  \times \boldsymbol{v} + \Omega_0 \partial_{{\phi}} \boldsymbol{v}  ,
\end{equation}
the first-order interior flow equations at the order $\epsilon E^0$ are then
\begin{equation} 
(\partial_t+\Omega_0 \partial_{{\phi}}) \, \boldsymbol{U}_0^1 + 2\, \Omega_s^0\, \widehat{\boldsymbol{z}}_R \times \boldsymbol{U}_0^1  + 2\, {\boldsymbol{\Omega}}_c^1  \times \boldsymbol{U}_0^0+\dot{\boldsymbol{\Omega}}_c^1 \times \boldsymbol{r} = -\nabla P_0^1, 
\label{eq:lf421df}
\end{equation}
together with the divergenceless condition $\boldsymbol{\nabla} \boldsymbol{\cdot} \boldsymbol{U}_0^1 = 0$ and $\Omega_s^0=\Omega_c^0+\Omega_0$, where the BC is $\left . \boldsymbol{U}_0^1 \boldsymbol{\cdot} \widehat{\boldsymbol{n}} \right |_{\Sigma} = 0$. 
Considering now the boundary-layer flows, we first integrate equation (\ref{eq:consmass}) using the BC $ \boldsymbol{u}_0^k(\zeta) \to \boldsymbol{0}$ when $\zeta \to +\infty$ and, since the first term is of order $E^{-{1}/{2}}$, we obtain the zeroth order condition for the boundary-layer flow $\boldsymbol{u}_0^k\boldsymbol{\cdot}{\widehat{\boldsymbol{n}}}=0$ at every order $\epsilon^k$ inside the boundary layer (since $\boldsymbol{V}_{\Sigma}^k \boldsymbol{\cdot}  \widehat{\boldsymbol{n}} = \boldsymbol{v}_{\Sigma}^k \boldsymbol{\cdot}  \widehat{\boldsymbol{n}}  =\boldsymbol{0}$). Then, the boundary-layer equation at the order $\epsilon E^0$ is
\begin{subequations}
\label{eq_CL01DF}
\begin{equation}
\mathcal{L} \boldsymbol{u}_0^1  = - \partial_{\zeta}{p^1_1} \, \widehat{\boldsymbol{n}}, \quad
\boldsymbol{u}_0^1+\boldsymbol{U}_0^1  =  \boldsymbol{V}_{\Sigma}^1 \quad \text{on} \quad \Sigma,
\tag{\theequation \emph{a,b}}
\end{equation}
\end{subequations}
with the linear operator $\mathcal{L}  \boldsymbol{u}_0^1 =  -(\partial_t+\Omega_0 \partial_{{\phi}}) \, \boldsymbol{u}_0^1 - 2 \,  { \Omega}_s^0 \widehat{\boldsymbol{z}}_R \times \boldsymbol{u}_0^1 + \partial_{\zeta \zeta}^2{\boldsymbol{u}_0^1}$.

At the order $\epsilon E^{1/2}$, the so-called Ekman circulation $\boldsymbol{U_1^1}$ is governed by
\begin{equation}
 ( \partial_t + \Omega_0 \partial_{\phi} )\boldsymbol{U}_1^1 +\partial_{\tau} \boldsymbol{U}_0^1 + 2 \, \Omega_s^0 \, \widehat{\boldsymbol{z}}_R   \times \boldsymbol{U}_1^1 = -\boldsymbol{\nabla}P_1^1, \quad
\boldsymbol{u}_1^1+\boldsymbol{U}_1^1 = \boldsymbol{V}_{\Sigma}^1 \quad \text{on} \quad\Sigma,
\label{eq:addon}
\end{equation}
and the divergenceless condition $\boldsymbol{\nabla} \boldsymbol{\cdot} \boldsymbol{U}_1^1 = 0$,
where we have anticipated that $\boldsymbol{U}_1^0$, and thus $\boldsymbol{u}_1^0$, can be set to zero without loss of generality (the flows are forced by the forcing of amplitude $\epsilon$, and we will see that all our equations can be verified with the solution $\boldsymbol{U}_1^0=\boldsymbol{u}_1^0=\boldsymbol{0}$). At this order, the mass conservation imposes
\begin{equation}
  \partial_{ \zeta} (\boldsymbol{u}_1^1
\boldsymbol{\cdot} \widehat{\boldsymbol{n}}) -\widehat{\boldsymbol{n}}
\boldsymbol{\cdot}\boldsymbol{\nabla}\times(\widehat{\boldsymbol{n}}\times\boldsymbol{u}^1_0) = 0, \label{eq:l8gnp}
\end{equation}
which allows us to obtain easily the Ekman pumping $\boldsymbol{u}_1^1={u}_1^1\, \widehat{\boldsymbol{n}}$ from $\boldsymbol{u}_0^1$.

At next order $\epsilon^2 E^0$, the bulk flow is divergenceless $\boldsymbol{\nabla} \boldsymbol{\cdot} \boldsymbol{U}_0^2 = 0$ and given by
\begin{align} 
(\partial_t+\Omega_0 \partial_{{\phi}}) \, \boldsymbol{U}_0^2  + 2\, \Omega_s^0\, \widehat{\boldsymbol{z}}_R \times \boldsymbol{U}_0^2 &= -\nabla P_0^2  - (\boldsymbol{U}_0^1\boldsymbol{\cdot}\boldsymbol{\nabla}) \, \boldsymbol{U}_0^1 -2 \boldsymbol{\Omega}_c^1 \times \boldsymbol{U}_0^1 \nonumber \\ &-2 \boldsymbol{\Omega}_c^2 \times \boldsymbol{U}_0^0 -\dot{\boldsymbol{\Omega}}_c^2 \times \boldsymbol{r}, \label{eq:lf421}
\end{align}
with the BC $\boldsymbol{U}_0^2 \boldsymbol{\cdot} \widehat{\boldsymbol{n}} = 0$ on $\Sigma$.
The boundary-layer equations are
\begin{subequations}
\allowdisplaybreaks
\label{plafplouf32}
\begin{eqnarray}
\mathcal{L}  \boldsymbol{u}_0^2 &=& -\widehat{\boldsymbol{n}}\, \partial_{ \zeta} p^2_1 +(\boldsymbol{u}_0^1\boldsymbol{\cdot}\boldsymbol{\nabla})\boldsymbol{u}_0^1-(\boldsymbol{u}_1^1+\boldsymbol{U}_1^1)\boldsymbol{\cdot}\widehat{\boldsymbol{n}}\, \partial_{ \zeta} \boldsymbol{u}_0^1  +2 \boldsymbol{\Omega}_c^1 \times \boldsymbol{u}_0^1   \nonumber \\
& & \quad +(\boldsymbol{U}_0^1\boldsymbol{\cdot}\boldsymbol{\nabla})\boldsymbol{u}_0^1 + (\boldsymbol{u}_0^1\boldsymbol{\cdot}\boldsymbol{\nabla})\boldsymbol{U}_0^1,  \label{eq:lm0jjhss1}   \\
\boldsymbol{u}_0^2+\boldsymbol{U}_0^2  & = & \boldsymbol{V}_{\Sigma}^2 \quad \text{on} \quad \Sigma . \label{eq:lm0jjhss} 
\end{eqnarray}
\end{subequations}
In the equations (\ref{eq:lf421})-(\ref{plafplouf32}) governing the order $\epsilon^2 E^0 $ , $\boldsymbol{U}_1^1$ and $\boldsymbol{u}_1^1$ only appear via their normal components in the boundary layer. 
Then, equation (\ref{eq:l8gnp}) gives directly $\boldsymbol{u}_1^1\boldsymbol{\cdot}\widehat{\boldsymbol{n}}$, and thus $\boldsymbol{U}_1^1\boldsymbol{\cdot}\widehat{\boldsymbol{n}} =-\boldsymbol{u}_1^1\boldsymbol{\cdot}\widehat{\boldsymbol{n}}|_{\zeta=0} $ via  the BC $(\boldsymbol{u}_1^1+\boldsymbol{U}_1^1) \boldsymbol{\cdot} \widehat{\boldsymbol{n}}=0 $  on $\Sigma$. 

Finally, the interior flow $\boldsymbol{U}_0^k$ at every order $\epsilon^k$ will be decomposed as $\boldsymbol{U}_0^k = {}_P\boldsymbol{U}_0^k  + {}_H\boldsymbol{U}_0^k$, where ${}_P\boldsymbol{U}_0^k$ is the particular solution forced by non-homogeneous terms, and ${}_H\boldsymbol{U}_0^k$ is the solution of the homogeneous part of the equation that is required to satisfy the BC for the total flow. 
We will actually see that ${}_H\boldsymbol{U}_0^1 = \boldsymbol{0}$ in certain cases (e.g. in the fast libration limit $\omega \gg E^{1/2}$). 
Moreover, in all the cases considered here, we will show that the theory gives $\langle \widebar{\boldsymbol{U}_0^2} \rangle= \langle \widebar{{}_P\boldsymbol{U}_0^2} \rangle  + \langle \widebar{{}_H\boldsymbol{U}_0^2} \rangle$ as an azimuthal flow, which provides the leading-order azimuthal component of $\langle \widebar{\boldsymbol{V}} \rangle_\phi$.
We will thus compare the values of $\langle \widebar{\boldsymbol{V}} \rangle_\phi$ obtained from DNS with the theoretical values of $\langle \widebar{\boldsymbol{U}_0^2 }\rangle$. 

\section{Asymptotic analysis}
\label{sec:asymp}
In this section, we aim at calculating the steady axisymmetric component $ \langle \overline{\boldsymbol{U}_0^2} \rangle$ of the interior flow $\boldsymbol{U}_0^2$, which requires the full mathematical expressions of $\boldsymbol{U}_0^1$ , $\boldsymbol{u}_0^1$ and $ \langle \overline{\boldsymbol{u}_0^2} \rangle$. 
To solve the corresponding equations, we employ the spheroidal orthogonal coordinates $(q_1,q_2,{\phi})$, associated with the orthogonal normal unit basis $(\widehat{\boldsymbol{q}}_1,\widehat{\boldsymbol{q}}_2,\widehat{\boldsymbol{\phi}})$ where $\widehat{\boldsymbol{\phi}}$ is the usual azimuthal unit vector. 
We introduce the change of variables
\begin{subequations}
\begin{equation}
    x_R = a \mathcal{T}_{(q_1)}\, \sin q_2 \cos {\phi}, \quad
    y_R = a \mathcal{T}_{(q_1)}\, \sin q_2 \sin {\phi}, \quad
    z_R = a \mathcal{T}^\prime_{(q_1)}\, \cos q_2,
    \tag{\theequation \emph{a--c}}
\end{equation}
\end{subequations}
where $(q_1,q_2,{\phi})$ are spheroidal coordinates, $a=|1-(r_{pol}/r_{eq})^2|^{1/2} = ({\mathcal{T}_{(q_1)}}^2+{\mathcal{T}^\prime_{(q_1)}}^2)^{1/2}$ is the distance between the centre and the foci of the ellipse. For later use, we also define the cylindrical radius $s=(x_R^2+y_R^2)^{1/2}=a \mathcal{T}_{(q_1)} \sin q_2$ and the scale factors $(h_1,h_2,h_{\phi})$ for the coordinates $(q_1,q_2,\phi)$ as
\begin{subequations}
\begin{equation}
    h_1=h_2=a \sqrt{{\mathcal{T}^\prime_{(q_1)}}^2+({\mathcal{T}_{(q_1)}}^2-{\mathcal{T}^\prime_{(q_1)}}^2)\cos^2 q_2} = a \tilde{h}, \quad h_{\phi}=a \mathcal{T}_{(q_1)}\, \sin q_2,
    \tag{\theequation \emph{a,b}}
\end{equation}
\end{subequations}
which gives $\tilde{h}=\sqrt{\sinh^ 2 q_1+\cos^2 q_2} $ and $\tilde{h}=\sqrt{\cosh^ 2 q_1-\cos^2 q_2} $ for oblate and prolate spheroidal coordinates, respectively. 
Note that we recover the usual spherical case with $q_1 \to \infty$, giving for instance $a \tilde{h} \to a \exp(q_1)/2 \approx r $ or $a \mathcal{T}_{(q_1)} \to r $, with the spherical radius $r$ \citep[see e.g.][]{schmitt2004numerical}. 
This definition of the spheroidal coordinates allows us to encompass both oblate and prolate spheroidal coordinates in a single framework, by using respectively $\mathcal{T}_{(q_1)}=\cosh q_1$  when $r_{eq}>r_{pol}$, and $\mathcal{T}_{(q_1)}=\sinh q_1$ otherwise. Here, we note $\mathcal{T}^\prime$ the derivative of $\mathcal{T}$ with respect to $q_1$. 
In spheroidal coordinates, the semi-axes $r_{eq}$ and $r_{pol}$ are given by $r_{eq}=a \mathcal{T}_{(Q_1)} $ and $r_{pol}=a \mathcal{T}^\prime_{(Q_1)}$, where $Q_1$ is the value of the radial-like coordinate $q_1$ at the boundary. 
In appendix \ref{sec:ans1}, we give various useful expressions related to the spheroidal coordinates used in this work.

\subsection{First-order flows}
Considering first the interior flow $\boldsymbol{U}_0^1={}_P\boldsymbol{U}_0^1+{}_H\boldsymbol{U}_0^1$, the particular solutions ${}_P\boldsymbol{U}_0^1$ are usually sought as uniform-vorticity flows because of the spatial dependency of the Poincar\'e term $\dot{\boldsymbol{\Omega}}_c^1 \times \boldsymbol{r}$. For instance, such solutions for ${}_P\boldsymbol{U}_0^1$ in ellipsoids have been successfully obtained for latitudinal libration \citep{vantieghem2015latitudinal}, precession  \citep[e.g.][]{noir2013precession}, and if we consider longitudinal librations in the mantle frame of reference, a natural solution is ${}_P\boldsymbol{U}_0^1 = - \epsilon \cos(\omega t) \, \widehat{\boldsymbol{z}}_R \times \boldsymbol{r}$. 
For the sake of our asymptotic analysis, we consider below a generic uniform-vorticity flow ${}_P\boldsymbol{U}_0^1$ (see equations \ref{eq:mld0cwm}-\ref{eq:besoin} in appendix \ref{sec:ans2}), which encompasses all the various cases. Naturally, ${}_P\boldsymbol{U}_0^1=\boldsymbol{0}$ in absence of non-homogeneous forcing terms, as this is for instance the case in the mean rotating frame for longitudinal librations in the spheroid or latitudinal librations in the sphere.

Considering now ${}_H\boldsymbol{U}_0^1$, the governing equations are then
\begin{subequations}
\label{eq:legi0}
\begin{equation} 
(\partial_t+\Omega_0 \partial_{{\phi}}) \, {}_H\boldsymbol{U}_0^1 + 2\, \Omega_s^0\, \widehat{\boldsymbol{z}}_R \times {}_H\boldsymbol{U}_0^1  = -\nabla P_0^1, \quad \boldsymbol{\nabla} \boldsymbol{\cdot} ( {}_H\boldsymbol{U}_0^1)  = 0,
  \tag{\theequation \emph{a,b}}
\end{equation}
\end{subequations}
which has to be integrated together with the BC $\boldsymbol{u}_0^1+\boldsymbol{U}_0^1  =  \boldsymbol{V}_{\Sigma}^1  $ on $\Sigma$. 
We assume in the mean flow computation below that the spin-up time scale $E^{-1/2}$ of the fluid is much longer than the characteristic time scale of the mechanical forcing (in the fluid rotating frame), which implies ${}_H\boldsymbol{U}_0^1 \to \boldsymbol{0}$ \citep[as in][]{busse2010,sauret2013libration}. 
In appendix \ref{sec:ans22}, we investigate the validity of this assumption, that is how this limit is approached when $\omega/E^{1/2}$ is increased (for the particular case of longitudinal librations). Note also that assuming ${}_H\boldsymbol{U}_0^1 =\boldsymbol{0}$, as in the following, is not valid when bulk flows are generated by the forcing at this order. Considering for instance longitudinal librations \cite[as in][]{aldridge1969axisymmetric}, we detail in appendix \ref{sec:ans23} how the excitation of an inertial mode flow ${}_H\boldsymbol{U}_0^1 \neq \boldsymbol{0}$ can indeed modify the mean zonal flow.

Then, since $\boldsymbol{U}_0^1={}_P\boldsymbol{U}_0^1$ is known, we can solve equations (\ref{eq_CL01DF}) to obtain $\boldsymbol{u}_0^1$. 
The computations of the first-order boundary-layer flow are detailed in appendix \ref{sec:ans3}, but here we only outline the essential steps. 
The pressure term in equation (\ref{eq_CL01DF}a) is usually removed by multiplying the equation by $\widehat{\boldsymbol{n}} \times (\dots)$ and $-\mathrm{i} \widehat{\boldsymbol{n}} \times (\widehat{\boldsymbol{n}} \times \dots)$, which gives $\mathcal{L} ( \widehat{\boldsymbol{n}} \times \boldsymbol{u}_0^1 +  \mathrm{i}\boldsymbol{u}_0^1  )  = 0$, with the imaginary number $\mathrm{i}$. 
While the no-penetration condition $\boldsymbol{u}_0^1  \boldsymbol{\cdot} \widehat{\boldsymbol{n}} =0 $ gives directly the first component of $\boldsymbol{u}_0^1$ as $\boldsymbol{u}_0^1 \boldsymbol{\cdot} \widehat{\boldsymbol{q}}_1=0$, the two other components $\boldsymbol{Y}=(\boldsymbol{u}_0^1 \boldsymbol{\cdot} \widehat{\boldsymbol{q}}_2, \boldsymbol{u}_0^1 \boldsymbol{\cdot} \widehat{\boldsymbol{\phi}})$ can then be obtained by integrating this equation. 
Given the spatio-temporal periodicity of the perturbation, $\boldsymbol{Y}$ is sought as the linear combination $\boldsymbol{Y} = \sum_k \boldsymbol{Y}_k$, where the individual terms $\boldsymbol{Y}_k$ are in the form $\exp[ \mathrm{i}(m_k {\phi} + \omega_k t)] $ to encompass the various sign possibilities.
Equation (\ref{eq_CL01DF}a) then reduces to
\begin{equation} 
\label{eq:mfx1}
\partial^2_{\zeta \zeta} \boldsymbol{Y}_k = \boldsymbol{M} \boldsymbol{Y}_k,
\end{equation}
where the anti-symmetric matrix $\boldsymbol{M}$ reads (noting $\gamma_k=(\omega_k+m_k  \Omega_0)/2$)
\begin{align} \label{eq:gtil}
 \boldsymbol{M} &= \begin{pmatrix} 
2\mathrm{i} \gamma_k & - 2 \gamma_1 \\
2 \gamma_1 & 2\mathrm{i} \gamma_k \\ 
\end{pmatrix} \\
    \gamma_1 &= \Omega_s^0\, \widehat{\boldsymbol{z}}_R \boldsymbol{\cdot} \widehat{\boldsymbol{q}}_1=(\boldsymbol{\Omega}_0+\boldsymbol{\Omega}_c) \boldsymbol{\cdot} \widehat{\boldsymbol{n}}  = \tilde{h}^{-1} \, \Omega_s^0 \, \mathcal{T}_{(q_1)}  \cos q_2 .
\end{align}
where the spherical geometry is recovered for $q_1 \to \infty$, with $\mathcal{T}_{(q_1)}/\tilde{h} \to 1$.

As detailed in appendix \ref{sec:ans3}, the linear system (\ref{eq:mfx1}) can be solved together with the no-slip BC to obtain $\boldsymbol{u}_0^1$. 
The resulting mathematical expression shows that the boundary-layer thickness is singular when $\gamma_1 \pm \gamma_{\pm}=0$, that is
\begin{equation}
\label{eq:lm0f82q}
 \widehat{\boldsymbol{z}}_R \boldsymbol{\cdot} \widehat{\boldsymbol{q}}_1=\pm \frac{\gamma_{\pm}}{\Omega_s^0} = \pm \frac{m \Omega_0 \pm \omega}{2\, \Omega_s^0} ,
\end{equation}
with  $\gamma_{\pm}=(m \Omega_0 \pm \omega)/2$. 
The presence of these singularities shows that boundary-layer theory is not valid at this order of approximation, and their description requires the introduction of new scalings near these so-called critical latitudes \citep[e.g.][]{kida2020steady}. 
The calculations performed in this work are thus strictly valid when $|\omega \pm m|>2 |\Omega_s^0|$, to prevent the generation of internal shear layers 
\citep[and the excitation of inertial waves or modes, e.g.][]{aldridge1969axisymmetric,sauret2013spontaneous}.

\subsection{Weakly nonlinear analysis: second-order bulk flows}
\label{sec:2bulk}
Using the decomposition $\boldsymbol{U}_0^2 = {}_P\boldsymbol{U}_0^2  + {}_H\boldsymbol{U}_0^2$, the average of equations (\ref{eq:lf421}) gives
\begin{subequations}
\label{eq:lmhhuuaa}
\begin{equation}
 2\, \Omega_s^0 \, \widehat{\boldsymbol{z}}_R \times  \langle \overline{{}_H \boldsymbol{U}_0^2} \rangle = -\nabla \langle \overline{{}_H {P}_0^2} \rangle, \quad
\boldsymbol{\nabla} \boldsymbol{\cdot} \langle \overline{{}_H \boldsymbol{U}_0^2} \rangle = 0,
\tag{\theequation \emph{a,b}}
\end{equation}
\end{subequations}
together with BC (\ref{eq:lm0jjhss}). 
Note that ${}_H \boldsymbol{U}_0^2 $ is related to viscous effects, and thus, contrary to ${}_P \boldsymbol{U}_0^2 $, it vanishes when $E=0$ (but is non-zero for $E \ll 1$). 
Equation (\ref{eq:lmhhuuaa}a) admits a solution of the form \citep[e.g.][]{busse1968}
\begin{eqnarray}
   \langle \overline{{}_H \boldsymbol{U}_0^2} \rangle   & = & s f(s)\, \widehat{\boldsymbol{\phi}},
\end{eqnarray}
where the rotation rate $f(s)$ of the mean zonal flow has to be determined. 
Considering now ${}_P\boldsymbol{U}_0^2$, the inhomogeneous forcing term in (\ref{eq:lf421}) is linear in the Cartesian coordinates $[x,y,z]$, such that we can seek ${}_P \boldsymbol{U}_0^2$ as a uniform-vorticity flow. For all the forcings considered in this work, the time average of $\langle {}_P \boldsymbol{U}_0^2\rangle$
can be written as $s g\, \widehat{\boldsymbol{\phi}}$, such that the mean zonal flow reduces to
\begin{equation}
   \langle \overline{\boldsymbol{U}_0^2} \rangle = s \, [f(s)+g ] \, \widehat{\boldsymbol{\phi}},
   \label{eq:Uphisfg}
\end{equation}
where $g$ is a constant, found to be $g=0$ in all cases studied here.

At the order $\epsilon^2 E^{1/2}$, the mean zonal component of the bulk flow $\boldsymbol{U}_1^2$ is governed by
\begin{eqnarray}
(\widehat{\boldsymbol{z}}_R \boldsymbol{\cdot} \boldsymbol{\nabla})  \langle \overline{\boldsymbol{U}_1^2}  \rangle & = & 0,
\end{eqnarray}
which is actually the Taylor-Proudman theorem \citep{greenspanbook}. 
It implies that the flux ejected out of the boundary layer through the interior (which is symmetric with respect to the axis and anti-symmetric with respect to the equatorial plane) vanishes at every distance from the axis, such that $\widehat{\boldsymbol{n}} \boldsymbol{\cdot} \boldsymbol{u}_1^2 |_{\zeta=0} = 0$ at the order $\epsilon^2\,E^{1/2}$. 
Moreover, the continuity equation at order $\epsilon^2 \, E^0$ reads
\begin{eqnarray}\label{conti}
\partial_{\zeta}(\boldsymbol{u}_1^2
\boldsymbol{\cdot} \widehat{\boldsymbol{n}})  =\widehat{\boldsymbol{n}}
\boldsymbol{\cdot}\boldsymbol{\nabla}\times(\widehat{\boldsymbol{n}}\times\boldsymbol{u}^2_0) .
\end{eqnarray}
Finally, integrating equation (\ref{conti}) between $\zeta=0$ and $\zeta \to +\infty$ yields \citep[e.g.][]{busse2010}
\begin{eqnarray}\label{condition}
\widehat{\boldsymbol{n}} \boldsymbol{\cdot} \boldsymbol{u}_1^2 |_{\zeta=0} & = & -\widehat{\boldsymbol{n}}
\boldsymbol{\cdot} \boldsymbol{\nabla}\times \int_{0}^{+\infty}\,\widehat{\boldsymbol{n}}\times\boldsymbol{u}^2_0 \, \text{d}\zeta = 0,
\end{eqnarray}
which is used to determine the unknown function $f(s)$.

\subsection{Weakly nonlinear analysis: second-order boundary-layer flows}
\label{sec:weakflows}
To obtain $ \langle \overline{\boldsymbol{u}}_0^2 \rangle$, we separate equation (\ref{plafplouf32}a) in three distinct problems by considering three distinct velocity fields $\boldsymbol{u}_{A}$, $\boldsymbol{u}_{B}$ and $\boldsymbol{u}_{C}$ such that $ \langle \overline{\boldsymbol{u}_0^2}  \rangle=\boldsymbol{u}_{A}+\boldsymbol{u}_{B}+\boldsymbol{u}_{C} $. 
In the first problem, we seek a velocity field $\boldsymbol{u}_{A}$ satisfying the homogeneous equations and the inhomogeneous BC, that is
\begin{subequations}
\label{eq:uA}
\begin{equation}
\boldsymbol{\mathcal{H}} \boldsymbol{u}_{A}+\widehat{\boldsymbol{n}}\, \partial_{\zeta} \Phi^2_1 = 0, \quad     \boldsymbol{u}_{A}+ \langle \overline{\boldsymbol{U}_0^2} \rangle =  \boldsymbol{0} \quad \text{on} \quad \Sigma,
\tag{\theequation \emph{a,b}}
\end{equation}
\end{subequations}
where we have defined the linear operator $\boldsymbol{\mathcal{H}} \boldsymbol{u}_0^2 = \langle \overline{ \mathcal{L} \boldsymbol{u}_0^2}  \rangle$, using the operator ${\cal L}$ defined below equation (\ref{eq_CL01DF}).
Then, we seek the velocity fields $\boldsymbol{u}_{B}$ and $\boldsymbol{u}_{C}$ that satisfy the homogeneous BC $\boldsymbol{u}_{B}  = \boldsymbol{u}_{C}  =  \boldsymbol{0}$ at $\zeta=0$ and the inhomogenous equations given by
\begin{eqnarray}\label{eq:uB}
      \boldsymbol{\mathcal{H}} \boldsymbol{u}_{B}+\widehat{\boldsymbol{n}}\,\partial_{\zeta} \Phi^2_1 &=&  \left \langle \overline{{(\boldsymbol{u}_0^1\boldsymbol{\cdot}\boldsymbol{\nabla}})\boldsymbol{u}_0^1+2 \boldsymbol{\Omega}_c^1 \times \boldsymbol{u}_0^1 +(\boldsymbol{U}_0^1\boldsymbol{\cdot}\boldsymbol{\nabla})\boldsymbol{u}_0^1 + (\boldsymbol{u}_0^1\boldsymbol{\cdot}\boldsymbol{\nabla})\boldsymbol{U}_0^1 }    \right \rangle, \\
    \boldsymbol{\mathcal{H}} \boldsymbol{u}_{C}+\widehat{\boldsymbol{n}}\,\partial_{\zeta} \Phi^2_1 &=&  \left \langle \overline{-(\boldsymbol{u}_1^1+\boldsymbol{U}_1^1)\boldsymbol{\cdot}\widehat{\boldsymbol{n}}\,\partial_{\zeta} \boldsymbol{u}_0^1} \right \rangle. \label{eq:uC}
   \end{eqnarray}
For tidal forcing, \citet{suess1971} claimed erroneously that the term $(\boldsymbol{u}_1^1 + \boldsymbol{U}_1^1)\boldsymbol{\cdot}\widehat{\boldsymbol{n}}\,\partial_{\zeta}{\boldsymbol{u}_0^1}$ vanishes in equation (\ref{eq:uC}), such that the contribution of $\boldsymbol{u}_C$ could be discarded. 
This would be correct if the normal velocity were zero at all orders, but this term only vanishes at the boundary and not everywhere in the boundary layer.
We will instead demonstrate that a non-zero $\boldsymbol{u}_C$ is required to balance the singularity of $\boldsymbol{u}_B$ on the rotation axis.

The equations governing $\boldsymbol{u}_A$ are formally similar to boundary-layer equations. 
Similarly, we obtain in the spheroidal coordinates (noting $\lambda=[1+\mathrm{i}\, \mathrm{sgn}(\gamma_1)]\,\sqrt{|\gamma_1|}$)
\begin{equation}
\label{defua}
\boldsymbol{u}_A
= -\frac{s}{2}\,f(s)
\begin{pmatrix}
 0  \\ 
\mathrm{i}\, ( \mathrm{e}^{-\lambda\,\zeta}-\mathrm{e}^{-\lambda^*\,\zeta} ) \\
 \mathrm{e}^{-\lambda\,\zeta}+\mathrm{e}^{-\lambda^*\,\zeta} \\
\end{pmatrix}.
\end{equation}

The calculation of $\boldsymbol{u}_{B}$ is more laborious. After some algebra, equation (\ref{eq:uB}) reduces to the following scalar equation 
   \begin{eqnarray}\label{plaf}
  ( \partial^2_{\zeta \zeta} - \lambda^2 ) F_0 &=&  \mathcal{E}
 ,  \end{eqnarray}
where
\begin{equation}
\mathcal{E} = \left \langle \overline{{(\boldsymbol{u}_0^1\boldsymbol{\cdot}\boldsymbol{\nabla}}) \, \boldsymbol{u}_0^1+2 \boldsymbol{\Omega}_c^1 \times \boldsymbol{u}_0^1 + (\boldsymbol{U}_0^1\boldsymbol{\cdot}\boldsymbol{\nabla}) \, \boldsymbol{u}_0^1 + (\boldsymbol{u}_0^1\boldsymbol{\cdot} \boldsymbol{\nabla}) \, \boldsymbol{U}_0^1 }    \right \rangle \boldsymbol{\cdot} ( \widehat{\boldsymbol{q}}_2  +\mathrm{i}\,  \widehat{\boldsymbol{\phi}}) ,
\end{equation}
with $F_0  = \boldsymbol{u}_{B} \boldsymbol{\cdot} ( \widehat{\boldsymbol{q}}_2  +\mathrm{i}\,  \widehat{\boldsymbol{\phi}} )$ and $ \lambda^2=2\,\mathrm{i}\, \gamma_1$. 
To calculate $\mathcal{E}$, we first consider each term separately, that is ${(\boldsymbol{u}_0^1\boldsymbol{\cdot}\boldsymbol{\nabla}}) \, \boldsymbol{u}_0^1$, $2 \boldsymbol{\Omega}_c^1 \times \boldsymbol{u}_0^1$, $(\boldsymbol{U}_0^1\boldsymbol{\cdot}\boldsymbol{\nabla}) \, \boldsymbol{u}_0^1$, and $ (\boldsymbol{u}_0^1\boldsymbol{\cdot}\boldsymbol{\nabla}) \, \boldsymbol{U}_0^1 $. Then, we decompose each term as a constant term, which contributes to the mean zonal average, and terms proportional to $\exp(\pm 2 \mathrm{i} \omega t)$, $\exp(\pm 2 \mathrm{i} m {\phi})$ and $\exp(\pm 2 \mathrm{i}(m {\phi} \pm \omega t))$ that only contribute to the average if $m=0$ or $\omega=0$. 

Considering each term of $\mathcal{E}$ separately, the problem is made simpler by making the $\zeta$ dependency explicit, that is by rewriting the equation as
   \begin{eqnarray} \label{eq:eurre}
   ( \partial^2_{\zeta \zeta} - \lambda^2 ) F_0 &=& \sum_k \left(\varkappa_k+\vartheta_k \, \zeta \right) \mathrm{e}^{-\varsigma_k \, \zeta},
  \end{eqnarray}
where the complex coefficients $(\varkappa_k,\vartheta_k)$ and $\varsigma_k$ (which is a linear combination of $\lambda_{\pm}$, $\kappa_{\pm}$ and their complex conjugates, see equation \ref{eq:dclmp2}) are independent of $\zeta$. We can then integrate equation (\ref{eq:eurre}) by considering each term of the sum separately, which gives
\begin{equation}
   \label{eq:eurre2}
   F_0 =  \sum_k \,  \frac{ \mathrm{e}^{- \varsigma_k \, \zeta} [(\varsigma_k^2-\lambda^2)\left(\varkappa_k+\vartheta_k \, \zeta \right)+2 \vartheta_k \varsigma_k]- \mathrm{e}^{-\lambda \, \zeta} [(\varsigma_k^2-\lambda^2) \varkappa_k+2 \vartheta_k \varsigma_k ]}{(\varsigma_k^2-\lambda^2)^2} ,
\end{equation}
and then $\boldsymbol{u}_B$ can be obtained using $ \boldsymbol{u}_{B} \boldsymbol{\cdot}  \widehat{\boldsymbol{q}}_2= \Re_e({F_0})$ and $\boldsymbol{u}_{B} \boldsymbol{\cdot}  \widehat{\boldsymbol{\phi}} = \Im_m({F_0}) $. 
   
Similarly, the calculation of $\boldsymbol{u}_C$ can be reduced to the integration of 
\begin{equation}
\label{kljq036}
    ( \partial^2_{\zeta \zeta} - \lambda^2 )  H_0 =  \mathcal{F}
\end{equation}
with $ H_0 = \boldsymbol{u}_{C} \boldsymbol{\cdot} ( \widehat{\boldsymbol{q}}_2  +\mathrm{i}\,  \widehat{\boldsymbol{\phi}} )$ and where the right-hand side is given by
\begin{equation}
   \mathcal{F}  = -  \left \langle \overline{ (\boldsymbol{u}_1^1 + \boldsymbol{U}_1^1) \boldsymbol{\cdot}\widehat{\boldsymbol{n}} \, \partial_{\zeta} [\boldsymbol{u}_0^1 \boldsymbol{\cdot} ( \widehat{\boldsymbol{q}}_2  + \mathrm{i}\,  \widehat{\boldsymbol{\phi}})]} \right \rangle .
\end{equation}
The calculation of $\mathcal{F}$ requires the expression of $(\boldsymbol{u}_1^1+\boldsymbol{U}_1^1)\boldsymbol{\cdot} \widehat{\boldsymbol{n}}$. 
The Ekman pumping $\boldsymbol{u}_1^1 \boldsymbol{\cdot} \widehat{\boldsymbol{n}}$ is obtained using the continuity equation
\begin{equation}
- \partial_{ \zeta} (\boldsymbol{u}_1^1
\boldsymbol{\cdot} \widehat{\boldsymbol{n}}) + \widehat{\boldsymbol{n}}
\boldsymbol{\cdot}\boldsymbol{\nabla}\times(\widehat{\boldsymbol{n}}\times \boldsymbol{u}_0^1) = 0. \label{eq:consmass2}
\end{equation}
Using the expression of $\boldsymbol{u}_0^1$, we then obtain $\boldsymbol{u}_1^1 \boldsymbol{\cdot} \widehat{\boldsymbol{n}} $ by integration. 
Together with the Ekman pumping $\boldsymbol{u}_1^1$, an Ekman (bulk) circulation $\boldsymbol{U}_1^1$ is generated via the no-penetration of the fluid at the boundary, such that $(\boldsymbol{u}_1^1+\boldsymbol{U}_1^1) \boldsymbol{\cdot} \widehat{\boldsymbol{n}}=0$ at $\zeta=0$. 
We thus obtain $\boldsymbol{U}_1^1\boldsymbol{\cdot}\widehat{\boldsymbol{n}} =-\boldsymbol{u}_1^1\boldsymbol{\cdot}\widehat{\boldsymbol{n}}|_{\zeta=0}$.
Using a similar procedure for $\boldsymbol{u}_B$, we can now calculate the analytical expression of $\mathcal{F}$ by considering the terms contributing to the average, in particular when $\omega=0$ or $m=0$. We obtain similarly $H_0$, and thus $\boldsymbol{u}_C$, by summing all the solutions.

\subsection{Mean axisymmetric zonal flow}
\label{sec:2mean}
Having explicitly obtained $\boldsymbol{u}_{B}$ and $\boldsymbol{u}_{C}$ in section \ref{sec:weakflows}, one now use equation (\ref{condition}), 
\begin{equation}
\label{lastleast0}
\widehat{\boldsymbol{n}} \boldsymbol{\cdot} \boldsymbol{\nabla}\times \int_0^{+\infty} \, \widehat{\boldsymbol{n}} \times (\boldsymbol{u}_{A} + \boldsymbol{u}_{B} + \boldsymbol{u}_{C}) \, \mathrm{d} \zeta = 0,
\end{equation}
to obtain the unknown rotation rate $f(s)$ present in $\boldsymbol{u}_{A}$. Therefore, using the expression of $\boldsymbol{u}_A$ given by equation (\ref{defua}), we obtain
\begin{eqnarray}\label{lastleast}
\frac{\partial}{\partial q_2} \left(-\frac{\mathrm{sgn}(\gamma_1) \, a \, \mathcal{T}_{(q_1)} \sin^2 q_2}{2\sqrt{|\gamma_1|}} f(s) + \int_0^{+\infty} \sin q_2 \, [\Re_e(F_0)+\Re_e(H_0)] \, \mathrm{d} \zeta  \right) = 0 ,
\end{eqnarray}
which gives
\begin{equation}
\label{moimoi}
f(s) = \frac{2\sqrt{|\gamma_1|}}{\mathrm{sgn}(\gamma_1) \, a \, \mathcal{T}_{(q_1)} \sin q_2} \, \Re_e \left(\int_0^{\infty} [ F_0 + H_0] \, \mathrm{d} \zeta \right)
\end{equation}
where the integration constant has to be taken equal to $0$ to avoid the divergence of the zonal flow when $s\to 0$. From a practical point of view, one can notice that the primitive function $\mathcal{G}$ of $F_0+H_0$ tends to $0$ for $\zeta \rightarrow \infty$ in order to ensure a zero flux at $\zeta=\infty$, such that equation (\ref{moimoi}) simplifies into
\begin{equation}
\label{moimoi2}
f(s) = - \frac{2\sqrt{|\gamma_1|}}{\mathrm{sgn}(\gamma_1) \, s} \, \Re_e \left( \mathcal{G}_{(\zeta=0)} \right),
\end{equation}
which gives the axisymmetric mean zonal flow through equation (\ref{eq:Uphisfg}).

\section{Results}
\label{sec:results}
In the DNS, we find that the geostrophic flows are produced in an $\mathcal{O}(E^{-1/2})$ interval of time, where $E^{-1/2}$ is the spin-up time scale \citep{greenspanbook}. 
Therefore, for the sake of numerical convergence, we have first integrated the nonlinear equations during a few spin-up times, and then time-averaged the flows over a few tens of forcing periods $2\pi/\omega$ to extract the mean geostrophic component from the three-dimensional velocity field. 

\subsection{Longitudinal librations}
\begin{figure}
	\centering
	\begin{tabular}{cc}
	\subfigure[]{\includegraphics[width=0.49\textwidth]{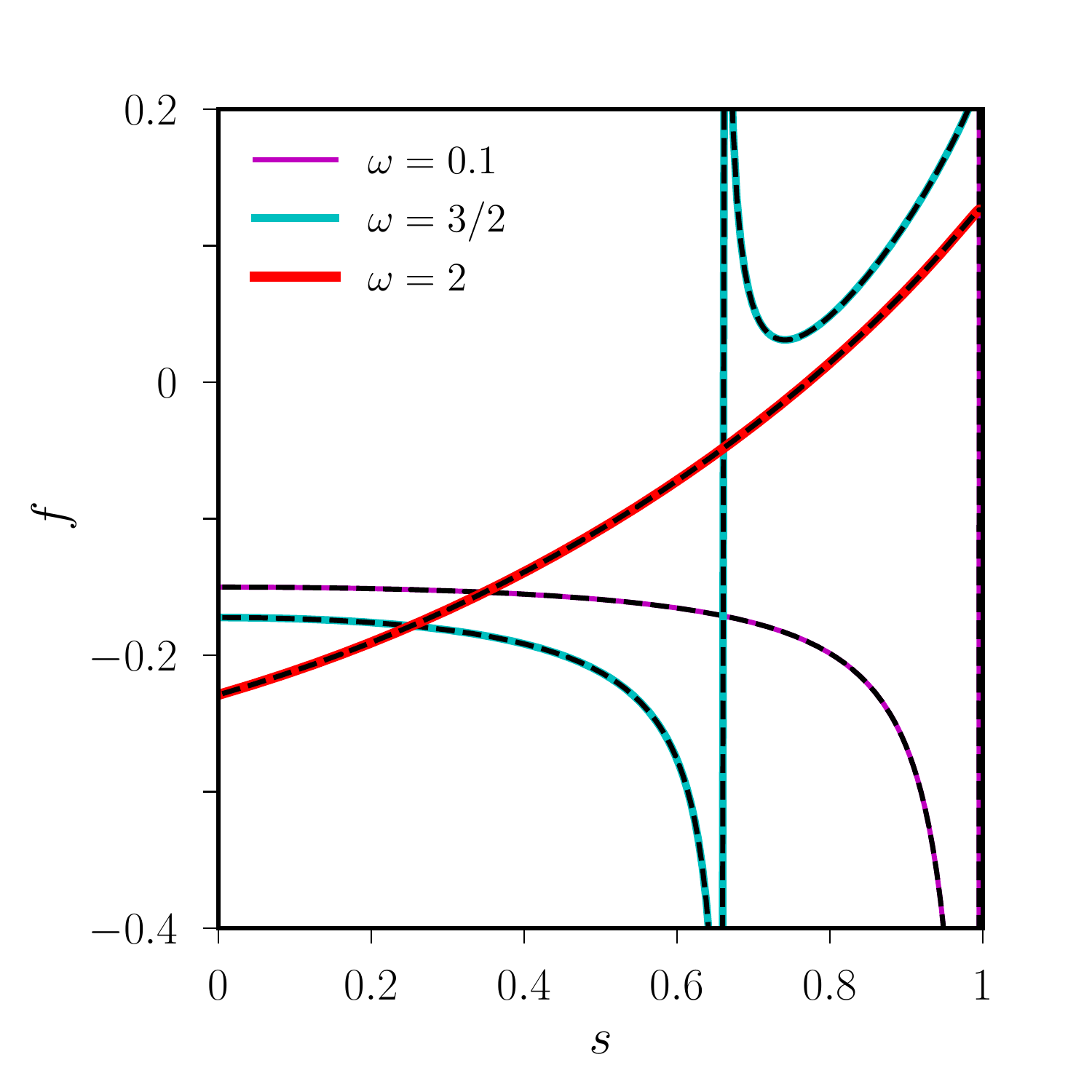}} &
	\subfigure[]{\includegraphics[width=0.49\textwidth]{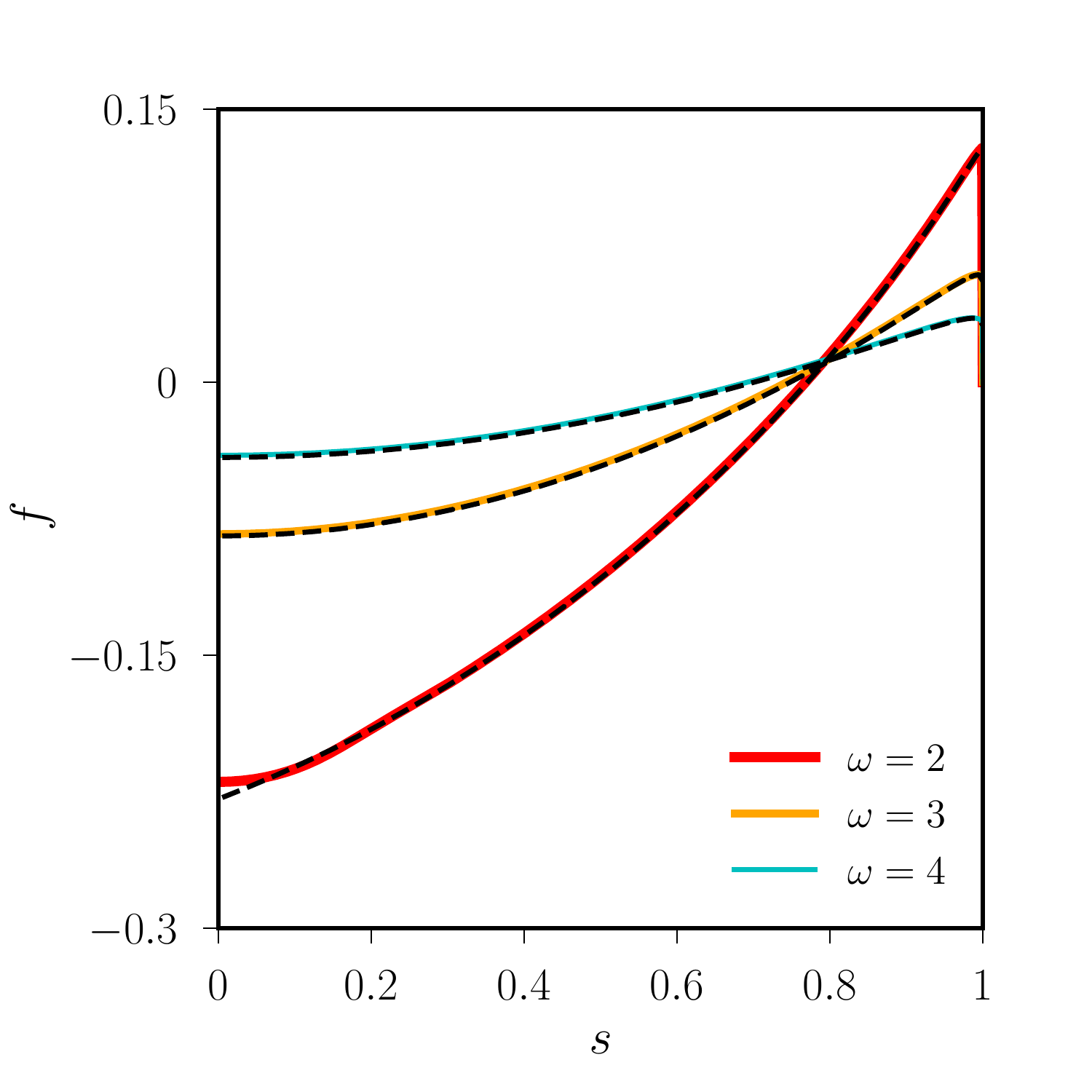}} \\
	\end{tabular}
	\caption{Rotation rate of the mean zonal flow, as a function of the cylindrical radius $s$, in rotating spheres subject to weak longitudinal librations (with $\Omega_0=0$, $m=0$). (a) Our analytical results (solid coloured curves) are compared with the theoretical results (black dashed curves) of \citet{sauret2013libration}. 
	(b). Comparison between the theory (black dashed curves) and DNS (solid coloured curves) at $E=10^{-7}$ and $\epsilon=10^{-4}$ when $\omega \geq 2$.}
	\label{fig:sauret}
\end{figure}

We consider weak longitudinal librations in the mean rotating frame with $\boldsymbol{\Omega}_c=\widehat{\boldsymbol{z}}_R$, $\boldsymbol{U}=\boldsymbol{0}$, and $\boldsymbol{V}_\Sigma = \epsilon \cos(\omega t) \,  \widehat{\boldsymbol{z}}_R \times \boldsymbol{r}$. 
In this reference frame, the zonal flow has first been studied theoretically in the sphere by \citet{busse2010} in the limit of vanishing libration frequency $\omega \rightarrow 0$. 
Using a mathematical description in terms of a stream function, \citet{sauret2013libration} extended the spherical theory to spherical shells and with an arbitrary libration frequency (but still neglecting the shear layers). 
To avoid the presence of critical shear layers, we only present here results for libration frequencies $|\omega| \geq 2$, 
The shear layers indeed excite inertial waves occurring when $|\omega| < 2$, as obtained from (\ref{eq:lm0f82q}), and modify the zonal flow \citep[][in the cylindrical geometry]{sauret2012fluid}. 
We obtain an excellent quantitative agreement with the results of \citet{sauret2013libration} in a full sphere, as shown in figure \ref{fig:sauret}(a), which validates our analytical theory. 
We naturally obtain the same results when calculating the theory in the mantle frame, in which the three last terms of equation \ref{eq:lm0jjhss1} are now non zero (their contributions to $f$ balance each other).

Since our theoretical approach closely follows the one of \citet{busse2010}, we aim at comparing our results with Busse's theoretical zonal flow, which surprisingly differs from the one obtained by \citet{sauret2013libration} for the full sphere librating at $\omega \to 0$. 
Indeed, \citet{busse2010} and \citet{sauret2013libration} obtained respectively in this regime
\begin{subequations}
\label{eq:busseZON}
\begin{equation}
    f(s)= \frac{51.8 \, s^2-72}{480\, (1-s^2)} \quad \text{and} \quad f(s) = \frac{59 \, s^2-72}{480\, (1-s^2)},
    \tag{\theequation \emph{a,b}}
\end{equation}
\end{subequations}
which were illustrated in figure \ref{fig:intro}(a). 
The two profiles are indistinguishable near the rotation axis, and are actually in overall good agreement with the experimental and numerical results of \citet{sauret2010} for small but finite values of $\omega \ll 1$ 
Nevertheless, as already noticed by \citet{sauret2013libration}, the two expressions differ significantly when $s>0.7$. 
The latter authors attributed this difference to their assumption $\epsilon \ll \omega \ll 1$, supposedly different from the assumption $\omega \ll \epsilon$ of \citet{busse2010}. 
Actually, our asymptotic theory follows closely \citet{busse2010}, but our results are in exact agreement with the zonal flow profile of \citet{sauret2013libration} as shown for $\omega=0.1$ in figure \ref{fig:sauret}(a). 
We have thus investigated the origin of this intriguing discrepancy by replicating step by step the calculations of \citet{busse2010}. 
We found that his equations are correct, contrary to his integration of the weakly nonlinear inhomogeneous equations, i.e. equations (A5)-(A7) are erroneous.
Performing the calculations of \citet{busse2010}
with a computer algebra system gives indeed (\ref{eq:busseZON}b) in the relevant limit $\omega \to 0$. To further assess the validity of the asymptotic theory, we have also performed DNS with \textsc{xshells} in the same frame of reference, rotating at $\boldsymbol{\Omega}_c=\widehat{\boldsymbol{z}}_R$. Considering extremely small viscosity and forcing amplitude (i.e. $E=10^{-7}$, $\epsilon=10^{-4}$), the numerical flows agree very well with the theoretical predictions (figure \ref{fig:sauret}b). 
This clearly confirms the agreement already obtained by \citet{sauret2013libration} at more moderate parameters. 
For $\omega=2$, note the small discrepancy at $s=0$, due to the presence of the critical latitude.

\begin{figure}
	\centering
	\begin{tabular}{cc}
		\includegraphics[width=0.49\textwidth]{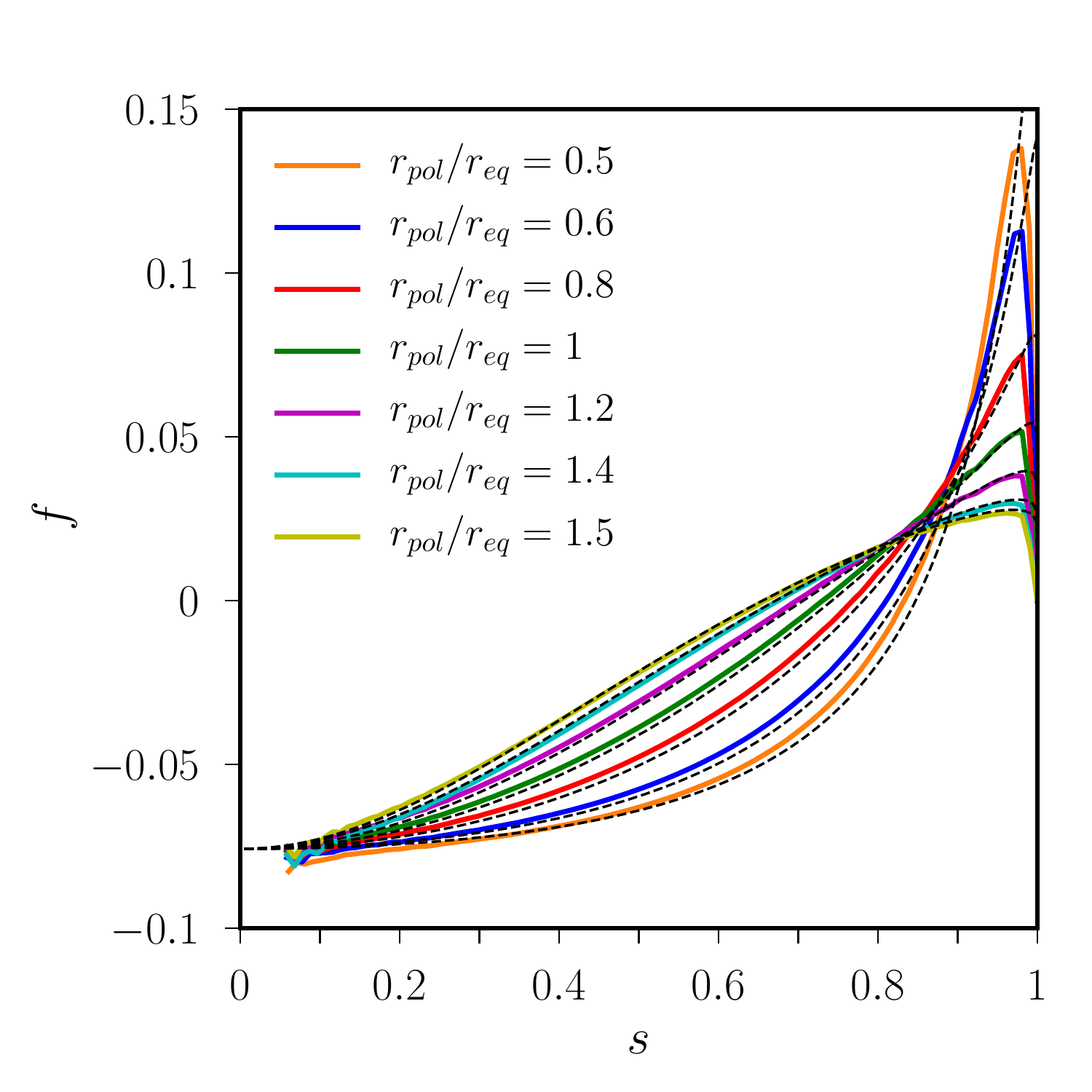} &
		\includegraphics[width=0.49\textwidth]{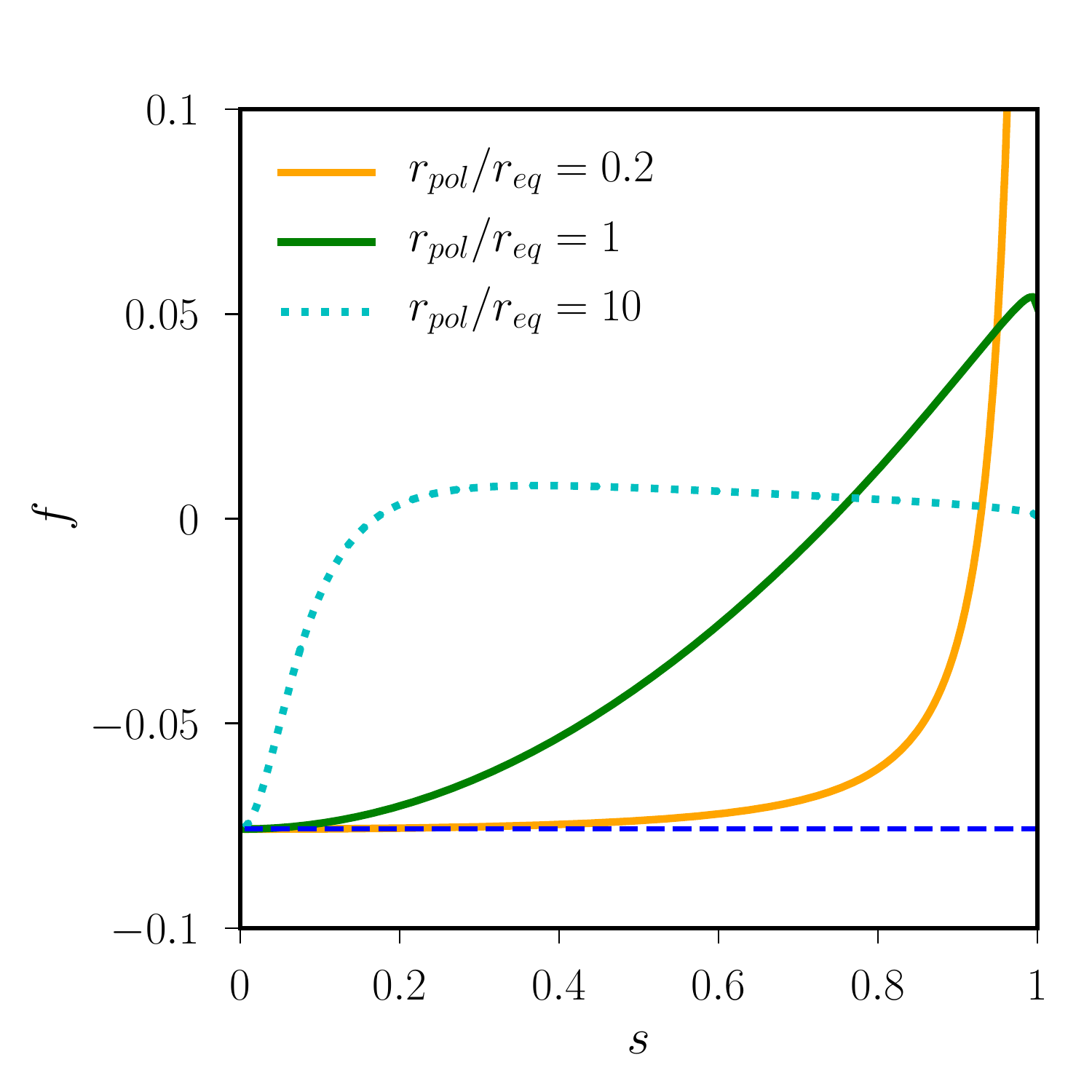} \\
		(a) & (b) \\
	\end{tabular}
	\caption{Rotation rate of the mean zonal flow, as a function of the cylindrical radius $s$, in rotating spheroids subject to longitudinal librations with $\omega = \pi$. (a) Comparison between DNS with $E=10^{-5}$ and $\epsilon=5 \times 10^{-4}$ (solid coloured curves) and theory (dashed black curves). 
	Numerical profiles have been truncated for $s\leq 0.05$, because there are not enough grid points near the centre to get reliable numerical averages with formula (\ref{eq:zonalNek5000}) for the rotation rate. 
	(b) Comparison of the theoretical rotation rate between the sphere, oblate spheroids ($r_{pol}/r_{eq}=0.2$), and prolate spheroids  ($r_{pol}/r_{eq}=10$). Horizontal dashed blue curve indicates the theoretical profile for the cylinder \citep{wang1970cylindrical,sauret2012fluid}.}
	\label{fig:libspheroid}
\end{figure}

Finally, we investigate how the zonal flows are modified in spheroids. Rapidly rotating planetary bodies are indeed deformed into ellipsoids due to centrifugal deformations, and several laboratory experiments have been designed such as the ZoRo experiment \citep{su2020acoustic,vidal2020compressible} with $r_{pol}/r_{eq} = 0.95$. 
We perform DNS with \textsc{Nek5000} in the mantle frame of reference (where the boundary velocity is zero), and present the mean zonal flows obtained from spheroidal DNS at $E=10^{-5}$ for various values of the ratio $r_{pol}/r_{eq}$ in figure \ref{fig:libspheroid}(a).
We also compare the results to the theoretical profiles that have been obtained in spheroidal coordinates. 
Overall, we find a good quantitative agreement, even if the DNS have not been performed in the regime $E \ll 1$. 
The numerical results convincingly validate our asymptotic theory of libration-driven zonal flows in spheroids. 
It is worth noting that significant departure from the spherical profile is found, even for moderate spheroidal deformations $r_{pol}/r_{eq} \leq 1$ as often considered experimentally \citep[e.g. $0.7$ in][]{grannan2016tidally}. 
The theory also allows us to explore more extreme spheroidal configurations that cannot be simulated numerically, as illustrated in figure \ref{fig:libspheroid}(b).
Two points are worthy of comments. 
We find that the mean zonal flow reaches a constant value at $s=0$.
The latter value actually corresponds to the constant profile obtained in the cylinder \citep{wang1970cylindrical,sauret2012fluid}, which gives a lower bound for $f$. 
In the interior $0 < s < 1$, $f$ tends again to the cylindrical value in the disc limit, that is $r_{pol}/r_{eq} \to 0$, whereas it vanishes in the infinite cylinder limit $r_{pol}/r_{eq} \to \infty$.  
Moreover, our results illustrate that the cylindrical geometry cannot be faithfully used as a reduced model of the spheroid. 
Therefore, results obtained in a cylindrical geometry should be interpreted with caution for planetary applications.

\subsection{Latitudinal librations}
We now consider the flows driven by latitudinal librations, which have only received scant attention so far
\citep[][]{chan2011simulationsliblat,zhang2012asymptotic,vantieghem2015latitudinal}.
In particular, the mean zonal flows have only been computed numerically at moderate values of $E$ \citep{chan2011simulationsliblat}, and never compared to theoretical predictions. 
Temporal and spatial perturbations must be indeed considered simultaneously, respectively at the frequency $\omega$ and at the azimuthal wavenumber $m=1$. 
This approach contrasts with previous theoretical studies of zonal flows, where only one kind of perturbations was considered, and fully justifies the generic theoretical framework presented in \S\ref{sec:asymp}.

\begin{figure}
	\centering
	\begin{tabular}{cc}
		\includegraphics[width=0.49\textwidth]{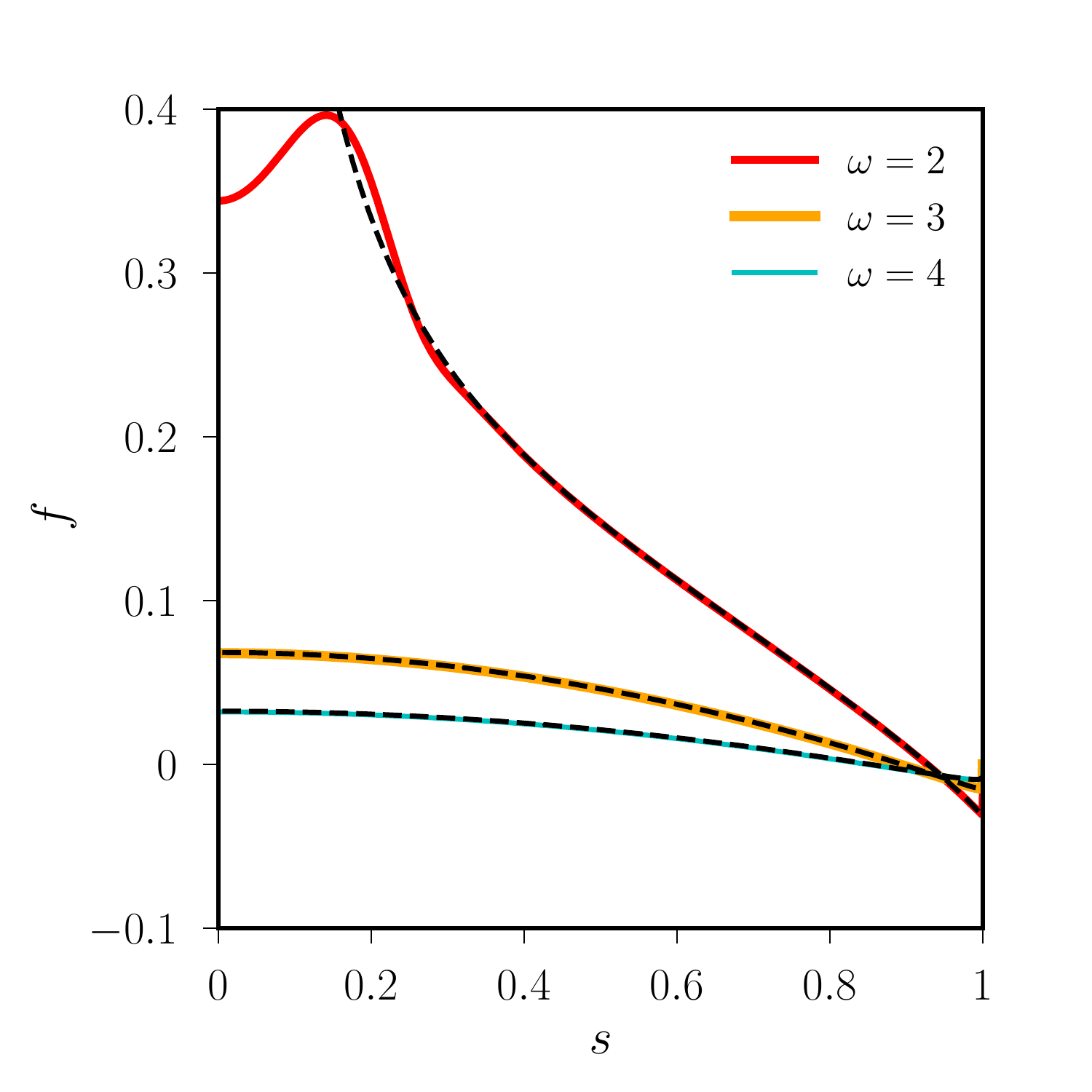} &
		\includegraphics[width=0.49\textwidth]{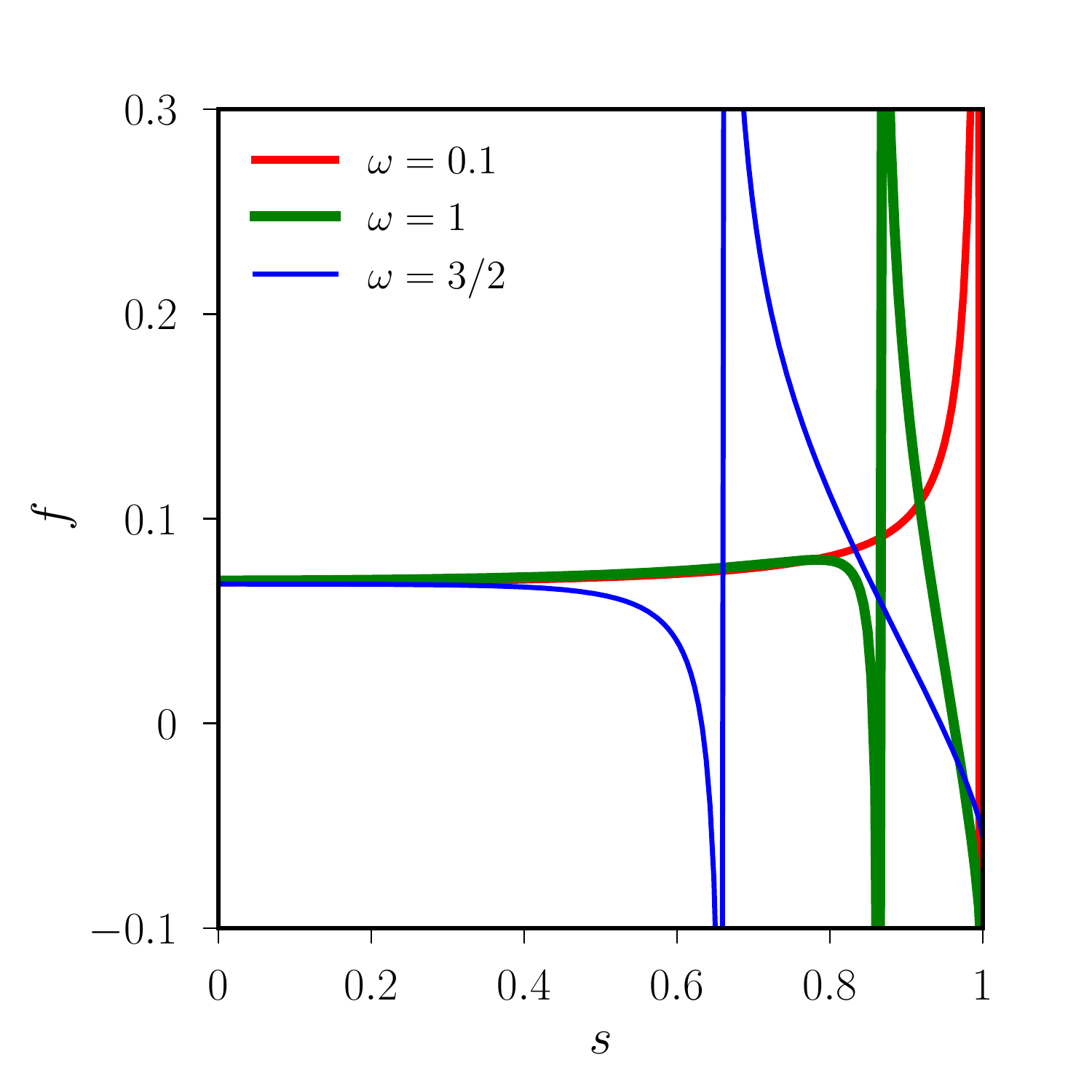} \\
		(a) & (b) \\
	\end{tabular}
	\caption{Rotation rate of the mean zonal flow, as a function of the cylindrical radius $s$, for a rotating sphere subject to weak latitudinal librations. 
	(a) Comparison between theory (black dashed curves) and DNS (coloured solid curves) at $E=10^{-7}$ and $\epsilon=10^{-4}$ in the regime $\omega \geq 2$.
	(a) Theoretical zonal flows when $\omega <2$  with divergent flows at the critical latitudes.}
	\label{fig:xshells_liblat}
\end{figure}

We consider for simplicity the spherical geometry, and we perform our analytical and numerical calculations in the mean rotating frame with $\boldsymbol{\Omega}_c = \widehat{\boldsymbol{z}}_R= \widehat{\boldsymbol{z}}_I$, $\boldsymbol{V}_\Sigma = \epsilon \cos(\omega t)\, \widehat{\boldsymbol{x}}_R \times \boldsymbol{r}$, and $\boldsymbol{U}=\boldsymbol{0}$. 
Using our asymptotic approach, we uncover the theoretical zonal flow associated with this forcing in the relevant limit of vanishing viscosity.
We compare the associated theoretical profiles with DNS in figure \ref{fig:xshells_liblat}(a).
We obtain an excellent agreement for the three different libration frequencies and for very small perturbation and viscosity ($E=10^{-7}, \epsilon=10^{-4}$) in the regime $\omega \geq 2$ (i.e. without critical latitudes).
Note that the rotation rate is always regular at $s=0$ in the DNS \citep[as mathematically expected from][see also appendix \ref{sec:jvreg}]{lewis1990physical}, but our theoretical profile diverges at $s=0$ for $\omega=2$.
Indeed, the mathematical singularity associated with the critical latitude is located on the rotation axis for $\omega=2$ (see equation \ref{eq:lm0f82q}). 
This mathematical singularity is smoothed out by viscosity in the DNS but, to regularise our asymptotic theory and obtain a regular rotation rate profile everywhere in space, additional viscous effects \citep[e.g.][]{kida2011steady} should be taken into account at the critical latitudes. 
Finally, we can explore with the theory how the rotation rate evolves with the libration frequency. 
Similarly to figure \ref{fig:sauret} for longitudinal librations, we illustrate in figure \ref{fig:xshells_liblat}(b) the theoretical zonal flows for various libration frequencies in the particular regime $\omega <2$ (where the theory may not be valid, which will be further discussed in \S\ref{sec:discussion}). 
Even if higher-order viscous effects are expected to smooth out the singularity, one can already notice that, at this order, the width of the divergence seems to increase as $\omega \to 2$.
The influence of the critical latitudes on the zonal flows is further discussed below.

\subsection{Precession-driven zonal flows}
\label{sec:precd}
In his seminal work, \citet{busse1968} considered a precessing sphere and found that the first-order bulk flow is a solid-body rotation $\boldsymbol{\omega}_f \times \boldsymbol{r}$, tilted from the boundary rotation vector $\Omega_0  \, \widehat{\boldsymbol{z}}_R$. 
Having shown that $\epsilon^2=(\Omega_0 \, \widehat{\boldsymbol{z}}_R-\boldsymbol{\omega}_f)^2=\Omega_0^2-\omega_f^2$, he showed that the component of $\boldsymbol{\Omega}_c$ normal to ${\boldsymbol{\omega}}_f$ is of the order $\epsilon E^{1/2}$, which can thus be neglected at the order of the mean zonal flow calculation. 
To calculate the mean zonal flow, he then neglected $\boldsymbol{\Omega}_c \boldsymbol{\cdot} {\boldsymbol{\omega}}_f $ with respect to $\omega_f^2=|{\boldsymbol{\omega}}_f |^2$ for simplicity, yielding finally $\boldsymbol{\Omega}_c = \boldsymbol{0}$ (his mean zonal flow results are thus obtained in the inertial frame of reference). 
However, only a crude agreement has been found experimentally \citep{malkus1968precession} and numerically \citep{noir2001} with his theoretical zonal flow (see figure \ref{fig:intro}).
To carefully compare theory and numerics, we have performed DNS in spheres to explore smaller values of $E$ than in ellipsoids, and also to consider very small precession angles.
Moreover, we have taken $\boldsymbol{\Omega}_c \boldsymbol{\cdot} \boldsymbol{\omega}_f$ into account in our theoretical calculations \citep[contrary to][]{busse1968}. 

\begin{figure}
	\centering
	\begin{tabular}{cc}
		\includegraphics[width=0.49\textwidth]{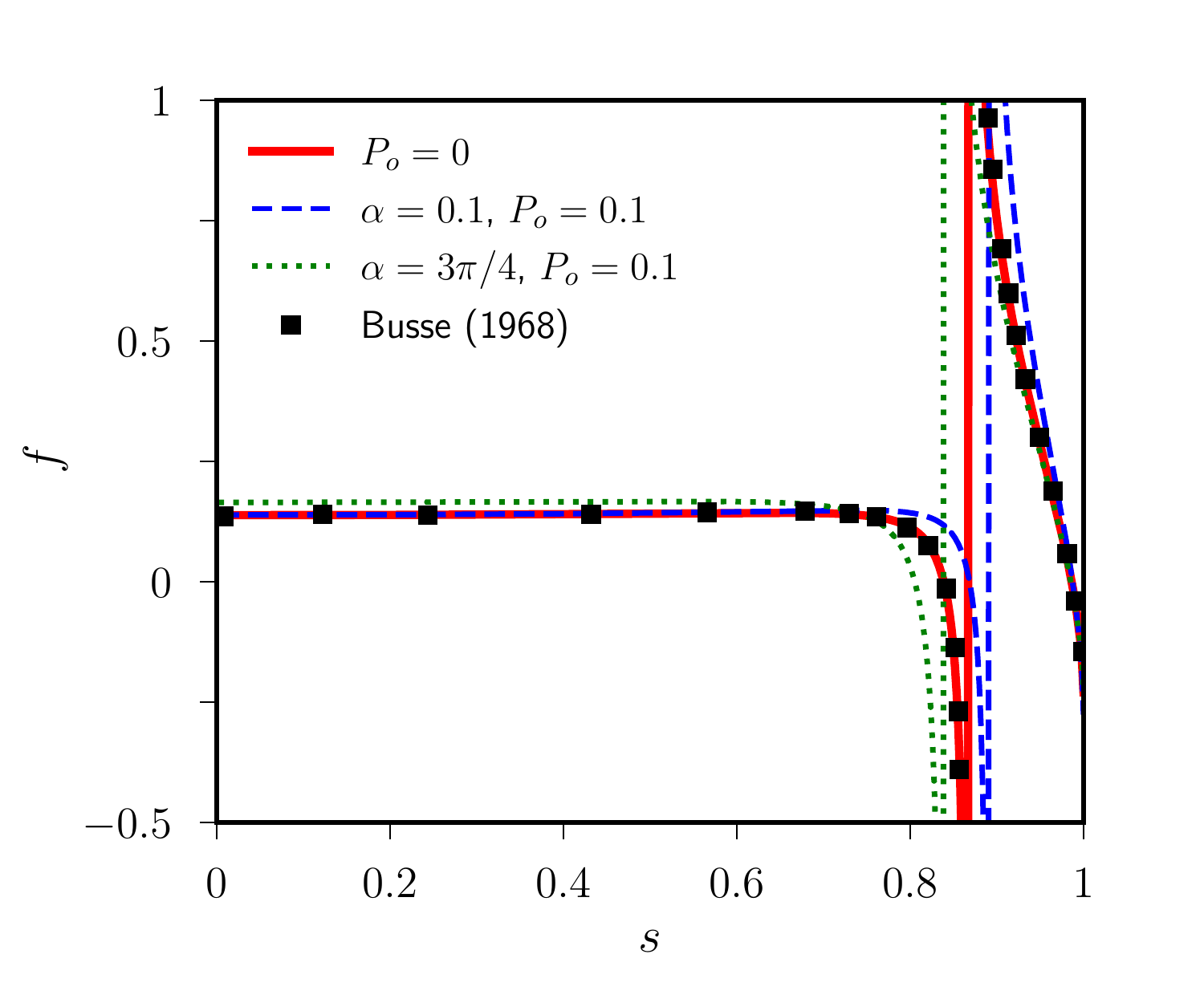} &
		\includegraphics[width=0.49\textwidth]{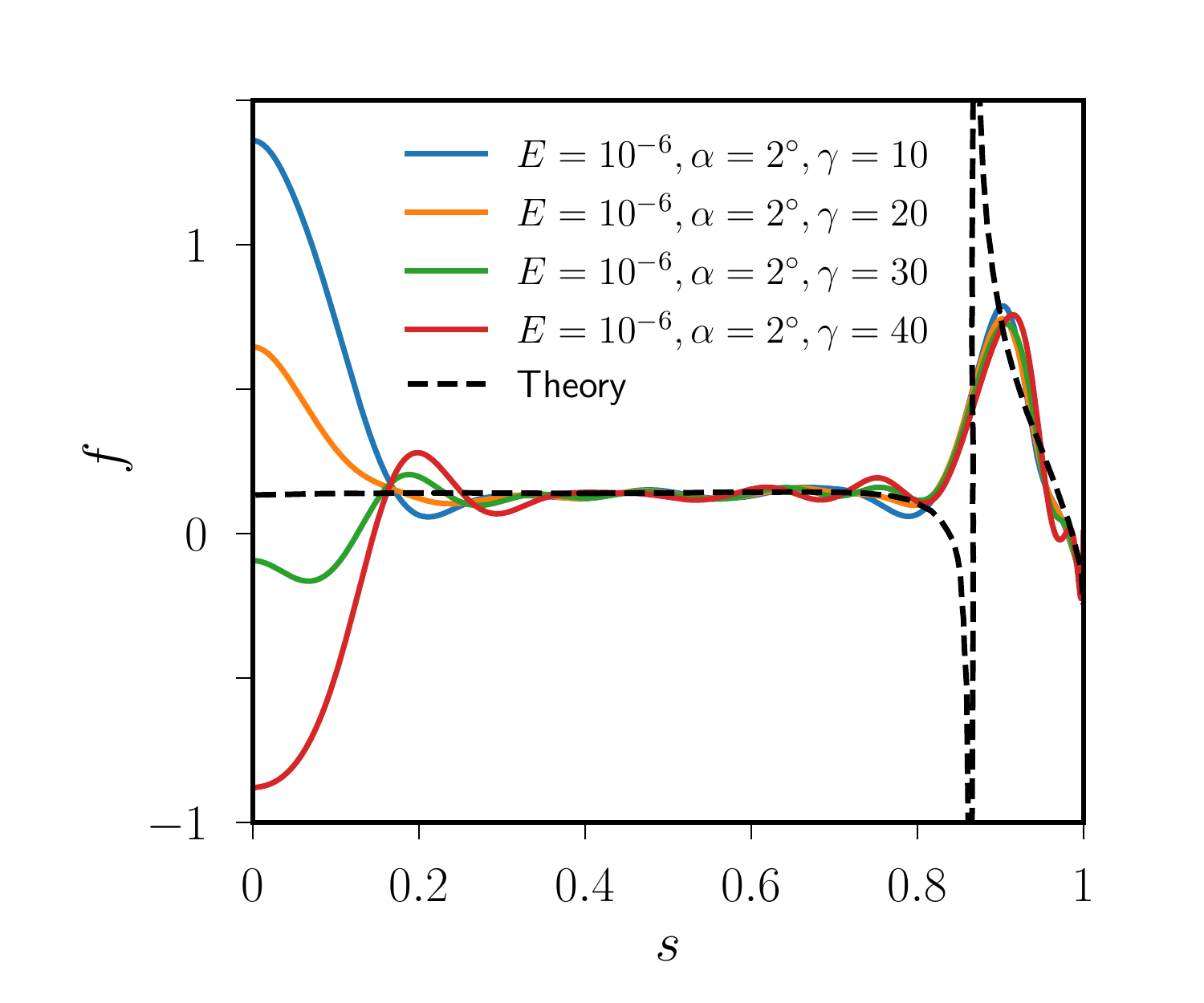} \\
		(a) & (b)
	\end{tabular}
	\caption{Rotation rate of the mean zonal flow, as a function of the cylindrical radius $s$, for a precessing sphere. (a) Comparison of theoretical profiles between our theory and Busse's predictions \citep[black squares, extracted from figure 1 in][]{busse1968}. (b) Comparison between our theory (black dashed curve) and DNS (solid curves) for different values of $P_o$ with $\gamma=P_o/E^{1/2}$.
	}
	\label{fig:busse68}
\end{figure}

While DNS are performed in the precessing frame described in section \ref{sec:mecha}, it is more convenient for theoretical calculations to consider the reference frame where the $z$-axis is along ${\boldsymbol{\omega}}_f$. 
Then, the boundary velocity can be written as $   \boldsymbol{V}_\Sigma =\Omega_0 \, \widehat{\boldsymbol{z}}_R \times \boldsymbol{r}+ \epsilon \,  \widehat{\boldsymbol{x}}_R \times \boldsymbol{r}$, with $\Omega_0=(1+P_o)^{-1}$. 
Since only the component of $\boldsymbol{\Omega}_c $ parallel to ${\boldsymbol{\omega}}_f$ is important at this order, equation (\ref{eq:skd9xx}) can be simplified to obtain $\boldsymbol{\Omega}_c \simeq  \Omega_0\,  P_o\, \cos (\alpha) \, \widehat{\boldsymbol{z}}_R.$ without loss of generality. We show in figure \ref{fig:busse68}(a) that the results of \citet{busse1968} are recovered in the limit $P_o=0$ (where the frame of reference reduces to the inertial frame). 
We also show the mean flow for small $P_o \neq 0$ (i.e. when $\boldsymbol{\Omega}_c \boldsymbol{\cdot} {\boldsymbol{\omega}}_f $ is taken into account). 
Considering $P_o \neq 0$ could be relevant for DNS performed at moderate values of $P_o$ and $\alpha$, but the mean flow is expected to be only marginally  modified in the planetary regime $P_o \ll 1$.

We have next performed DNS at low values of $E$ to validate the asymptotic theory of \citet{busse1968}, since previous studies have failed to recover the theoretical mean zonal flows (see figure \ref{fig:intro} above). 
The simulations have been computed in the precession frame, where the rotation vector $\boldsymbol{\omega}_f$ of the basic flow is accurately given by \citep{cebron2019precessing}
\begin{subequations}
\allowdisplaybreaks
\label{eq:vortfluidprec}
\begin{align}
\boldsymbol{\omega}_f \boldsymbol{\cdot} \widehat{\boldsymbol{x}}_R &= \frac{1}{1+P_o} \frac{[\lambda_i+\gamma \cos (\alpha)] \, \gamma \sin (\alpha) }{\gamma [ \gamma +2 \lambda_i \cos ( \alpha) ] + |\underline{\lambda}|^2}, \label{eq:sphereXX} \\
\boldsymbol{\omega}_f \boldsymbol{\cdot} \widehat{\boldsymbol{y}}_R &= -\frac{1}{1+P_o} \frac{\gamma  \lambda_r \sin (\alpha)}{\gamma [\gamma +2 \lambda_i \cos (\alpha) ]+|\underline{\lambda}|^2}, \\
\boldsymbol{\omega}_f \boldsymbol{\cdot} \widehat{\boldsymbol{z}}_R &= \frac{1}{1+P_o}  \frac{\gamma [ \gamma \cos^2 (\alpha) +2 \lambda_i \cos (\alpha)]+|\underline{\lambda}|^2}{\gamma [\gamma +2 \lambda_i \cos (\alpha) ]+|\underline{\lambda}|^2}, \label{eq:sphereZZ}
\end{align}
\end{subequations}
with $\gamma=P_o/E^{1/2}$ and where $\underline{\lambda}=\lambda_r+\textrm{i} \lambda_i$ is the spin-over damping factor given by \citep{hollerbach1995oscillatory,noir2001}
\begin{subequations}
\begin{equation}
    \lambda_r \approx -2.62 -1.36 \, E^{0.27}, \quad \lambda_i \approx 0.258 +1.25 \, E^{0.21}. 
    \tag{\theequation \emph{a,b}}
\end{equation}
\end{subequations}
To compare the simulations with the theory, which assumes that $\boldsymbol{\omega}_f$ is along $\widehat{\boldsymbol{z}}_I$ at leading order \citep{busse1968}, 
we post-process the data as follows.
We introduce a new reference frame, called the fluid frame, where the new $z-$axis is along the axis of rotation given by equation (\ref{eq:vortfluidprec}). 
Then, we rotate the velocity field into that frame and, to isolate the mean zonal component of order $\epsilon^2$ from the leading-order steady uniform-vorticity flow given in equation (\ref{eq:vortfluidprec}), we compute the rotation rate of the mean zonal flow as
\begin{equation}
    f(s) = \frac{V_{\phi}(z=0) - \omega_f s}{\epsilon^2 s},
    \label{eq:fantonin}
\end{equation}
where $V_{\phi}(z=0)$ is the azimuthal velocity in the equatorial plane $z=0$ of the fluid frame of reference considered in this section, $\omega_f = ||\boldsymbol{\omega}_f||$ is the norm of the fluid rotation vector $\boldsymbol{\omega}_f$ given by equation (\ref{eq:vortfluidprec}), $\epsilon^2 = \Omega_0^2-\omega_f^2$, and $s$ is the cylindrical radius measured from the fluid rotation axis. 
A typical DNS is illustrated in figure \ref{fig:busse68}(b). 
We observe that the axial value $f(0)$ is generally non zero in the DNS, which agrees with previous findings of \citet{noir2001}, but the numerical profile is in good agreement with the theory far from the axis (when $0.3 \leq s\leq 0.7$). We now investigate how these axial values $f(0)$ vary with the control parameters, and the oscillations of $f$ are discussed further in section \ref{sec:discussion}.

\begin{figure}
	\centering
	\begin{tabular}{cc}
	\includegraphics[width=0.49\textwidth]{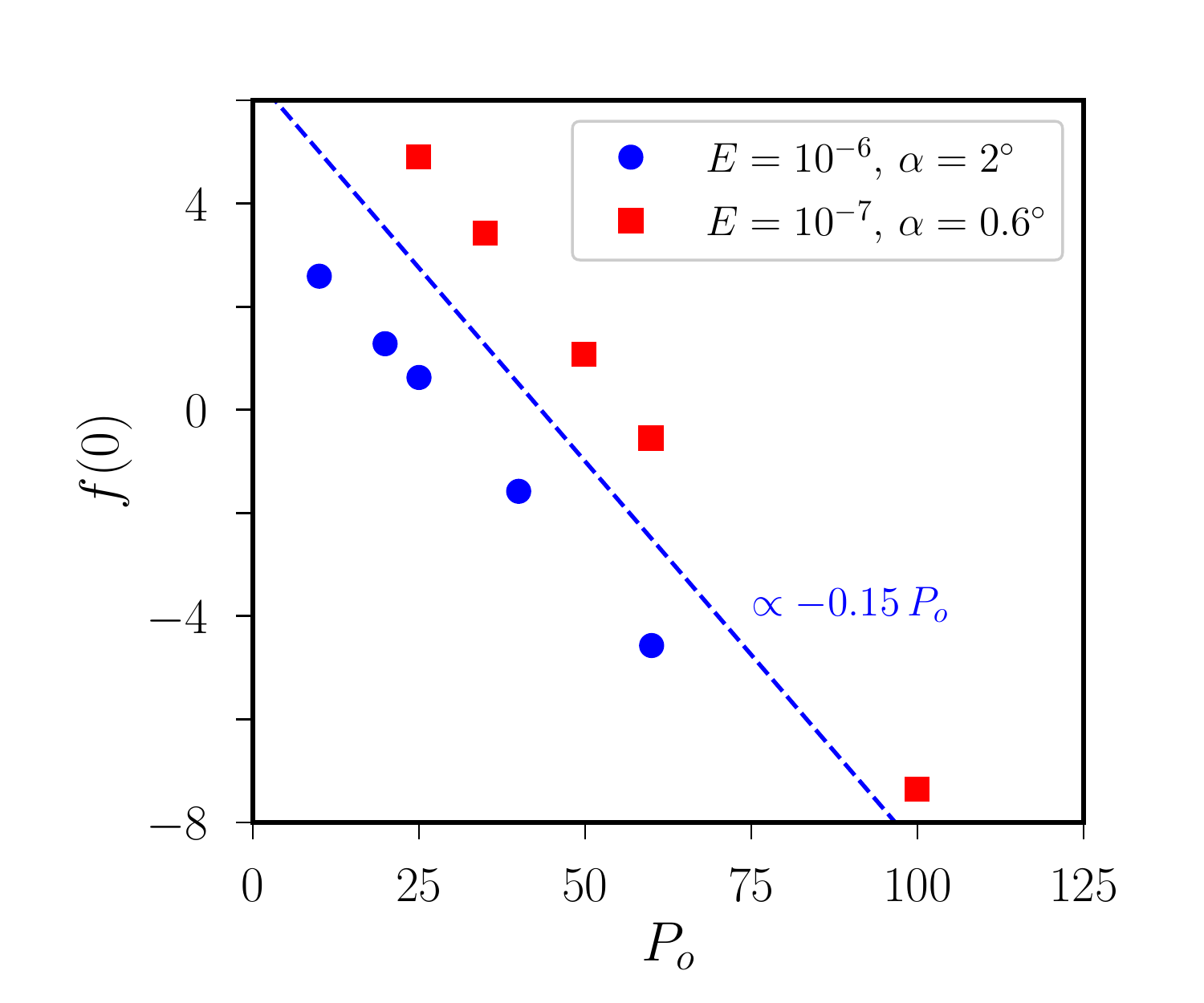} & 
	\includegraphics[width=0.49\textwidth]{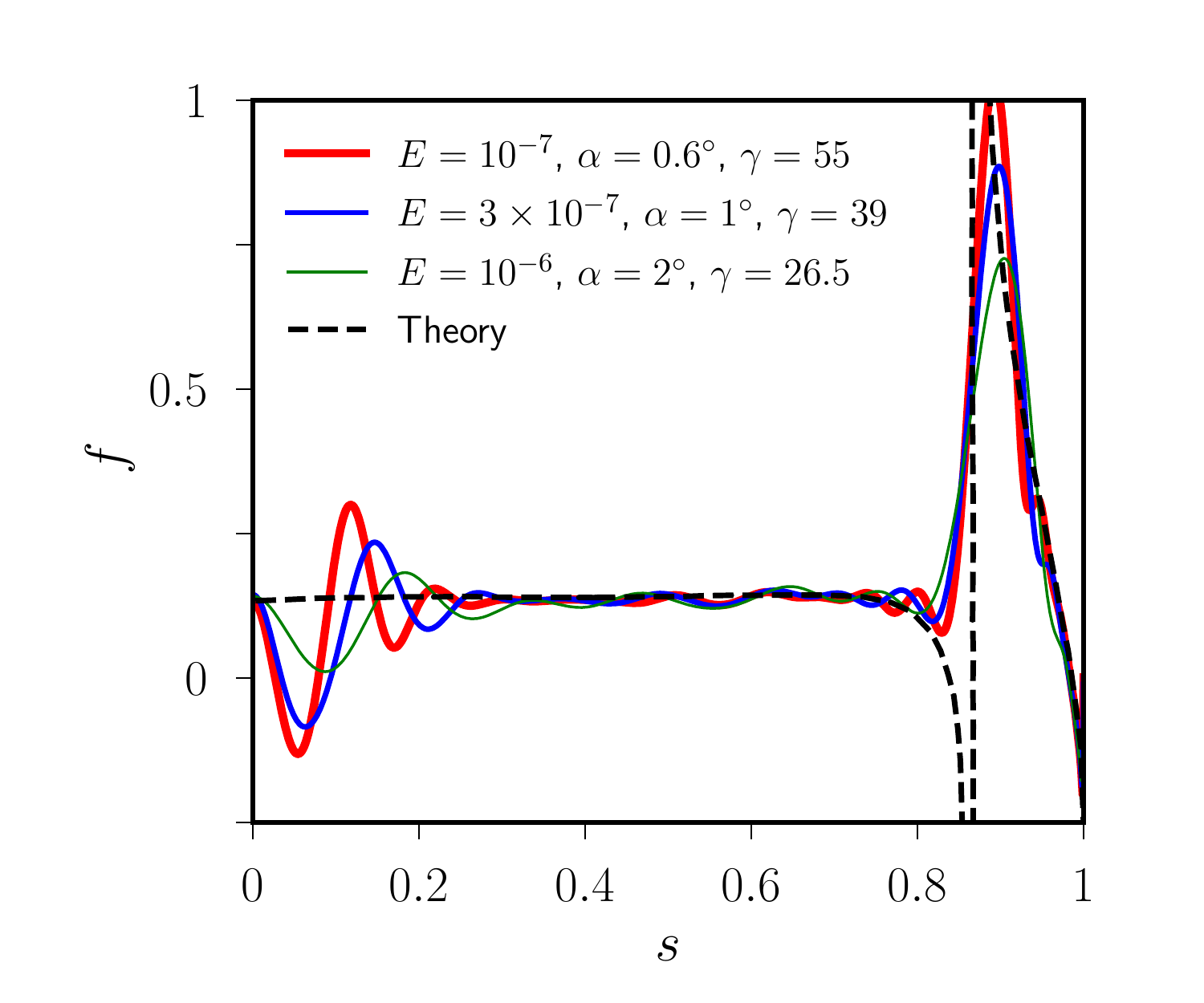} \\
	(a) & (b) \\
	\end{tabular}
	\caption{Mean zonal flows driven by precession in spheres. (a) Value at $s=0$ of $f$, as a function of $\gamma = P_o/E^{1/2}$ in DNS (symbols). Dashed blue line shows a slope $-0.15$. 
	(b) Rotation rate of the mean zonal flow, as a function of the cylindrical radius $s$. Comparison between theory (black dashed curve) and DNS (solid coloured curves) choosing values of $\gamma$ such that $f(0)=0$.}
	\label{fig:precDNS}
\end{figure}

Our DNS show that the axial value $f(0)$ depends linearly on $\gamma$, as illustrated in figure \ref{fig:precDNS}(a). However, note that the azimuthal velocity $sf(s)$ always vanishes at $s=0$.  
Thus, the value $f(0)$ only governs the weak slope of the velocity at $s=0$, which can vary in the DNS.
In the following, for every $E$, we have varied $P_o$ to obtain the value of $\gamma$ which cancels out $f(0)$. 
The corresponding DNS are shown in figure \ref{fig:precDNS}(b).
Similarly to figure \ref{fig:busse68}(b), the numerical profiles are in good agreement with the theory for $E \leq 10^{-6}$ for $0.3 \leq s\leq 0.7$.
Near the critical latitude located at $s=\sqrt{3}/2 \simeq 0.866$, the mathematical singularity is smoothed out by viscous effects. However, the lower the viscosity, the better the agreement with the theory on both sides of the singularity, with the mean zonal flow converging towards the theory when $E$ is reduced (see figure \ref{fig:precDNS}b).

\subsection{Revisiting the tidal-like forcing of \citet{suess1971}}
We finally consider the tidal-like forcing, first considered analytically by \citet{suess1971}, assuming $\omega=0$ and $m=2$ in boundary flow (\ref{cond_limite567}) together with $\boldsymbol{\Omega}_c=\boldsymbol{0}$. 
\citet{suess1971} investigated theoretically and experimentally the mean zonal flow generated by this forcing. 
\citet{suess1971} predicted theoretically the generation of a strong retrograde vortex along the rotation axis, and his prediction was surprisingly in broad agreement with his experimental results as reproduced in figure \ref{fig:suess}(a). However, the experimental profile cannot be singular at $s=0$ (see appendix \ref{sec:jvreg}), and the reported broad agreement with theory has thus to be erroneous.
Actually, a thorough analysis reveals that even his theory is incorrect because (i) he erroneously discarded the contribution from $\boldsymbol{u}_C$ and, (ii) because he made some errors in his calculation (e.g. his equation 42 is incorrect).
A possible singularity at $s=0$ may actually be expected from the divergence of the first-order boundary-layer solution $\boldsymbol{u}_0^1$ at $s=0$, and figure \ref{fig:suess}(b) shows that the contributions of $\boldsymbol{u}_B$ and $\boldsymbol{u}_C$ are indeed singular at $s=0$. 
Nevertheless, figure \ref{fig:suess}(b) clearly shows that the total zonal flow, given by the sum of the contributions, is smooth because they exactly balance each other on the axis at $s=0$. 
This confirms the crucial role of the flow $\boldsymbol{u}_C$, which was discarded by \citet{suess1971}.

\begin{figure}
	\centering
	\begin{tabular}{cc}
		\includegraphics[width=0.48\textwidth]{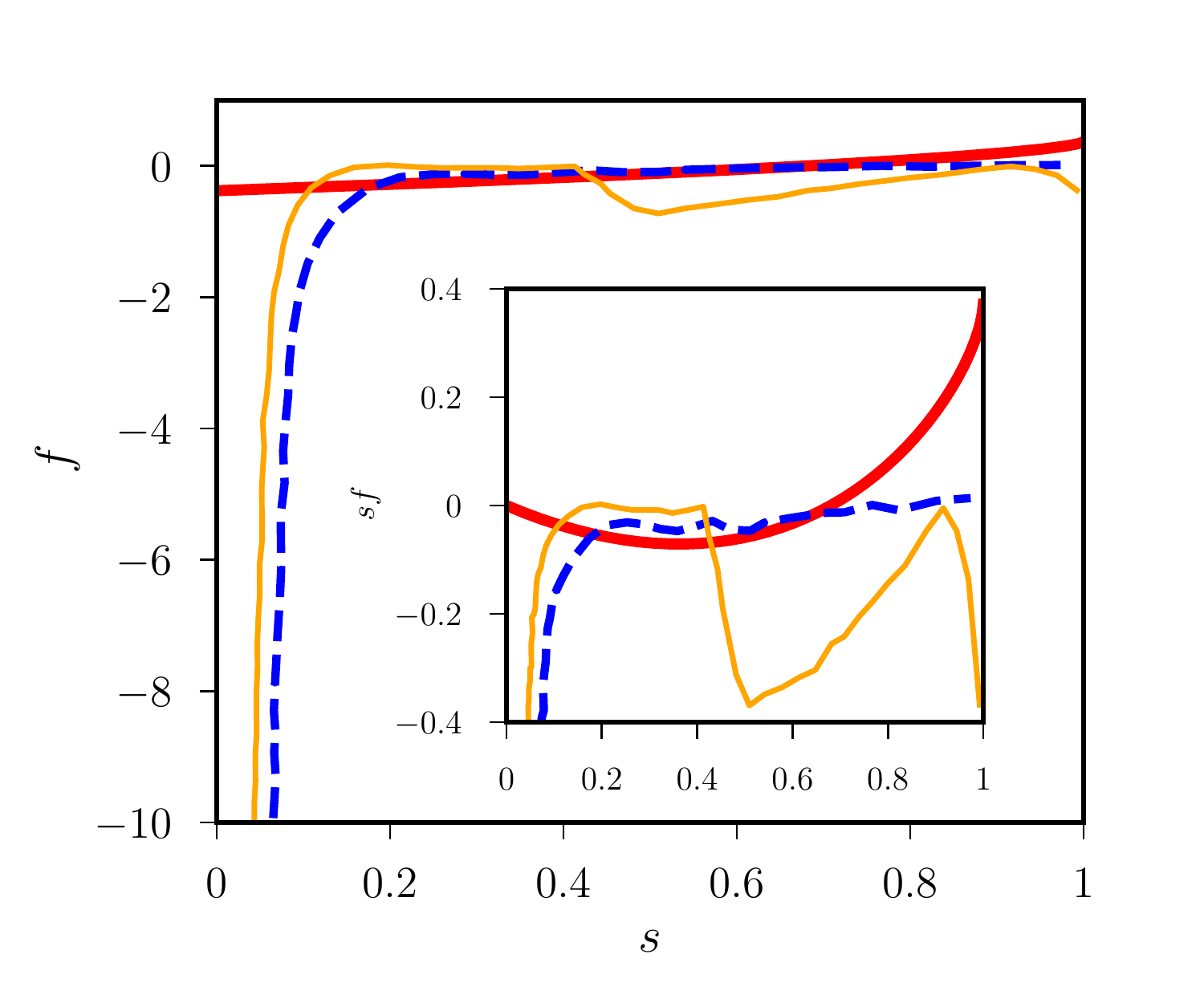} &
		\includegraphics[width=0.48\textwidth]{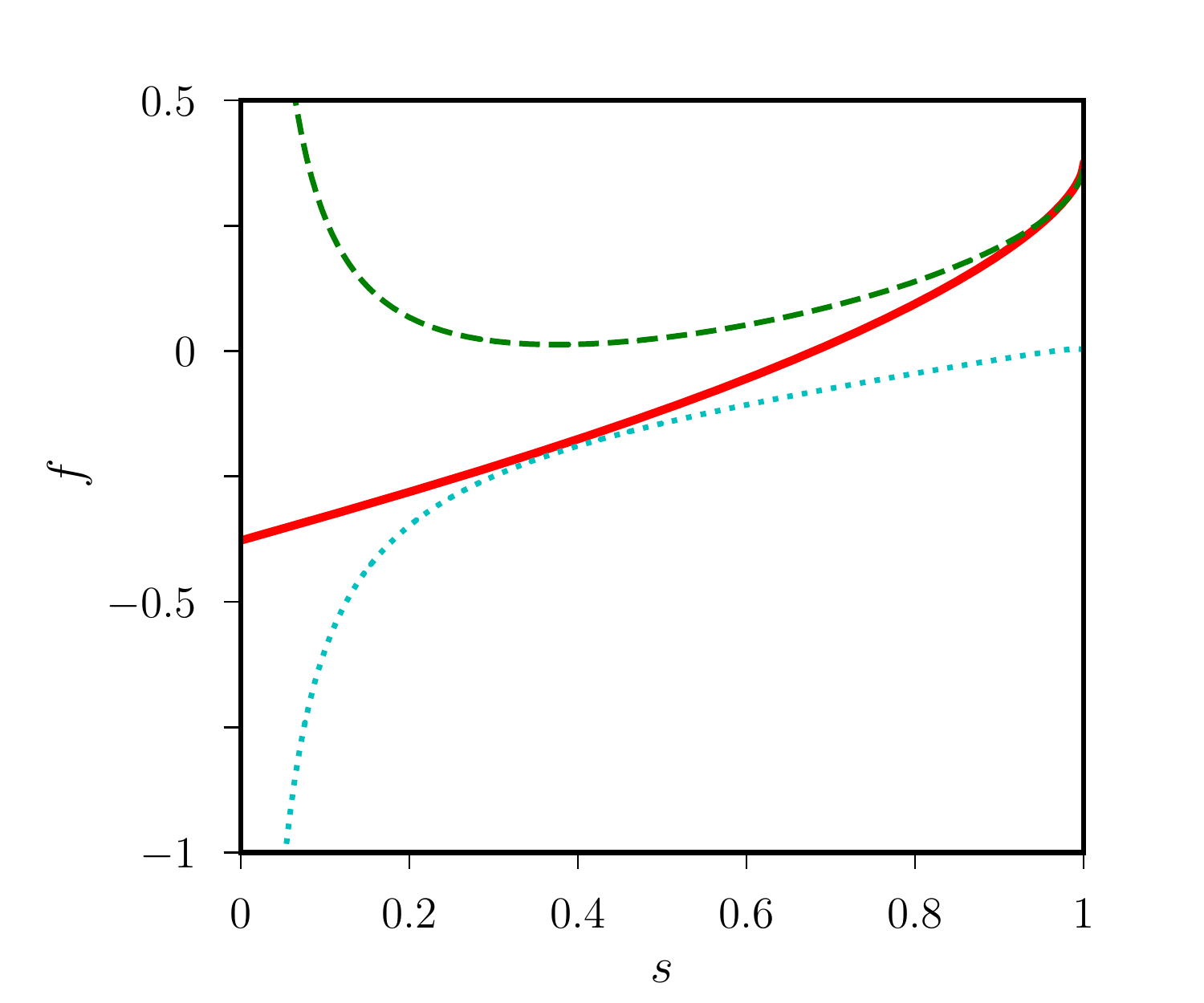} \\
		(a) & (b) \\
	\end{tabular}
	\caption{Rotation rate of the mean zonal flows driven by the tidal-like forcing of \citet{suess1971}, obtained with $\Omega_c=\omega=0$ and $m=2$ in expression (\ref{cond_limite567}). (a) Thick solid red curve: present theory. Solid orange curve: experimental results of \citet{suess1971}. Dashed blue curve: erroneous theory of \citet{suess1971}. Inset shows the azimuthal velocity $sf$.
	(b) Various contributions to $f$ (solid red curve), made of the sum of two contributions: the one due to $\boldsymbol{u}_B$ (dotted curve), which was mistakenly considered to be $f$ by \citet{suess1971}, and the one due to $\boldsymbol{u}_C$ (dashed curve).}
	\label{fig:suess}
\end{figure}

We have drawn the correct theoretical solution in figure \ref{fig:suess}(a), clearly showing  that it does not agree with the experimental measurements (especially at $s=0$). 
One may wonder whether the observed strong retrograde flow near the axis of rotation is reminiscent of the non-zero values of $f(0)$ reported above for precession and libration at $\omega \leq 2$. 
However, since the axial value $sf(0)$ of the azimuthal velocity does not seem to vanish (contrary to the corresponding value for precessing flows), this flow may have a different origin. 
Indeed, the experimental results might instead exhibit the saturation of an elliptical instability in the bulk of the fluid, which is not taken into account in the theory \citep[as it was not known at this time, see the review in][]{kerswell2002elliptical}. 
The elliptical instability can be excited when the streamline ellipticity $\beta$ is large compared to the viscous dissipation \cite[with $\beta=2 \epsilon$ in][]{suess1971}. 
More quantitatively, the growth rate $\sigma$ of the elliptical instability can be estimated as $\sigma = 9 \beta/16-K E^{1/2}$, with the typical values $1\leq K \leq 10$ related to viscous dissipation in the Ekman boundary layer \cite[e.g.][]{cebron2012elliptical}. 
Since the experimental results of \citet{suess1971} shown in figure \ref{fig:suess}(a) have been obtained for $E^{1/2}=5.3 \times 10^{-3}$ and $\beta =2.5 \times 10^{-2}$, the expected growth rate is $-0.04 \leq \sigma \leq 10^{-2}$.
Thus, an elliptical instability might have been excited in the experiment, which would certainly modify the observed zonal flow \citep[see e.g. figure 5 in][obtained for $\Omega_c=-1$]{grannan2016tidally}.

\section{Discussion}
\label{sec:discussion}
\subsection{Mean zonal flows when $\omega \leq 2$}
\label{sec:IW2}
We have so far successfully validated the theoretical mean zonal flows driven by librations when $\omega > 2$, that is in the absence of inertial waves and critical latitudes. 
A successful validation was, however, less straightforward to obtain for the precession forcing.
Precession-driven flows are indeed strongly affected by the presence of inertial waves and conical shear layers spawned from the critical latitudes \citep{noir2001experimental}. 
However, we have still found an overall good quantitative agreement with the analytical model, even if the latter has been obtained by neglecting these two additional effects. 
By analogy, more complicated flow structures are also expected for libration-driven flows when $\omega \leq 2$, as previously reported in cylinders \citep{sauret2012fluid} and spheres \citep {lin2020libration}. 
Critical latitudes indeed also exist for libration-driven flows when $\omega \leq 2$ but, based on our findings for precession, one may still expect a rather good agreement between the analytical libration-driven mean zonal flows and DNS. 

\begin{figure}
	\centering
	\begin{tabular}{cc}
		\includegraphics[width=0.49\textwidth]{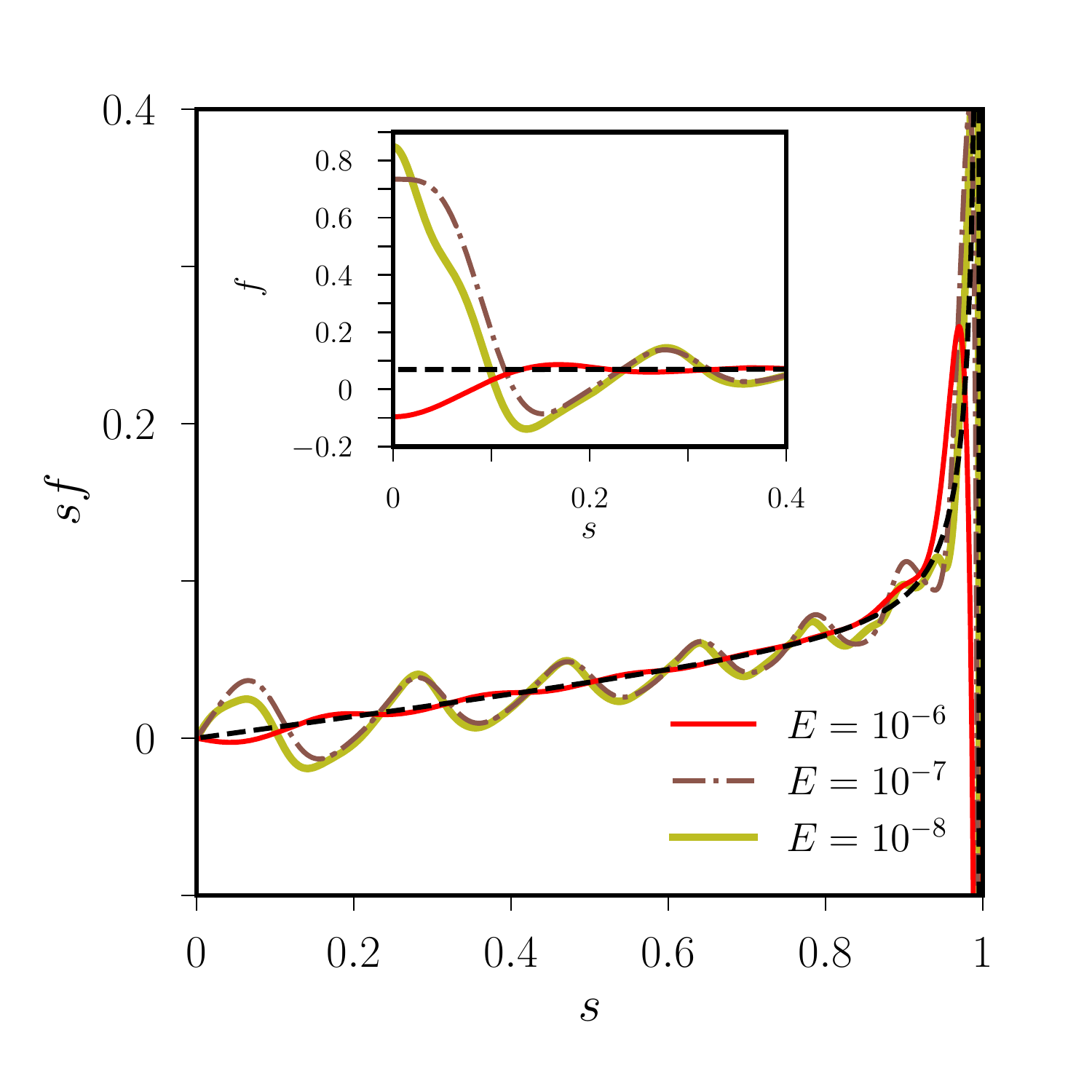} &
        \includegraphics[width=0.49\textwidth]{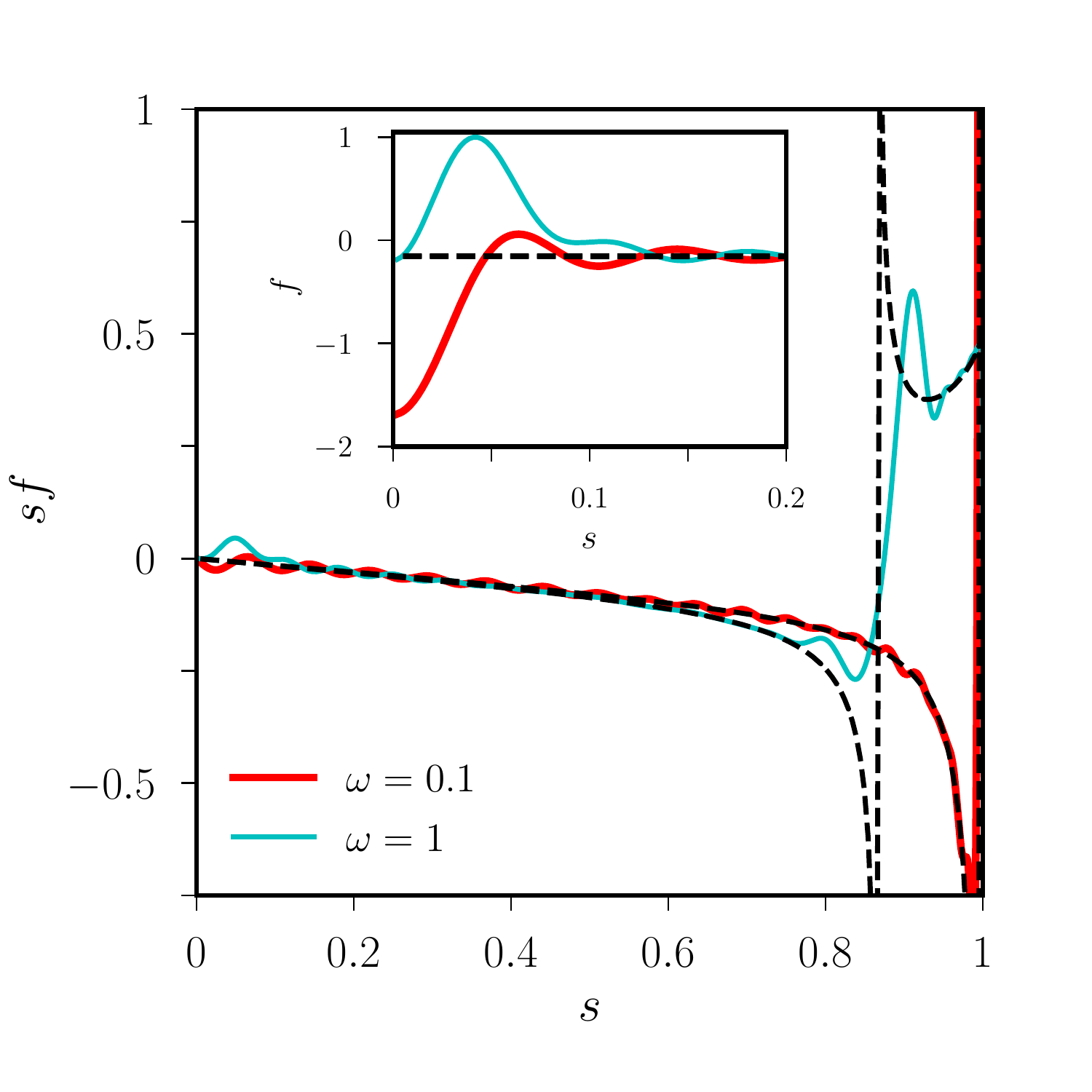} \\
		(a) Latitudinal librations & (b) Longitudinal librations \\
	\end{tabular}
	\caption{Geostrophic velocity $sf$ (or angular velocity $f$ in the insets) of the mean zonal flow, as a function of the cylindrical radius $s$, for a sphere subject to (a) latitudinal librations at $\omega=0.1$ and (b) longitudinal librations for two $\omega$. Comparison between theory (black dashed curves) and DNS (solid coloured curves) at $\epsilon=10^{-4}$. The curve at $E=10^{-7}$ is indistinguishable from the one obtained with $\epsilon=10^{-5}$ (not shown). In (b), DNS have been performed at $E=10^{-7}$, and we have checked that the numerical profile $f$ at $\omega=1$ is unchanged for $\epsilon=10^{-3}$ and $\epsilon=10^{-6}$ (not shown). 
	Insets show that $f(0)$ 
	remains finite in DNS.}
	\label{fig:oscillation}
\end{figure}

We illustrate in figure \ref{fig:oscillation} the mean zonal flows in rotating spheres subject to latitudinal librations in panel (a) and longitudinal librations in panel (b). 
Outside the region affected by the critical latitudes, the rotation rate $f$ exhibits oscillations in the bulk, which are very similar to the ones reported for precession in \S\ref{sec:precd}, that are superimposed onto the theoretical profile. 
In particular, the fluid rotation rate at the axis $f(0)$ can be quite different than the theoretical profile.
Because the finite difference method loses some accuracy at $r=0$ (see appendix \ref{sec:jvreg}), we have carefully checked numerical convergence at $E=10^{-7}$ by refining the grid (especially at the centre) and also by increasing the maximum spherical harmonic degree up to $l_{\max}=339$ (and up to $l_{\max}=425$ at $E=10^{-8}$).
The variations are within the thickness of the curve.

In the case of the latitudinal librations shown in figure \ref{fig:oscillation}(a), the same stationary oscillations are also present when only $|m|\leq1$ are considered in the DNS.
Careful inspection of the instantaneous flow reveals that forced inertial modes (equatorially anti-symmetric $m=\pm 1$ for latitudinal librations, and equatorially symmetric  $m=0$ for longitudinal librations) are present, which have a number of radial zeros. 
We conjecture that these modes have almost the same frequency as the excitation (here $\omega \simeq 0.1$) and produce the multiple jets by nonlinear interaction, which carry the signature of inertial modes.
In addition, we have computed the same case at a lower viscosity ($E=10^{-8}$) also shown in figure \ref{fig:oscillation}a. 
It highlights that once viscosity is low enough for the oscillations to appear, their amplitude is nearly independent of the Ekman number. 
Having changed $\epsilon$ in figure \ref{fig:oscillation}, we obtain that the amplitude of such oscillations does not depend on $\epsilon$ when $\epsilon \ll 1$. 
The zonal jet velocities scale thus as $\epsilon^2 E^0$, exactly as the theoretical mean zonal flow. 
Such scaling law is consistent with a zonal flow associated with a libration-excited mode, which has an amplitude of the order $\epsilon E^0$ (see appendix \ref{sec:ans23}).
The precise mechanism is beyond the scope of this paper, but it is worth noting that \cite{lin2020libration} also observed an imprint of the excited inertial mode on the mean zonal flow in shells. 
Despite the presence of these jets, our asymptotic theory fairly reproduces the mean zonal flows found in the DNS (see appendix \ref{sec:ans23} for further details). 

\begin{figure}
    \centering
    \begin{tabular}{cc}
    \includegraphics[width=0.49\textwidth]{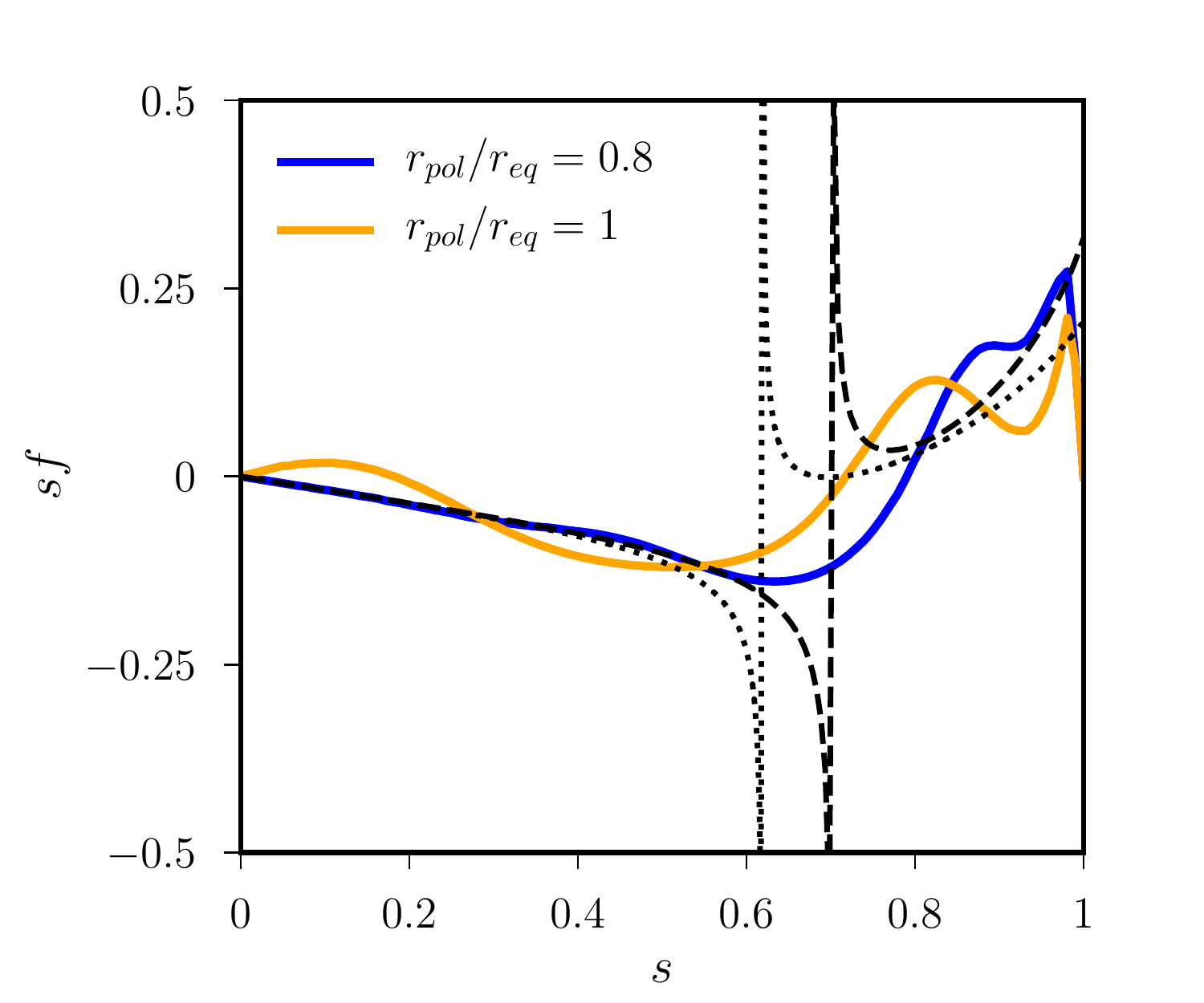} &
    \includegraphics[width=0.49\textwidth]{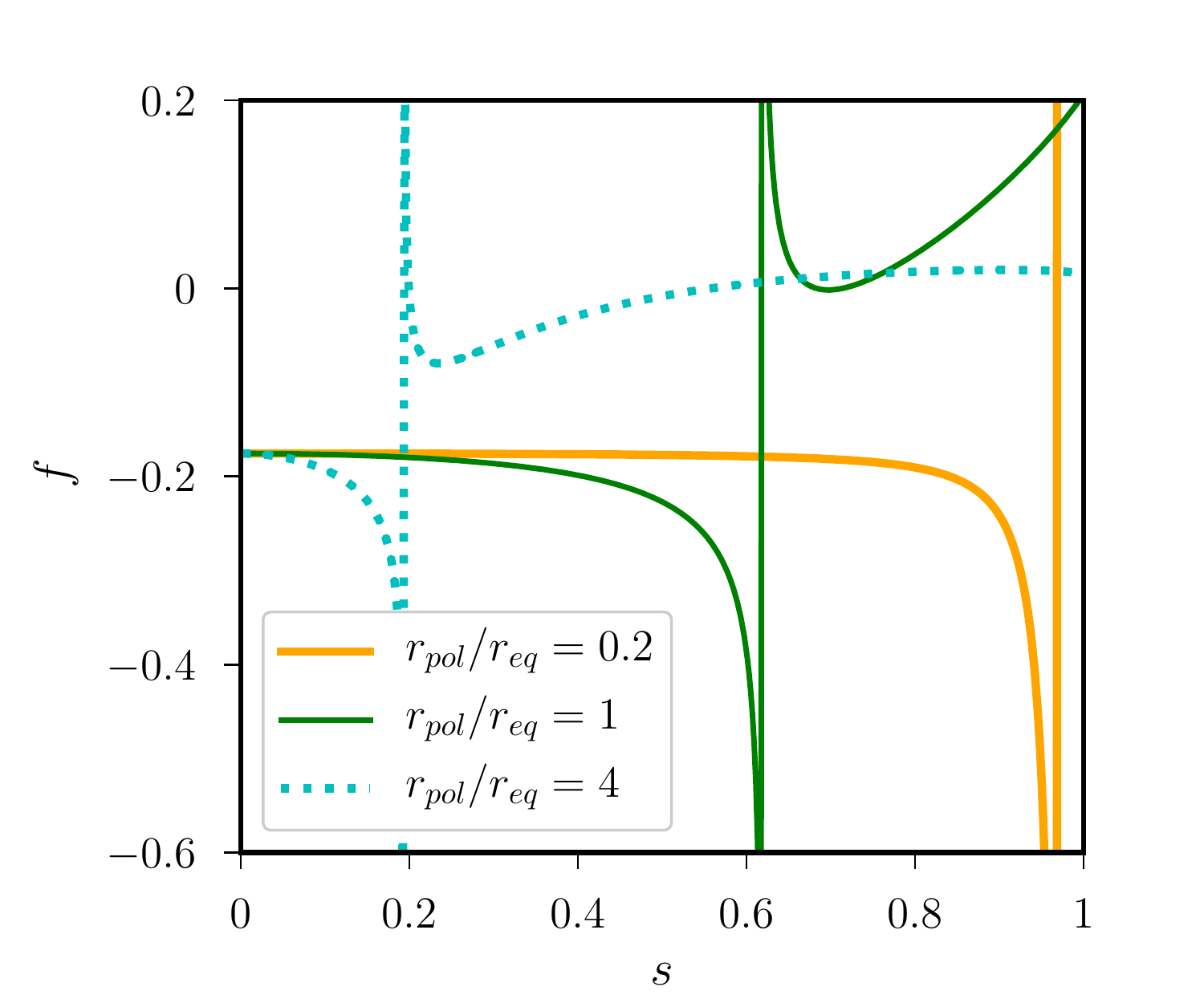} \\
    (a) & (b) \\
    \end{tabular}
    \caption{Mean zonal flows in rotating spheroids subject to longitudinal librations in the presence of critical latitudes ($\omega =\pi/2 \leq 2$). 
	(a) Comparison of the mean zonal velocity  $sf(s)$ given by the theory (black dotted curve for the sphere, black dashed curve for $r_{pol}/r_{eq}=0.8$) and DNS (coloured solid curves) at $E=10^{-5}$ and $\epsilon=10^{-4}$.
	(b) Comparison of the theoretical rotation rate between the sphere, an oblate ($r_{pol}/r_{eq}=0.2$) or prolate ($r_{pol}/r_{eq}=4$) spheroid.}
    \label{fig:liblong_pi2}
\end{figure}

The same conclusion is drawn in spheroids, as reported in figure \ref{fig:liblong_pi2}(a), even if the DNS in spheroidal geometries have been performed for moderate values of $E \geq 10^{-5}$ (compared to DNS in spheres with $E \geq 10^{-7}$). 
Another striking point in the figure is that the location of the critical latitude varies with the ellipsoidal deformation.  
Indeed, the unit vector $\boldsymbol{e}_1$ in the left-hand side of equation (\ref{eq:lm0f82q}) depends on the spheroidal geometry, such that the spatial position of the critical latitude is modified. 
This phenomenon is further illustrated with more deformed spheroidal geometries in figure \ref{fig:liblong_pi2}(b). 
This effect has direct consequences for the numerical profiles obtained.
Only a small part of the volume is affected by the shear layer when the critical latitude is close to the boundary (see in figure \ref{fig:liblong_pi2}a for $r_{pol}/r_{eq}=0.8$), such that the oscillations are rather localised around the position of the critical latitude. 
However, the oscillations can spread in the volume when the critical latitude is far from the boundary (see $r_{pol}/r_{eq}=1$ in figure \ref{fig:liblong_pi2}a).

\subsection{Geostrophic shear spawned from the critical latitudes when $E \ll 1$}
\label{sec:critcen}
The analytical solutions show that the mean zonal flows are singular at the cylindrical radius $s_c$ when $\omega \leq 2$
(see e.g. figure \ref{fig:liblong_pi2}b), Nevertheless, this singularity is regularised by viscosity such that the mean zonal flow takes the form of a shear layer near $s_c$, as first noticed by \citet{busse1968}. 
Moreover, it is known that the amplitude of the geostrophic shear increases as $E$ is reduced, by contrast with the typical horizontal length scale of the shear \citep[e.g.][for precession]{noir2001}.
However, the corresponding scaling laws with the Ekman number have been disputed, and no conclusive answer has been obtained yet. 
We aim at revisiting here that problem numerically with DNS in spheres, to hopefully capture the correct asymptotic behaviour in the relevant regime $E \ll 1$.

\begin{figure}
	\centering
	\begin{tabular}{cc}
		\includegraphics[width=0.49\textwidth]{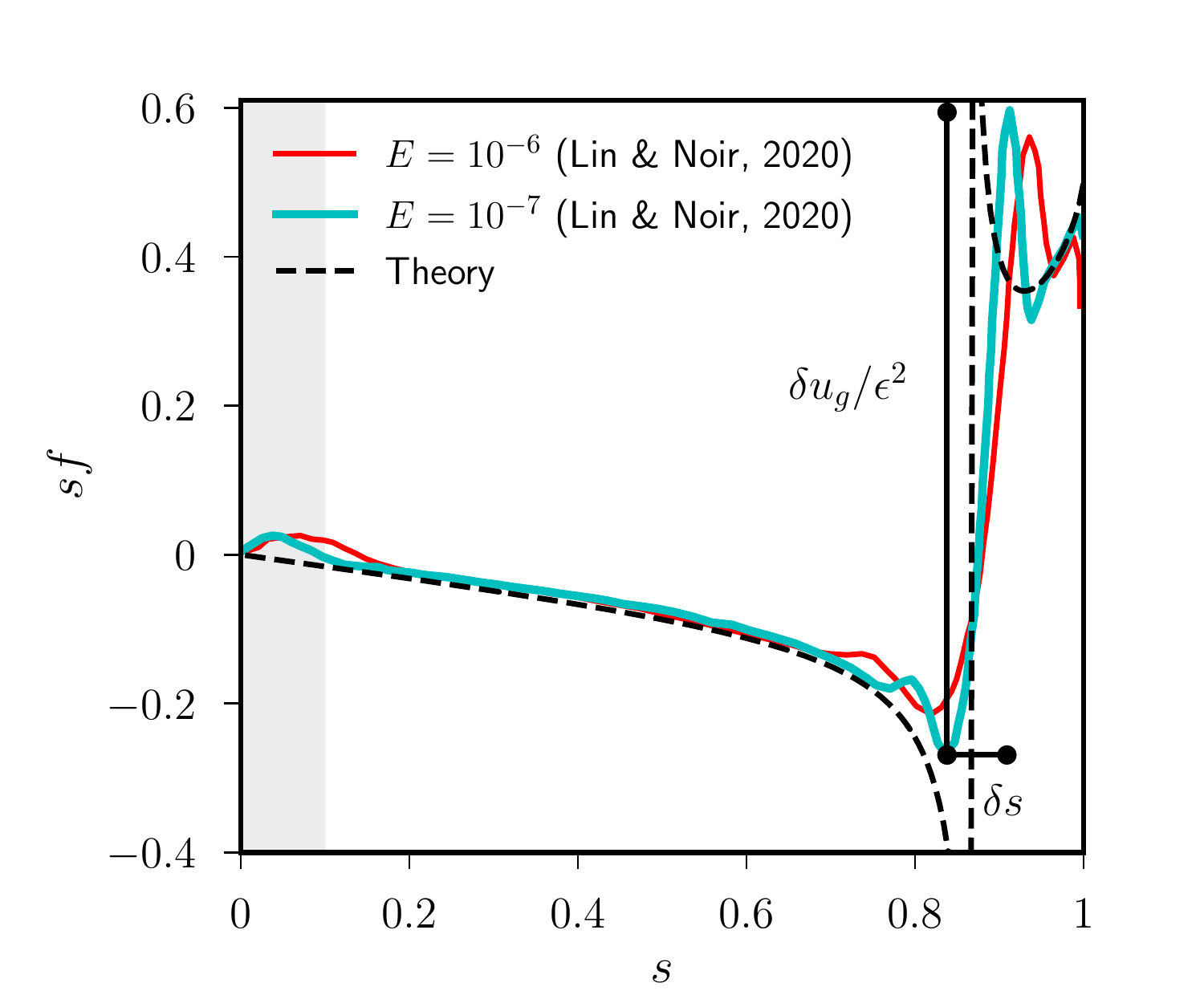} &
        \includegraphics[width=0.49\textwidth]{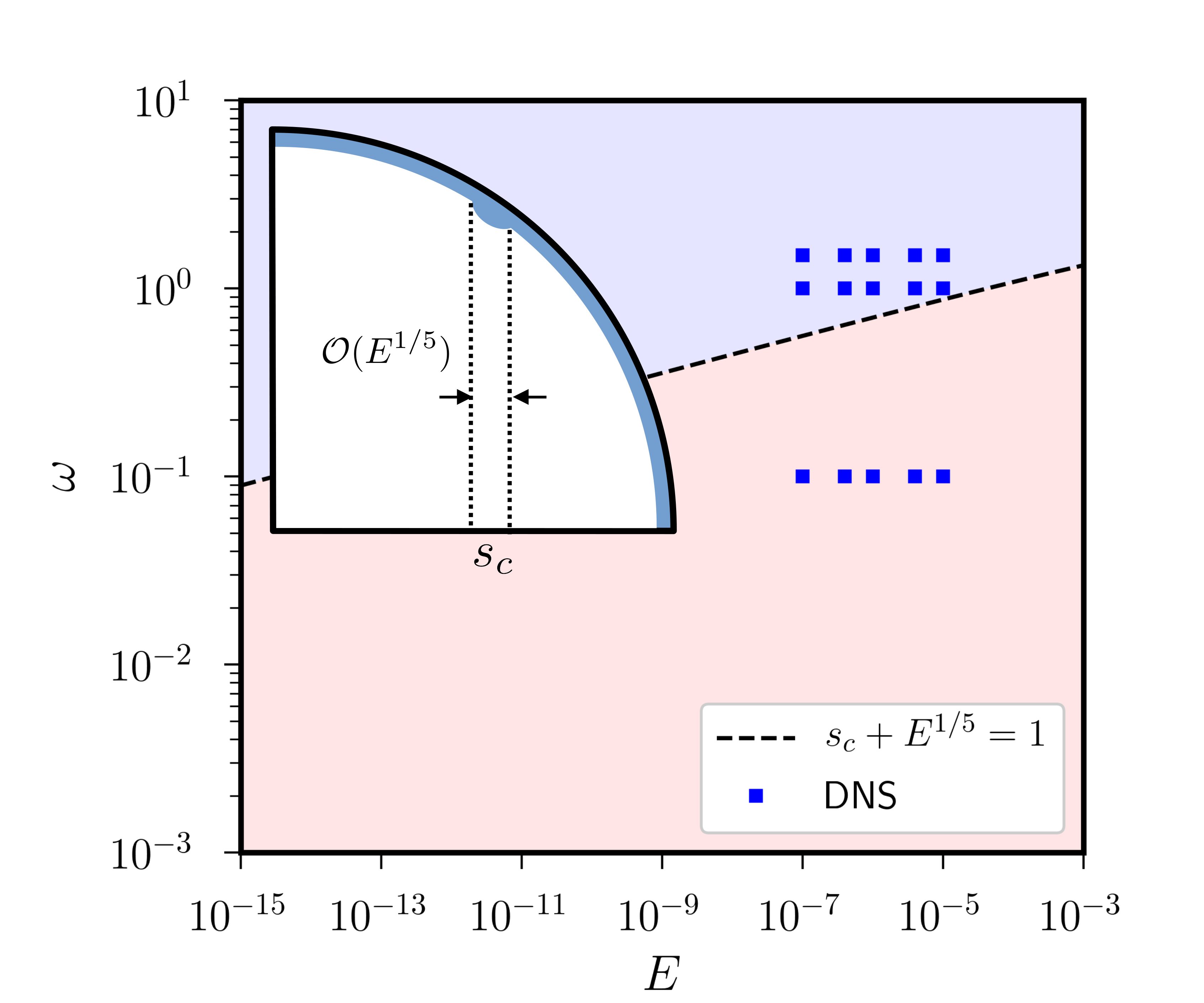} \\
		(a) $\omega=1$ & (b) \\
	\end{tabular}
	\caption{Geostrophic shear associated with the critical latitudes of longitudinal librations with $\omega \leq 2$. 
	(a) Comparison between theory and DNS \citep{lin2020libration} in a spherical shell for libration-driven zonal flows computed with $\epsilon=10^{-2}$. The grey area indicates the tangential cylinder $s\leq 0.1$ associated with the inner core in \citet{lin2020libration}. 
	(b) Schematic regime diagram for the evolution of $\delta u_g$ and $\delta s$, as a function of $\omega$ and $E$. 
	Pink area indicates the regime dominated by the scalings of the Ekman boundary layer. In the blue area the dominant scalings are given by (\ref{eq:scalingliblong}).
	Inset illustrates the geostrophic shear centred on $s_c$ in a meridional section, where the blue area represents the Ekman boundary layer of thickness $E^{1/2}$.}
	\label{fig:shear_liblong}
\end{figure}

\citet{lin2020libration} recently explored the geostrophic shear generated by longitudinal librations with $\omega = 1$ in spherical shells, but with a small inner core ($s\leq 0.1$). 
We reproduce their results in figure \ref{fig:shear_liblong}(a), and find that their mean zonal flows are in very good agreement with the theoretical solution at $\omega=1$ in the full sphere. 
Considering now the width $\delta s$ and the peak-to-peak amplitude $\delta u_g$ of the geostrophic shear (as defined in figure \ref{fig:shear_liblong}a), 
\citet{lin2020libration} proposed the scaling laws
\begin{subequations}
\label{eq:scalingliblong}
\begin{equation}
     \delta u_g/\epsilon^2 \propto E^{-1/10}, \quad \delta s \propto E^{1/5},
    \tag{\theequation \emph{a,b}}
\end{equation}
\end{subequations}
using DNS of longitudinal librations at $\omega=1$ and order-of-magnitude arguments. 
We confirm numerically scaling laws (\ref{eq:scalingliblong}) for various libration-driven flows at different $\omega$, but only in a certain parameter range (see figure \ref{fig:shear_liblong}b). 

\begin{figure}
	\centering
	\begin{tabular}{cc}
		\includegraphics[width=0.49\textwidth]{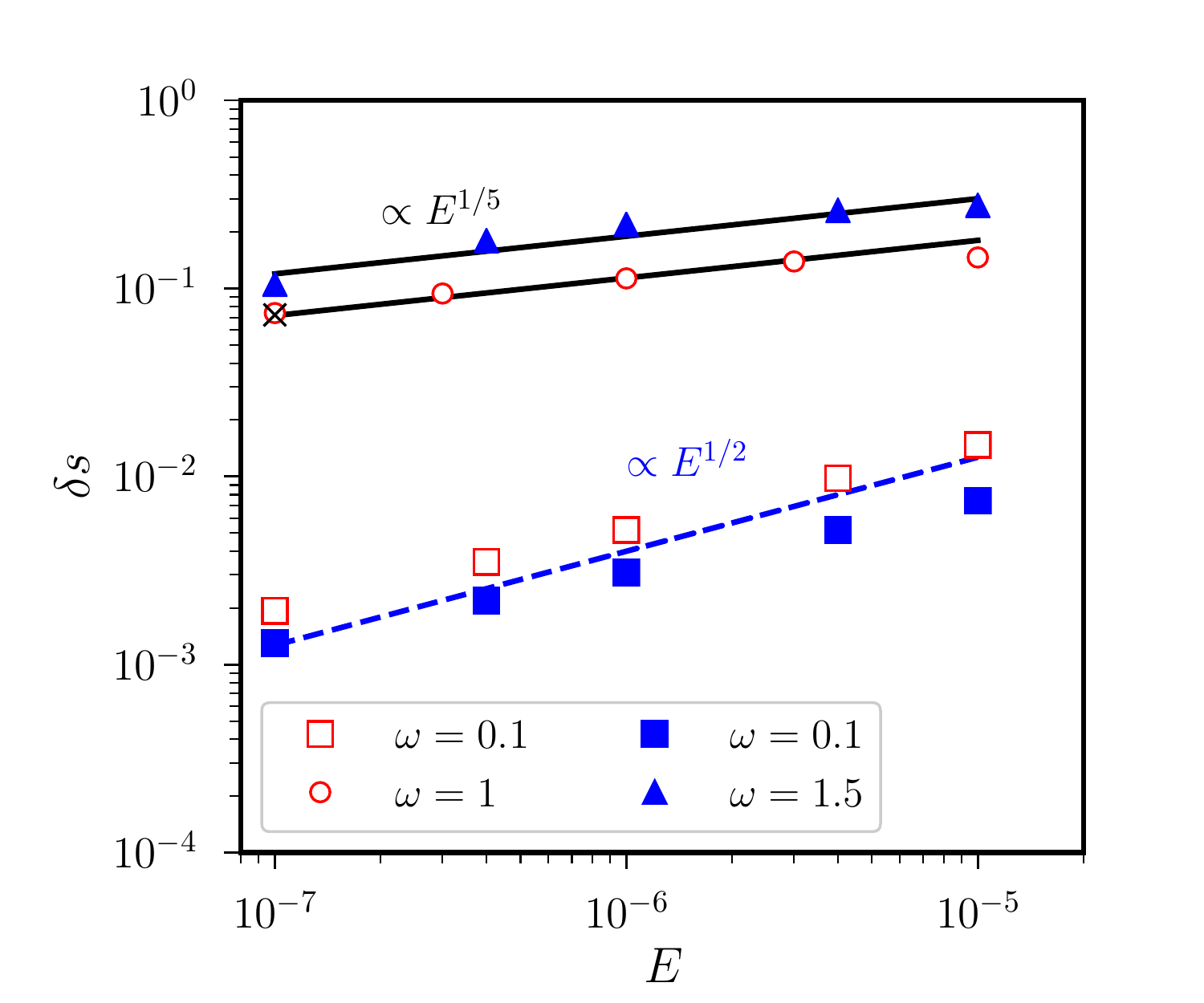} &
        \includegraphics[width=0.49\textwidth]{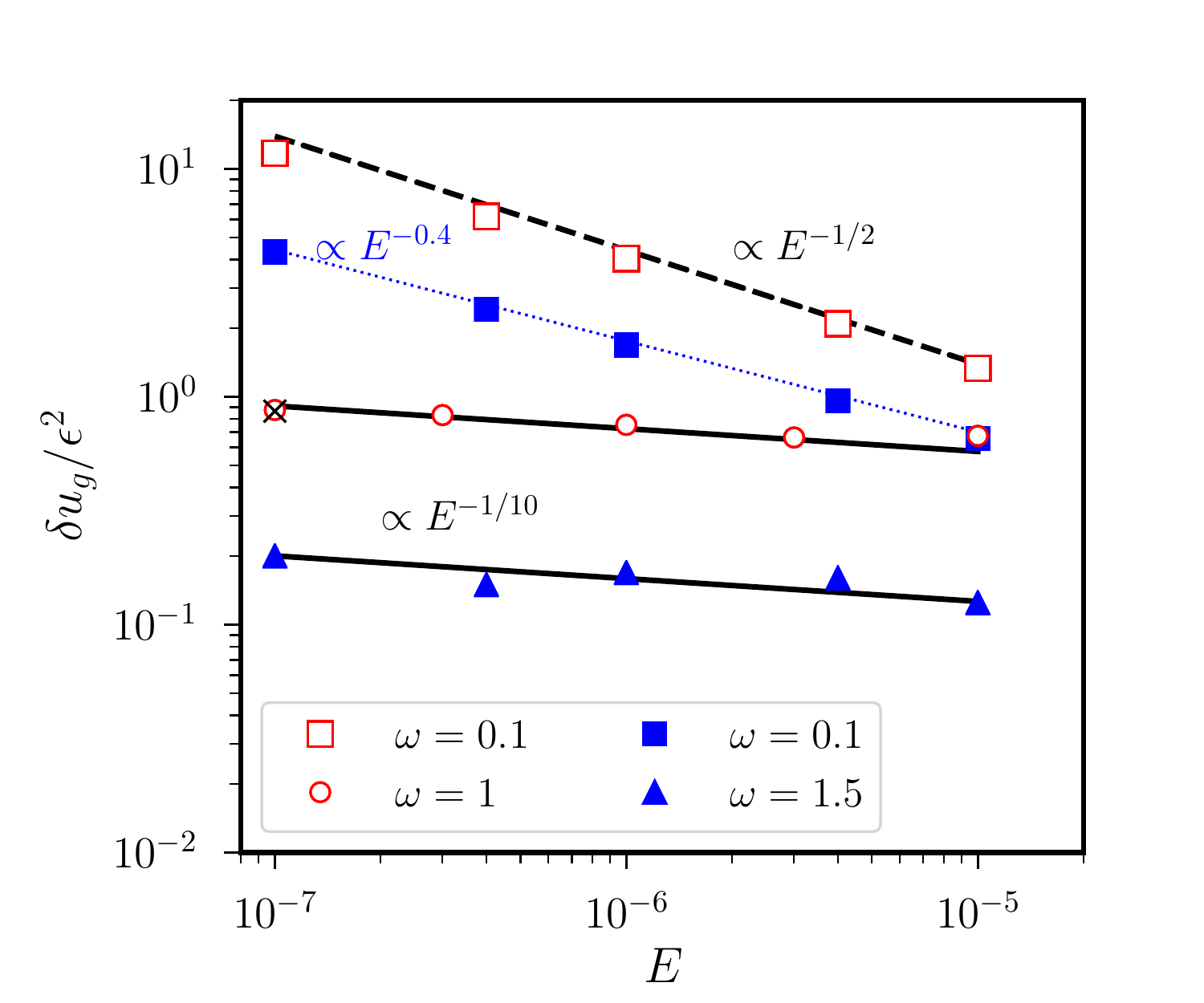} \\
		(a) & (b) \\
	\end{tabular}
	\caption{Geostrophic shear associated with the critical latitudes of longitudinal librations (empty points) and latitudinal librations (filled points) in a sphere. The symbols $\omega=1$ performed for longitudinal librations have been extracted from figure 10 in \citet{lin2020libration}. 
	The black cross indicates our DNS in a full sphere at $\omega=1$ and $\epsilon=10^{-4}$, which agrees with \citet{lin2020libration}. Distance $\delta s$ between peaks in (a), and peak-to-peak amplitude $\delta u_g/\epsilon^2$ in (b), as a function of $E$.}
	\label{fig:shear_liblong2}
\end{figure}

One can indeed anticipate a possible change of regime when the geostrophic shear, of typical thickness $E^{1/5}$ \citep{kerswell1995,noir2001} and centred on the cylindrical radius $s_c$ (see inset in figure \ref{fig:shear_liblong}b), interacts with the equatorial boundary layer at $s\simeq1$. 
Because the typical Ekman layer thickness $E^{1/2}$ is negligible with respect to $E^{1/5}$ when $E \ll 1$, we expect a different behaviour for a certain value $\omega=\omega_c$ given by
\begin{equation}
    s_c +\mathcal{O}(E^{1/5}) \simeq 1,
    \label{eq:cjolie}
\end{equation}
with the cylindrical radius of the critical latitude $s_c=\sin \cos^{-1} (\omega_c/2) \approx 1-\omega_c^2/8 $ for small values of $\omega_c$ in the sphere. 
Equation (\ref{eq:cjolie}) gives $\omega_c = \mathcal{O}(E^{1/10}) \ll 1$ in the regime $E \ll 1$ (see figure \ref{fig:shear_liblong}b). 
When $\omega \ll E^{1/10}$, we thus expect the Ekman layer scaling laws
\begin{subequations}
\begin{equation}
    \delta u_g/\epsilon^2 \propto E^{-1/2}, \quad \delta s \propto E^{1/2}.
    \tag{\theequation \emph{a,b}}
\end{equation}
\end{subequations}
In the opposite regime $\omega \gg E^{1/10}$, the relevant scaling laws should be (\ref{eq:scalingliblong}) as proposed by \citet{lin2020libration}. 
In order to validate these theoretical considerations, we have performed DNS in both regimes at different libration frequencies, especially near the transition between these two configurations (figure \ref{fig:shear_liblong}b). 
The numerical results, obtained for various libration forcings and frequencies, are summarised in figure \ref{fig:shear_liblong2}.
The various scaling laws are numerically recovered, as well as the change of regimes. 
Note that the typical frequency of most planetary bodies subject to longitudinal librations is $\omega \geq \mathcal{O}(1)$ \citep[e.g.][]{noir2009experimental,sauret2013libration}, such that scaling laws (\ref{eq:scalingliblong}) are expected to be relevant for most planetary applications.

Finally, the aforementioned scaling laws differ from the ones that have been proposed for precession-driven flows \citep{noir2001}, that is
\begin{subequations}
\label{eq:scalingprec}
\begin{equation}
    \delta u_g / \epsilon^2 \propto E^{-3/10}, \quad \delta s \propto E^{1/5}.
    \tag{\theequation \emph{a,b}}
\end{equation}
\end{subequations}
We have checked that we also recover scaling laws (\ref{eq:scalingprec}) for our DNS of precession-driven flows. 
Note that the scaling law (\ref{eq:scalingprec}a) has also been experimentally observed in a rotating sphere subject to a tidal deformation \citep{morize2010experimental}. 
As outlined in \cite{lin2020libration}, the origin of the different scaling laws between libration and precession remains puzzling.

\subsection{Influence of a solid inner core}
\label{sec:shell}
Planetary fluid layers are often bounded by two solid layers (e.g. the Earth's liquid core is surrounded by the upper mantle and a solid inner core). 
One can thus wonder how our results, obtained in coreless geometries, could be modified by the presence of an inner boundary. 
We focus here on libration-driven zonal flows, which have already received attention \citep[e.g.][]{calkins2010axisymmetric,sauret2013libration,lin2020libration}.

We first consider the case where the critical latitudes and inertial waves are absent.
This regime has been theoretically investigated for longitudinal librations in \citet{sauret2013libration}, showing that the mean zonal flows in a spherical shell can be entirely deduced from the solutions in the full sphere \citep[if we exclude the Stewartson layers associated with the presence of the inner boundary, see][]{stewartson1966almost}.
Here, we consider a (possibly non-homoeoidal) spheroidal shell, where the inner boundary is spheroidal, with respectively the (dimensional) inner equatorial $r_{eq}^{in} = \eta_{eq} \, r_{eq}$ and polar $r_{pol}^{in} =  \eta_{pol} \, r_{pol}$ axes, where $[\eta_{eq},\eta_{pol}]$ are the equatorial and polar shell aspect ratios ($\eta_{eq}=\eta_{pol}$ in homoeoidal shells).
We impose on the inner boundary a harmonic tangential velocity of magnitude $\epsilon_{in}$ and angular frequency $\omega_{in}$, which may differ from the forcing at the outer boundary (with the amplitude $\epsilon$ and angular frequency $\omega$ as above). 
Following \citet{sauret2013libration}, we find that the rotation rate of the mean zonal flow is given in dimensionless form by
\begin{equation} 
\frac{\langle \overline{V}_\phi \rangle}{s} = \begin{cases}
\dfrac{\epsilon_{in}^2 \, (1-s^2)^{1/4} \, f_{sp}^{icb} (s/r_{eq}^{in}) + \epsilon^2 \, [1-(s/r_{eq}^{in})^2]^{1/4} \, f_{sp}^{cmb} (s)}{(1-s^2)^{1/4}+[1 - (s/r_{eq}^{in})^2]^{1/4} }  &\text{for} \quad s<r_{eq}^{in}, \\
\epsilon^2 \, f_{sp}^{cmb} &\text{for} \quad s>r_{eq}^{in},
\end{cases}
\label{eq:coshell}
\end{equation}
where $f_{sp}^{icb}$ (respectively $f_{sp}^{cmb}$) is the rotation rate profile of the mean zonal flow in a coreless geometry when the forcing at the inner (respectively outer) boundary is considered. 
According to equation (\ref{eq:coshell}), the presence of an inner boundary at $s=r_{eq}^{in}$ is not expected to modify the mean zonal flow for $s>r_{eq}^{in}$ (i.e. outside the tangent cylinder). 

\begin{figure}
    \centering
    \begin{tabular}{cc}
    \includegraphics[width=0.49\textwidth]{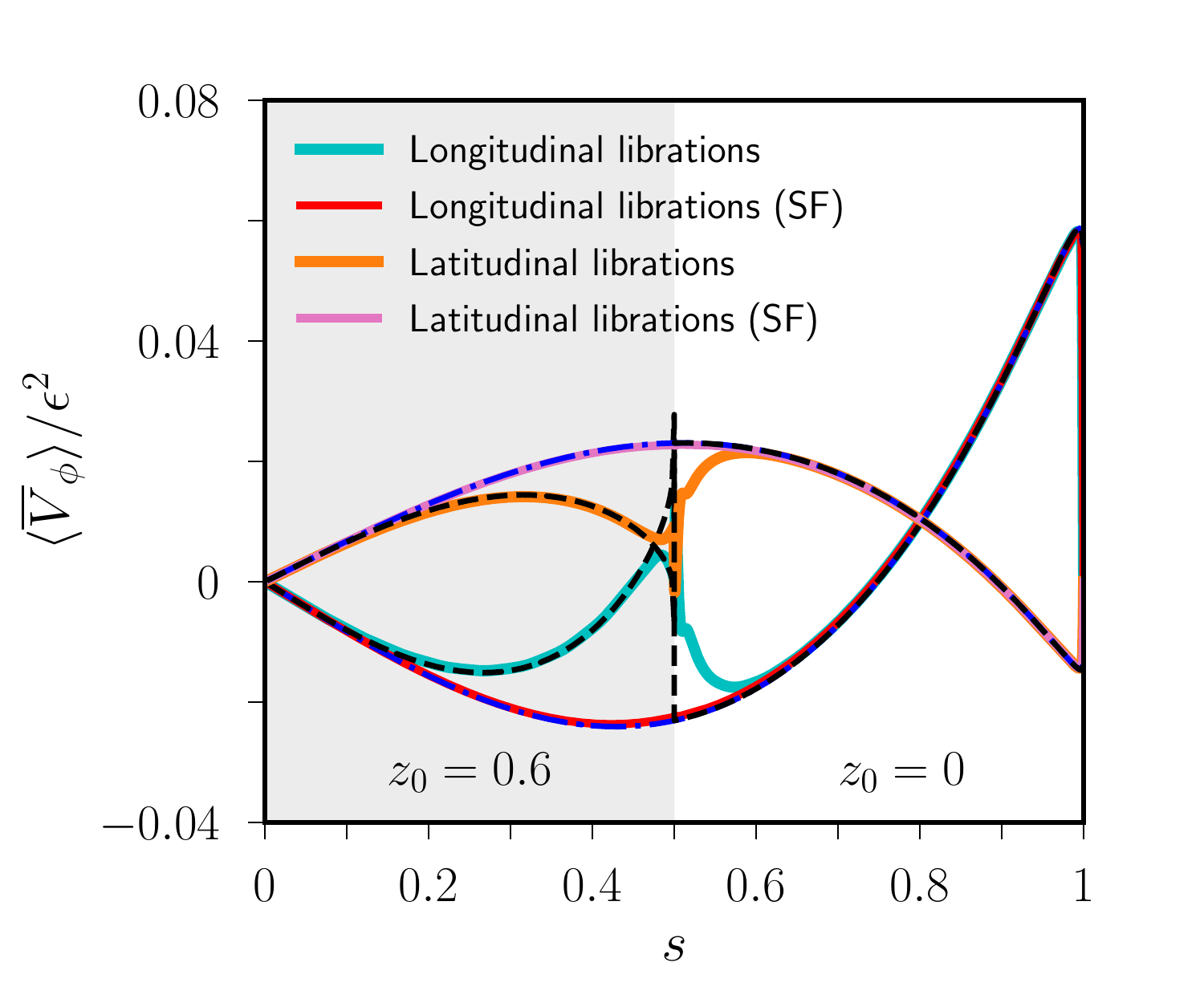} &
    \includegraphics[width=0.49\textwidth]{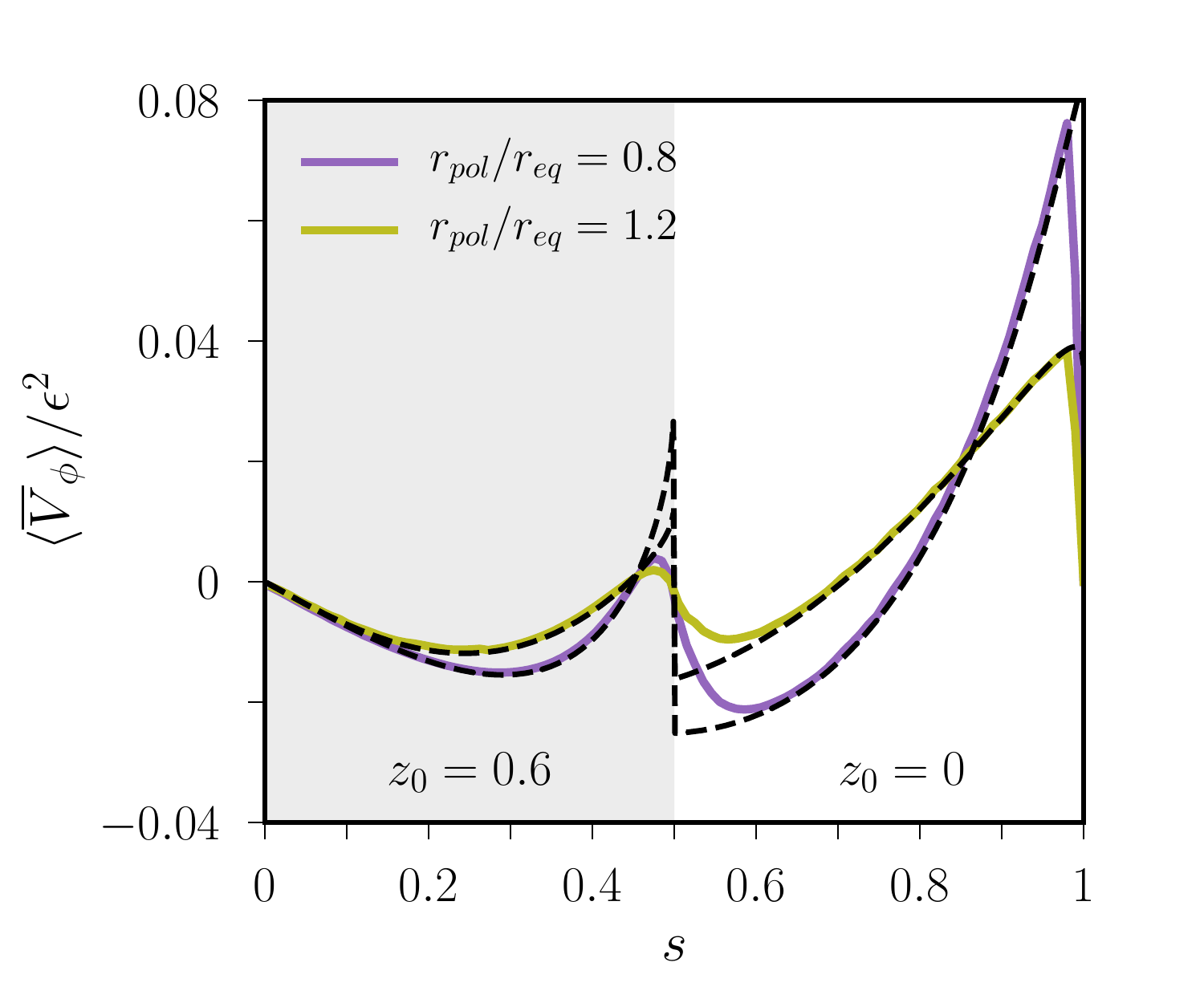} \\
    (a) Spherical shell $(r_{pol}/r_{eq}=1)$ &
    (b) Spheroidal shells $(r_{pol}/r_{eq}\neq1)$ \\
    \end{tabular}
    \caption{Normalised mean zonal velocity $\langle \overline{V}_\phi \rangle/\epsilon^2$ at various dimensionless heights $z=\sqrt{1-s^2}$ in homoeoidal shell geometries (with aspect ratios $\eta_{pol}=\eta_{eq}=0.5$) for libration forcings. 
    The DNS profiles are computed at $z=z_0$ for $s \leq \eta_{eq}$, and at $z_0 = 0$ for $s > \eta_{eq}$, giving a single profile for each DNS. 
    Grey area shows the tangent cylinder $s \leq \eta_{eq}$.
    Forcings with $\omega_{in}=\omega$ and $\epsilon_{in}=\epsilon$ on the no-slip inner boundary, which is subject to the same forcing as the outer boundary (except for the stress-free case, labelled SF). 
    Black dashed curves indicate the theoretical profiles in the shell geometry.
    (a) Spherical shell with a no-slip (cyan and orange solid curves) or a stress-free (red and pink solid curves) inner boundary.
    Blued dotted dashed curves illustrate the full-sphere analytical profiles.
    DNS performed at $E=10^{-6}$, $\omega=3$, and $\epsilon=10^{-6}$ for the two forcings.
    (b) Homoeoidal (i.e. $\eta_{pol}=\eta_{eq}$) spheroidal shells subject to longitudinal librations (solid coloured curves). 
    DNS performed at $E=2.5 \times 10^{-6}$, $\omega=\pi$, and $\epsilon=5 \times 10^{-4}$.}
    \label{fig:shellco}
\end{figure}

Note that \citet{sauret2013libration} only considered the particular situation $\omega_i=\omega$ for inner and outer boundaries subject to longitudinal librations, our expression (\ref{eq:coshell}) naturally agrees with their formula (4.24) in this case. 
DNS in spherical shells are in excellent agreement with formula (\ref{eq:coshell}) as shown in figure \ref{fig:shellco}(a), even for latitudinal librations not considered in \citet{sauret2013libration}. 
Note that the observed discontinuity at $s=r_{eq}^{in}$ is related to the presence of the Stewartson layers due to the velocity mismatch between the zonal bulk flow and the inner boundary (these layers are absent for a stress-free inner boundary, as found in figure \ref{fig:shellco}a). 
Introducing nested viscous layers would be required to smooth out the singularity at the Stewartson layer \citep{sauret2013libration}. 
Another striking point in figure \ref{fig:shellco}(a) is that considering a stress-free inner boundary does not modify the mean zonal flow when $s \leq r_{eq}^{in}$ (red curve).
Indeed, if the flow obeys stress-free conditions on the inner boundary, the corresponding mean flow in the tangent cylinder is only generated by nonlinear interactions within the Ekman layer at the outer boundary in formula (\ref{eq:coshell}), and so we recover the coreless solution when $s \leq r_{eq}^{in}$. 
The agreement with DNS is also very good in homoeoidal shells (i.e. $\eta_{pol}=\eta_{eq}$), see in figure \ref{fig:shellco}(b).

\begin{figure}
    \centering
    \begin{tabular}{cc}
    \includegraphics[width=0.49\textwidth]{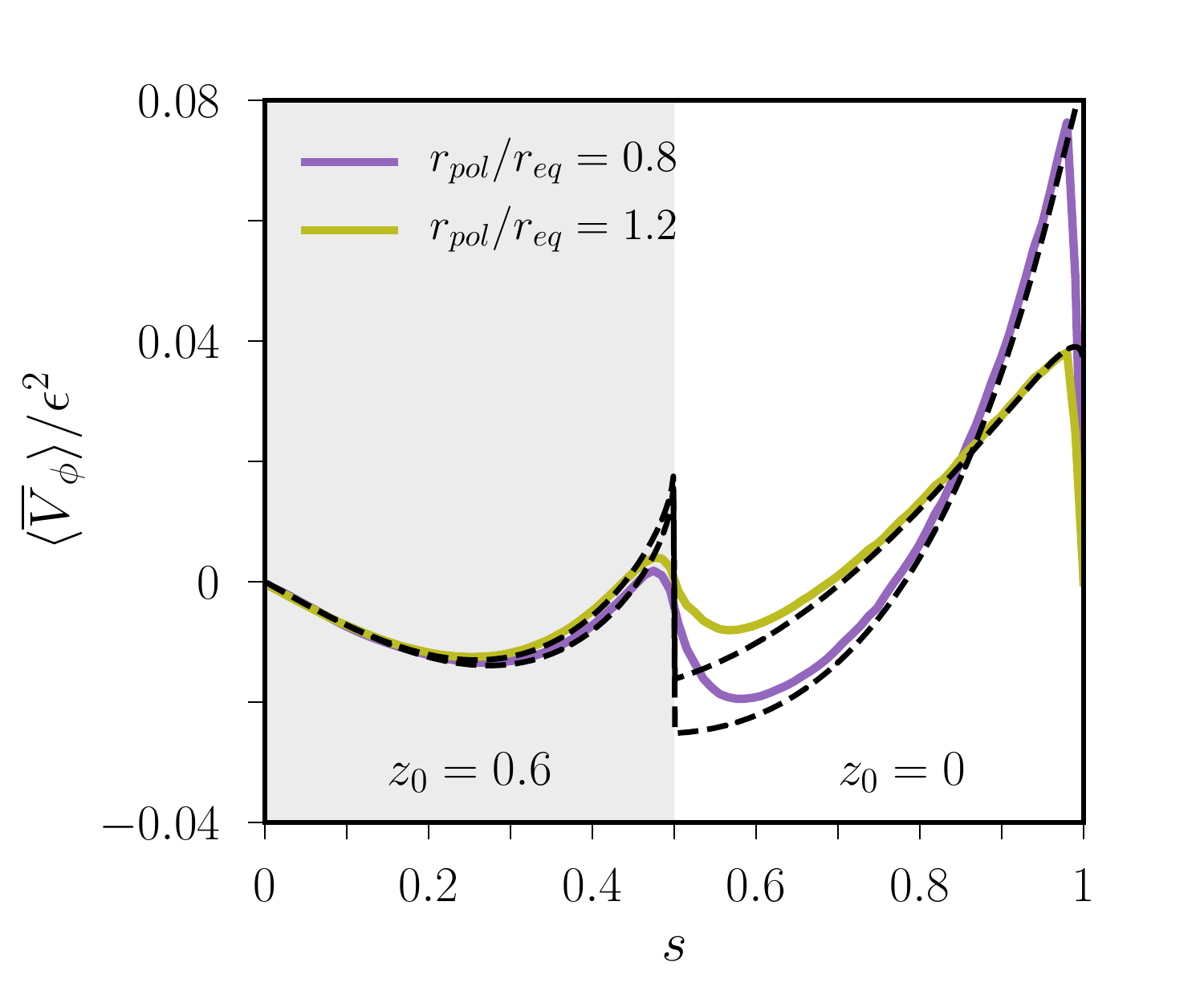} &
    \includegraphics[width=0.49\textwidth]{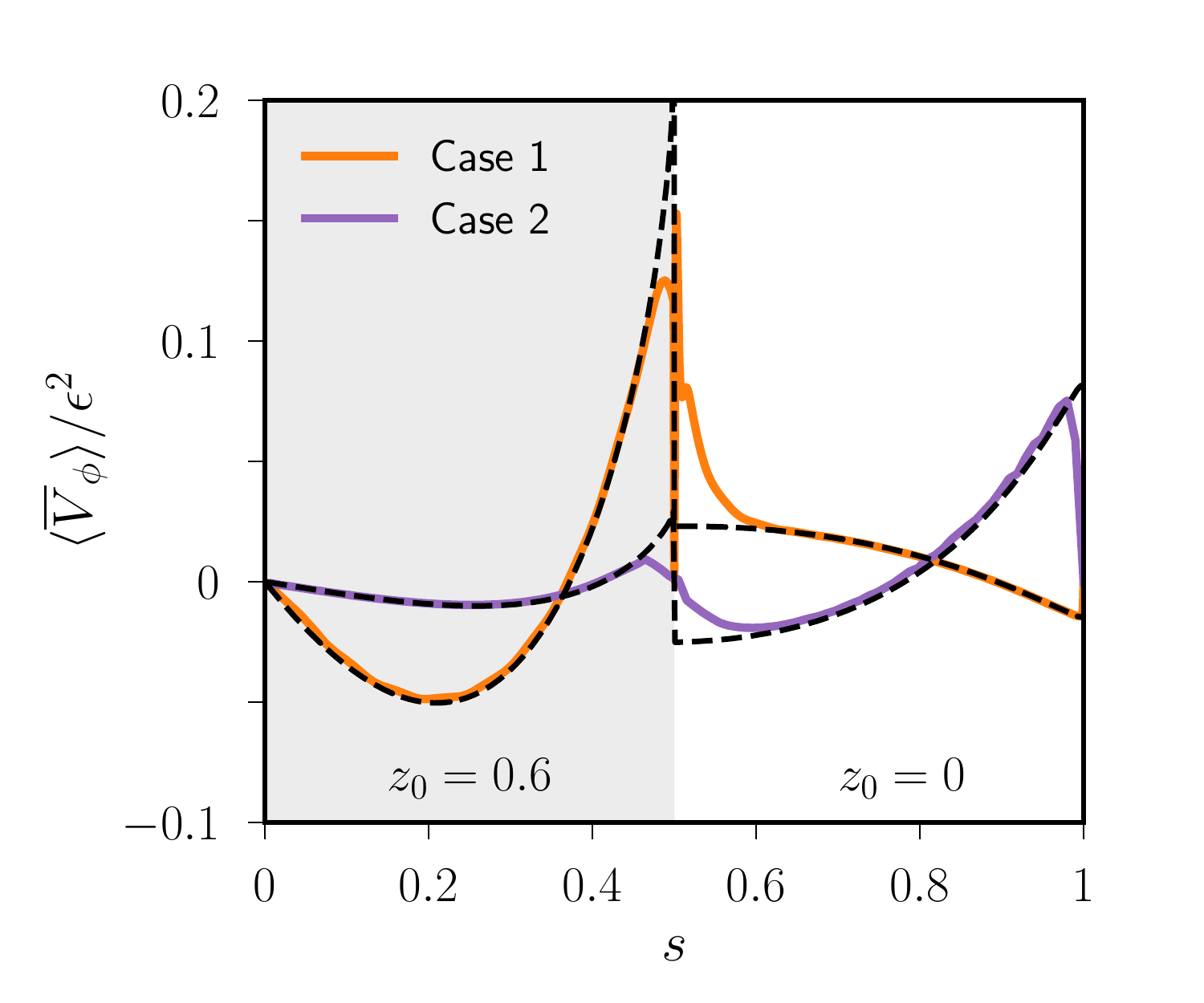} \\
    (a) Non-homoeoidal shells & (b) Different inner and outer forcings
    \end{tabular}
    \caption{Normalised geostrophic mean velocity $\langle \overline{V}_\phi \rangle/\epsilon^2$ at various heights $z=\sqrt{1-s^2}$ in shells with $\eta_{eq}=0.5$.
    Black dashed curves indicate the theory in the shell. 
    The DNS profiles computed at $z=z_0$ for $s \leq \eta_{eq}$ and at $z_0 = 0$ for $s > \eta_{eq}$, giving a single profile for each DNS.
    Grey area shows the tangent cylinder $s \leq \eta_{eq}$. 
    (a) Non-homoeoidal shells subject to longitudinal librations (solid coloured curves), with a spherical inner boundary (i.e. $r_{pol}^{in}=r_{eq}^{in}$) and a spheroidal outer boundary (i.e. $r_{pol} \neq r_{eq}$). 
    DNS performed at $E=2.5 \times 10^{-6}$, $\omega=\omega_{in}=\pi$, $\epsilon=\epsilon_{in}=5 \times 10^{-4}$.
    (b) Homoeoidal shells with $\eta_{pol}=\eta_{eq}=0.5$. 
    Case 1: DNS at $E=10^{-6}$ in a spherical shell subject to latitudinal librations at outer boundary (with $\omega=3$ and $\epsilon=10^{-6}$), and to longitudinal librations at inner boundary (with $\omega_{in}=2$, $\epsilon_{in}/\epsilon=2$).  
    Case 2: Spheroidal shell with $r_{pol}/r_{eq}=0.8$ subject to longitudinal librations.
    DNS at $E=5 \times 10^{-6}$ with $\epsilon=5 \times10^{-4}$ and $\omega=\pi$ at outer boundary, and with $\epsilon_{in}/\epsilon=\omega_{in}/\omega=2$ at inner boundary.}
    \label{fig:shellco2}
\end{figure}

Figure \ref{fig:shellco2} shows that formula (\ref{eq:coshell}) is also valid for other configurations.
An excellent agreement is found in non-homeoidal spheroidal shells (i.e. $\eta_{pol} \neq \eta_{eq}$ with $r_{eq}\neq r_{pol}$) as shown in figure \ref{fig:shellco2}(a) or, as illustrated in figure \ref{fig:shellco2}(b), in homoeoidal shells with distinct angular frequencies $\omega_{in} \neq \omega$ and magnitudes $\epsilon_{in} \neq \epsilon$ (purple curve), or in the presence of different kind of mechanical forcings at inner and outer boundaries (orange curve, when the inner boundary undergoes longitudinal librations and the outer one latitudinal librations). This confirms that formula (\ref{eq:coshell}) is valid even for such complicated cases.

\begin{figure}
    \centering
    \begin{tabular}{cc}
    \includegraphics[width=0.49\textwidth]{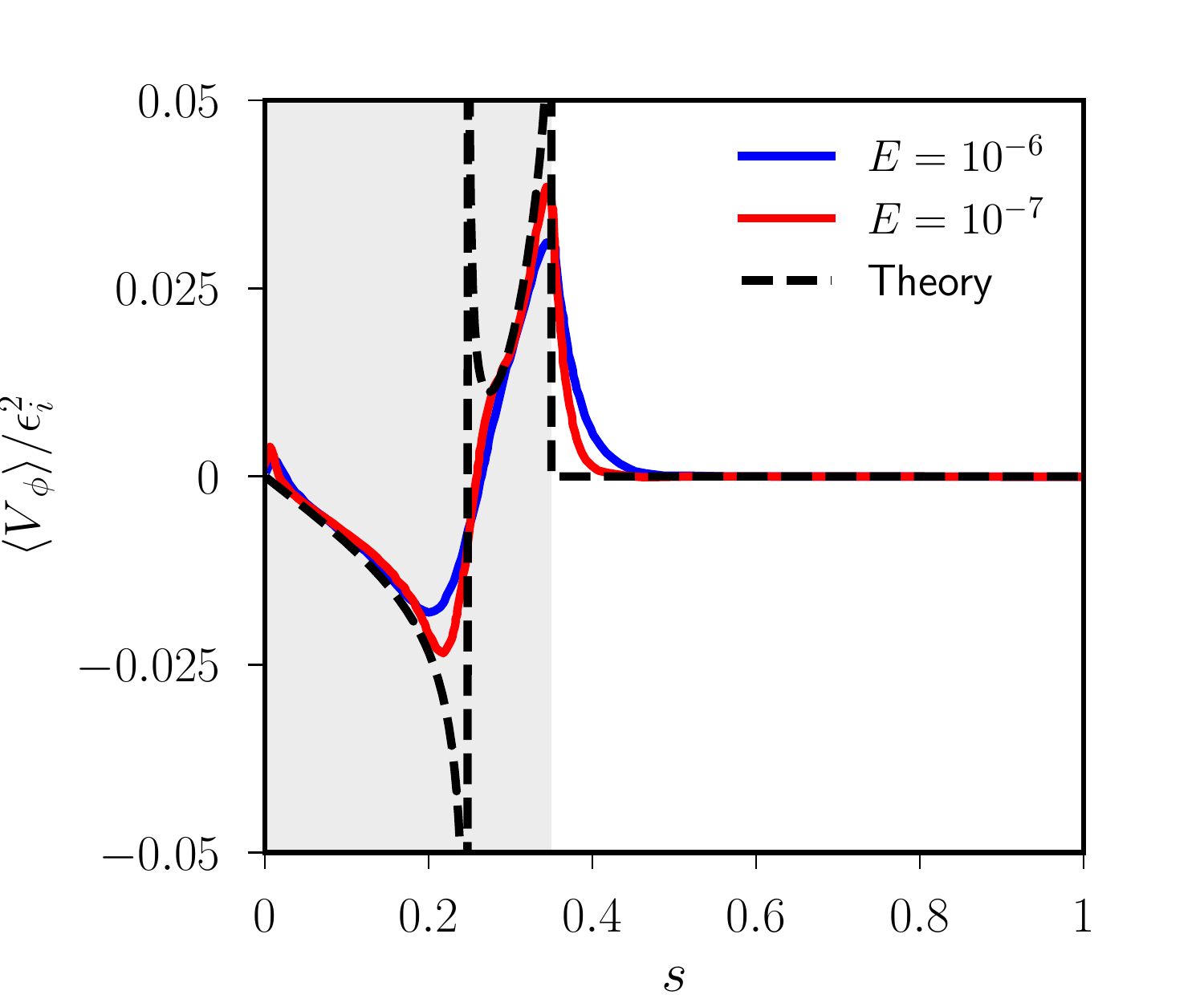} &
    \includegraphics[width=0.49\textwidth]{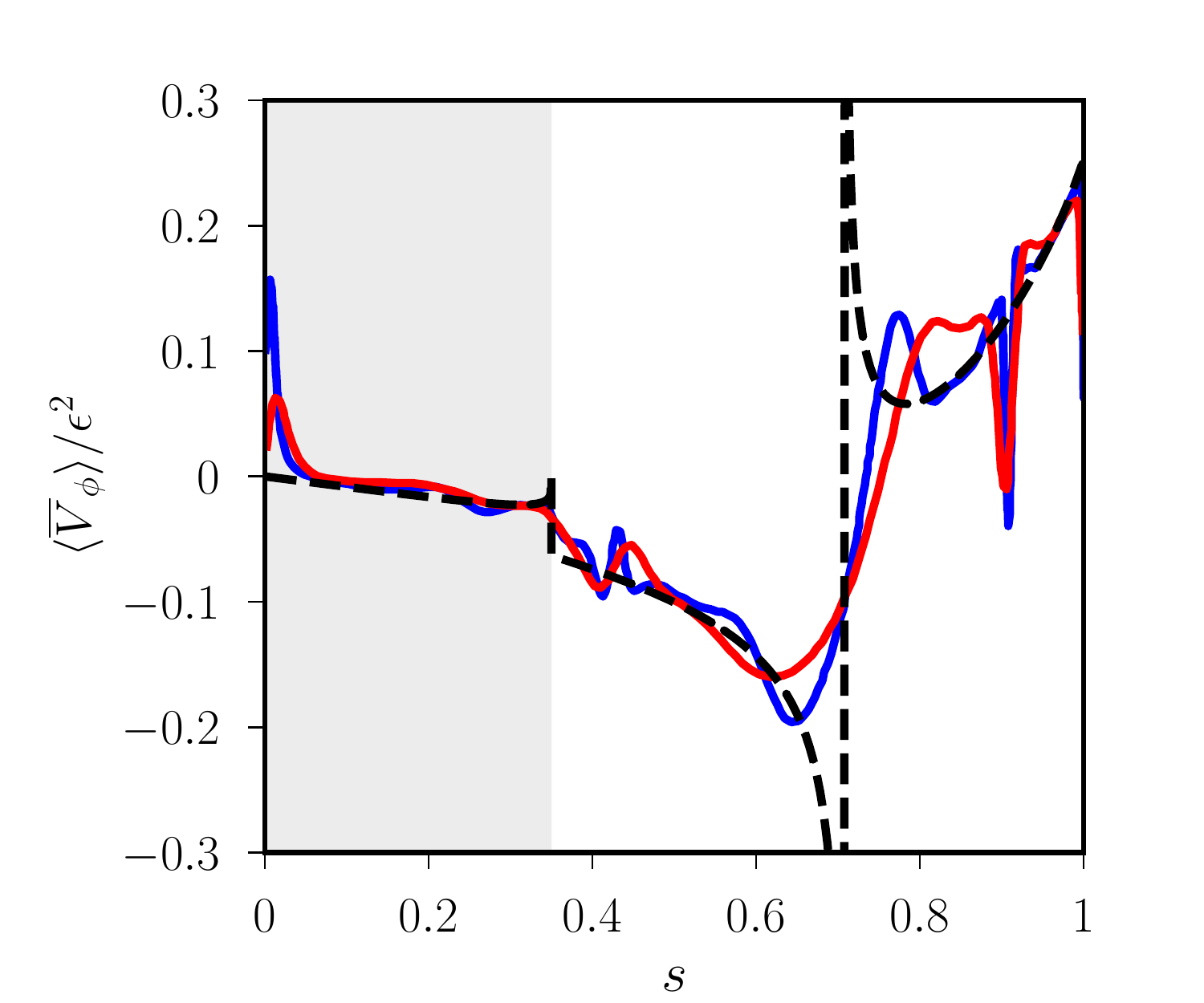} \\
    (a) & (b) \\
    \end{tabular}
    \caption{Normalised geostrophic mean velocity in a spherical shell ($\eta_{pol}=\eta_{eq}=0.35$) subject to longitudinal librations at $\omega=\sqrt{2}$. In panel (a), $\epsilon_{in}=\sqrt{2}/100$ and $\epsilon=0$, whereas, in panel (b), $\epsilon_{in}=0$ and $\epsilon=\sqrt{2}/100$.
    Legend in panel (b) is identical to the one in panel (a). 
    Coloured curves have been reproduced from figures 13(a) and 17(a) in \citet{lin2020libration}, where $\langle \overline{V}_\phi \rangle$ is computed as in formula (\ref{eq:zonalNek5000}), with $z_{\max}=1$, but only considering only the $m=0$ component. 
    Theory is given by formula (\ref{eq:coshell}).
    Grey area indicate the tangent cylinder $s \leq 0.35$.}
    \label{fig:fig1317lin2020}
\end{figure}

We have obtained and validated so far formula (\ref{eq:coshell}) for libration angular frequencies larger than $2$ \citep[see also in][]{sauret2013libration}. 
One can thus wonder how this formula compares with DNS when critical latitudes are present. 
Such a situation has been recently considered for longitudinal librations in \citet{lin2020libration}, for the particular libration frequency $\sqrt{2}$ which is associated with conical shear layers spawned form the critical latitudes leading to a simple closed trajectory for the forced inertial wave in a spherical shell with $\eta_{pol}=\eta_{eq}=0.35$ \citep[e.g.][]{rieutord2001inertial}.
We revisit their results in figure \ref{fig:fig1317lin2020}, reproducing their figure 13(a) in our panel (a), for which only the inner boundary is subject to librations, and their figure 17(a) in our panel (b) that corresponds to the opposite situation.
While it is very challenging to analytically tackle properly this problem, 
we find that formula (\ref{eq:coshell}) provides a reasonably good agreement with DNS, by capturing the essential features of the mean zonal flow profile. 
Therefore, even in the presence of additional complicated flow structures and waves in shell geometries, nonlinear interactions within the Ekman boundary layers still make a significant contribution to the mean zonal flows.
This agrees with previous findings in librating cylinders \citep[see figure 17 in][]{sauret2012fluid}, which showed that analytical theory obtained in the regime $\omega>2$ provides the general trend for $\langle \overline{V}_\phi \rangle$ at $\omega<2$, on which additional mean flow contributions can be superimposed.

\subsection{Planetary applications}
We have shown that our theory fairly predicts the mean zonal flows in rotations ellipsoids and shells, and for various mechanical forcings. 
The relevance of these mean zonal flows ought now to be addressed for planetary applications. 
First, the Ekman boundary layers must be laminar for our theory to be valid.
Various mechanisms are known to destabilise laminar Ekman boundary layers, such as Taylor-G\"ortler instability \citep{noir2009experimental,calkins2010axisymmetric} or local shear instabilities \citep[e.g.][]{lorenzani2001fluid}. 
The transition between laminar and turbulent Ekman boundary layers would here occur when $\epsilon \sim K \, E^{1/2}$ \citep{cebron2019precessing}, where $K$ is a numerical prefactor.
The first boundary-layer instabilities would occur when $K=20-55$ \citep[e.g.][]{lorenzani2001fluid,noir2009experimental,calkins2010axisymmetric}, and fully turbulent boundary layers are expected when $K \simeq 150$ \citep[e.g.][]{caldwell1970characteristics,sous2013friction}. 
For precession and latitudinal librations, note that the conical shear layers spawned from the critical latitudes (either at inner or outer boundaries) can also be prone to shear instabilities  \citep[e.g. the conical shear instability, see in][]{lin2015shear,horimoto2020conical}. 

\begin{figure}
    \centering
    \begin{tabular}{cc}
    \includegraphics[width=0.49\textwidth]{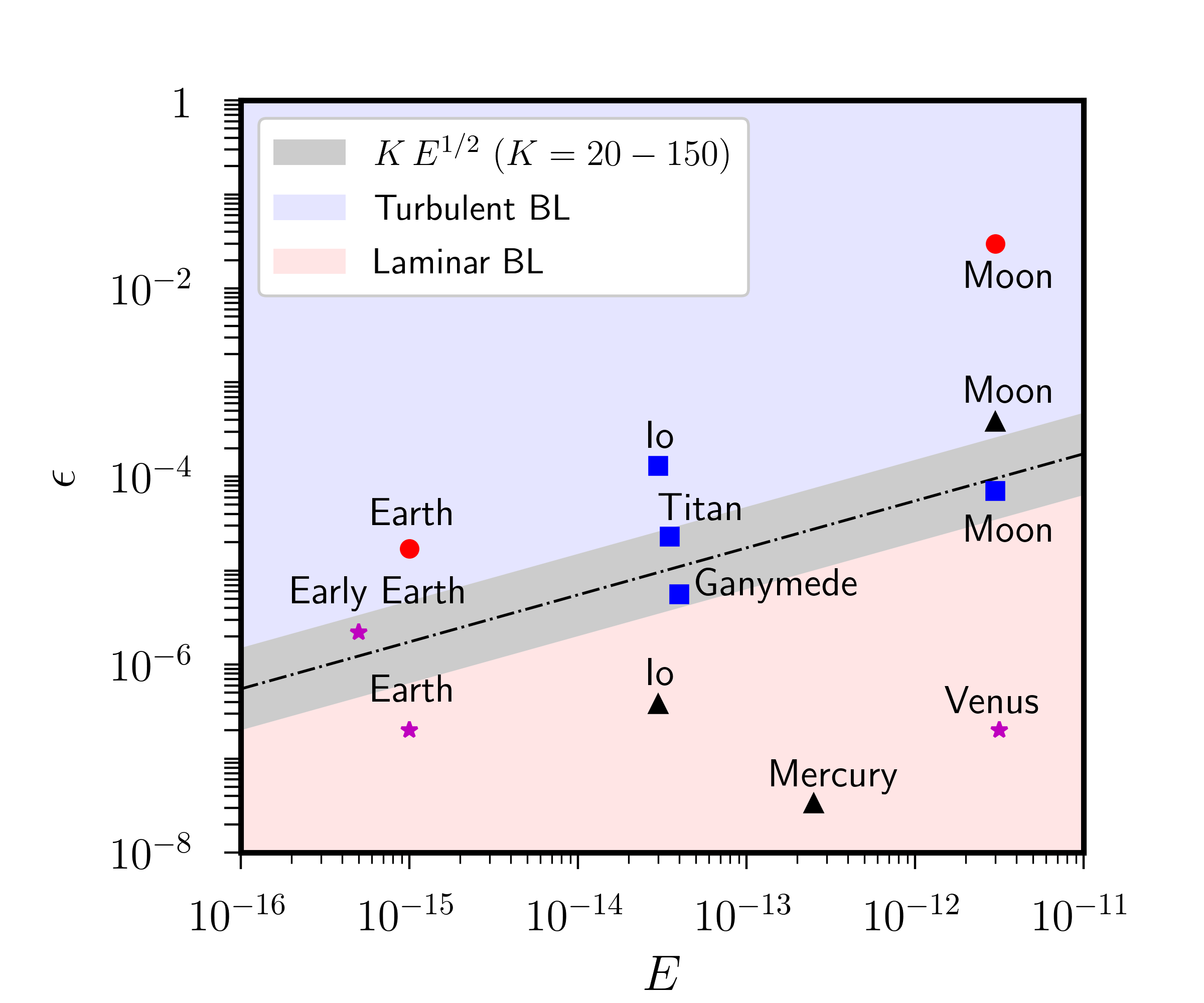} & 
    \includegraphics[width=0.49\textwidth]{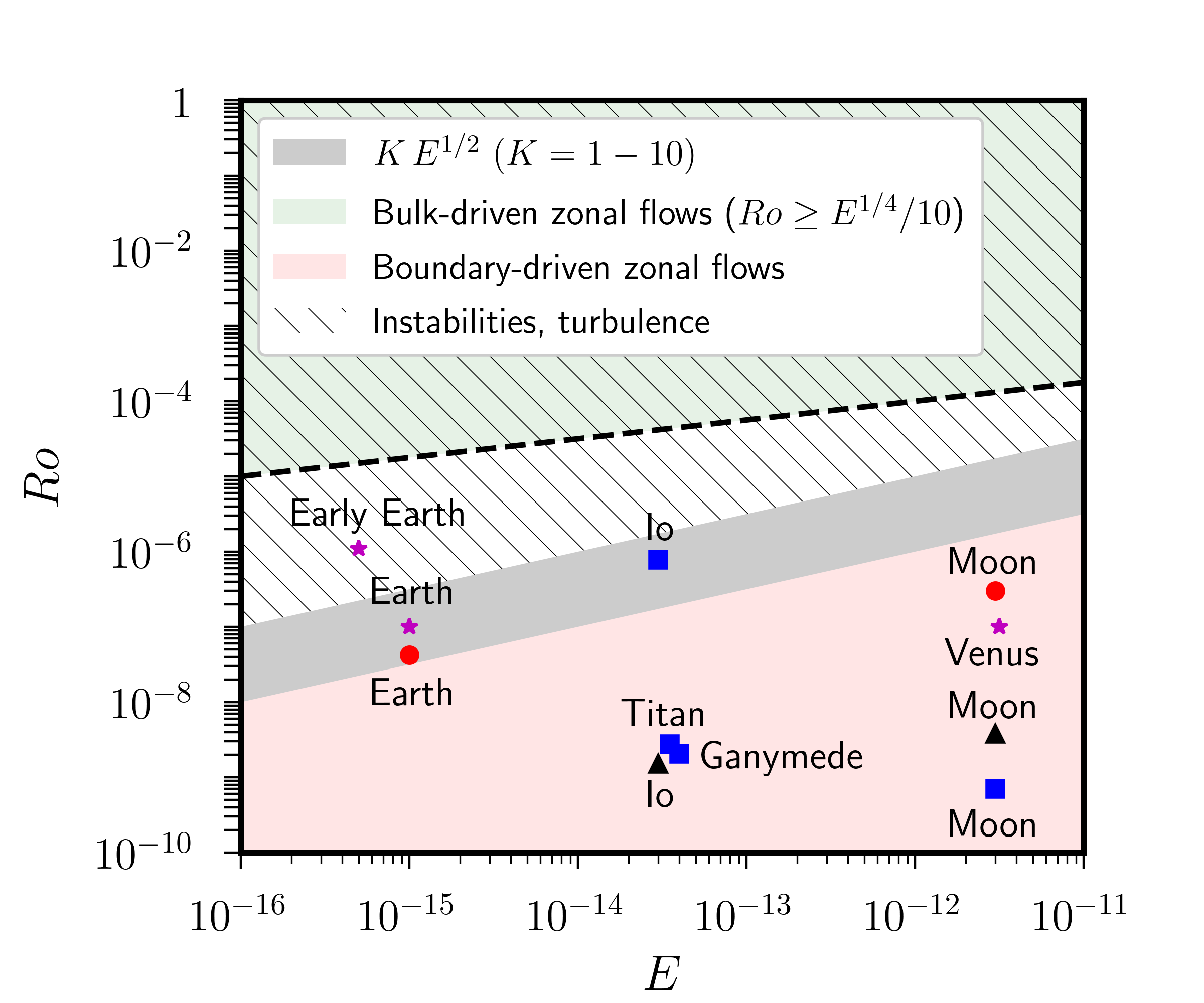} \\ 
    (a) & (b) \\
    \end{tabular}
    \caption{Schematic regime diagrams for the existence of mean zonal flows driven by mechanical forcings
    \citep[planetary values estimated from][]{noir2009experimental,cebron2012elliptical,lin2015shear,vantieghem2015latitudinal}.
    Red circles: precession. Blue squares: longitudinal librations. Black triangles: latitudinal librations. Magenta stars: tides.
    (a) Transition between laminar and turbulent boundary layers (BL). Transition occurs when $\epsilon \sim K \, E^{1/2}$ (with the typical values $K = 20-150$ shown by the grey area, $K = 55$ by the
    dashed dotted line).
   (b) Competition between bulk and boundary driven generation of mean zonal flows. 
   Input Rossby number $Ro = \mathcal{O}(\beta)$ for tidal forcing or $Ro = \mathcal{O}(\epsilon \beta)$ for precession and libration forcings, where $\beta$ is the typical boundary (equatorial or polar) ellipticity.
   Instabilities and bulk turbulence (hatched area) onsets when $Ro \gtrsim K \, E ^{1/2}$ (with the values $K=1-10$ shown by the grey area). 
   Thick dashed line shows $Ro = E ^{1/4}/k$, with $k=10$.}
    \label{fig:extrapolation}
\end{figure}

However, the Ekman boundary layers are expected to become turbulent before the onset of such instabilities when $E \ll 1$ \citep[see figure 6 in][]{cebron2019precessing}.
Next, in addition to laminar boundary layers, our theory also assumes laminar bulk flows. 
Bulk turbulence may indeed alter the mean zonal flows, as previously reported for strong tidal or libration forcings \citep[e.g.][]{favier2015generation,grannan2016tidally}.
To characterise the forcings, we introduce the dimensionless Rossby number
$Ro = \mathcal{U}/(\Omega_s r_{eq})$, where $\mathcal{U}$ is the typical amplitude of the nonlinear flows driven by the mechanical forcings (based on control parameters).
Laminar bulk flows are known to be destabilised by several instabilities when $Ro \geq K E^{1/2}$  (e.g. the elliptical instability), where $Ro$ is also here the typical inviscid growth rate of the instability and $K =1-10$ is a numerical pre-factor due to Ekman pumping \citep{lemasquerier2017libration}. 
In ellipsoids, we have at leading order $Ro \sim \beta$ for tides \citep[e.g.][]{grannan2016tidally,vidal2017inviscid}, and $Ro \sim \epsilon \beta$ for topographic precession \citep{kerswell1993instability} or libration forcings \citep{vantieghem2015latitudinal,vidal2019fossil}, where $\beta$ is here a typical measure of the boundary (equatorial or polar) ellipticity. 
Several secondary instability mechanisms could then occur to sustain bulk-driven zonal flows. 
Although different in nature, these various scenarios are due to nonlinear bulk interactions and apparently all operate on the dimensionless time scale of order $(k Ro)^{-2}$ in the rapidly rotating planetary regime $Ro \ll 1$ \citep[e.g.][]{kerswell1999secondary,brunet2020shortcut,lereun2020greenspan}, where $k$ is a typical wavenumber of the flow. 
On the contrary, boundary-layer interactions establish geostrophic flows on the spin-up time scale $E^{-1/2}$.
The bulk mechanisms should thus operate faster than our boundary-driven mechanism when $k^2 Ro^2 \gg E^{1/2}$, giving $Ro \gg E^{1/4}/k$
\citep[which is also the threshold onset for the secondary instabilities, see in][]{kerswell1999secondary,le2019experimental}. 
The typical wavenumber for the aforementioned bulk mechanisms is poorly constrained from previous studies (as these mechanisms have only been explored without taking the beta effect into account),
such that a rigorous scaling law is still unknown. 

For a direct comparison with the boundary-driven zonal flows, we assume $k = 1-10$ (i.e. to focus on the large-scale components of the geostrophic flows) and illustrate the corresponding stability diagrams for the boundary-driven and bulk-driven mechanisms in figure \ref{fig:extrapolation}. 
Typical planetary values are $E = 10^{-15}- 10^{-12}$ and $\epsilon = 10^{-7}-10^{-3}$, depending on the considered forcing.  
We thus expect laminar Ekman layers in several planetary bodies, as observed in figure \ref{fig:extrapolation}(a). 
Figure \ref{fig:extrapolation}(b) then clearly indicates that the bulk-driven mechanisms are likely irrelevant to explain the occurrence of mean zonal flows in planetary bodies.
On the contrary, several planetary bodies may have simultaneously laminar boundary layers and no bulk-driven turbulence, such that nonlinear interactions within the laminar Ekman layers could be important to generate mean zonal flows. 

\section{Conclusion}
\label{sec:ccl}
\subsection{Summary}
In this work, we have investigated the mean zonal geostrophic flows in rapidly rotating spheres and spheroids subject to weak mechanical forcings (librations, precession and tides).
Geostrophic flows are indeed often encountered in geophysical or astrophysical systems, which are usually attributed to nonlinear interactions occurring at a small scales \citep[e.g.][]{christensen2002zonal,aubert2002observations}.
However, the external mechanical forcings can generate large-scale geostrophic flows in the bulk by nonlinear viscous effects, as considered here. 
We have presented a generic asymptotic theory accounting for the various forcings, in the double limit of small Ekman numbers $E$ and small (dimensionless) forcing amplitude $\epsilon$, and we have analytically considered simultaneously azimuthal and temporal variations of the forcings.  
We have also assessed the range of validity of the analytical profiles as a function of the forcing frequency $\omega$, using targeted DNS. 

For all the forcings, we have shown that the leading-order mean zonal flows in the bulk scale as $\epsilon^2$, and are independent of the Ekman number when $\omega$ is greater than twice the rotation rate in the absence of inertial waves (i.e. $\omega\geq 2$ in dimensionless spin units), as previously found in spherical geometries for precession \citep{busse1968} and longitudinal librations \citep{busse2010,calkins2010axisymmetric,sauret2010,sauret2013libration,lin2020libration}.
Moreover, we have shown that these flows can be significantly modified in spheroids subject to longitudinal librations. 
Our asymptotic theory provides thus a reliable point of comparison for forthcoming experimental measurements, for instance in the ZoRo experiment \citep[Zonal jets in Rotating fluids, see in][]{su2020acoustic,vidal2020compressible} that is currently used to investigate libration-driven zonal flows.

Then, the existence of critical latitudes and inertial waves when $\omega < 2$ is known to lead to more complicated mean zonal flows in terms of amplitude and structure. 
Indeed, the critical shear layers spawned from the critical latitudes \citep[e.g.][]{kerswell1995} are responsible for zonal geostrophic shears at the singular points of the theoretical profiles.
We have numerically confirmed that the geostrophic shear driven by longitudinal and latitudinal librations has a typical width $\delta s \propto E^{1/5}$ and a characteristic amplitude $\delta u_g / \epsilon^2 \propto E^{-1/10}$ when $\omega \gg E^{1/10}$, which contrasts with the scaling law $\delta u_g / \epsilon^2 \propto E^{-3/10}$ for the geostrophic shear driven by precession.

Finally, we have investigated how the mean zonal flows are modified in the presence of a solid inner core. 
We have focused on libration-driven flows to revisit previous numerical findings at low $E$ in shells \citep{lin2020libration}.
Interestingly, we have shown that the mean zonal flows in homoeoidal shells can be fairly estimated from the coreless solutions, in agreement with previous analytical works \citep{sauret2012fluid,sauret2013libration}.

\subsection{Perspectives}
Further work remains to be done to get a more complete description of the generation of mean zonal flows in rapidly rotating bodies. In particular, the competition between bulk-driven and boundary-driven zonal flows should be quantitatively investigated in future studies, to go beyond the qualitative picture discussed above.
The forcing amplitude was indeed set here to be small enough, to filter out any fluid instabilities that can grow in the bulk \citep[e.g.][]{kerswell2002elliptical,lin2015shear,vantieghem2015latitudinal,vidal2017inviscid,nobili2021hysteresis}. 
Only a few experimental or numerical works have hitherto studied mean zonal flows in the presence of bulk turbulence \citep[e.g.][]{favier2015generation,grannan2016tidally,le2019experimental}. 
Yet, it is difficult to draw robust planetary conclusions from these studies, which only explored the dynamics for values of $\epsilon$ and $E$ that were not representative of the planetary regime. 
Therefore, the competition between bulk and boundary mechanisms remains to be explored in the geophysically relevant regime of small forcing amplitudes $\epsilon \ll 1$ and small Ekman numbers $E \ll 1$.
To do so, note that the curvature of the boundaries (i.e. the beta effect) should be included to obtain realistic large-scale zonal flows for planetary systems. 
However, this effect cannot be consistently taken into account in any local Cartesian models that are commonly used in turbulence \citep[e.g.][]{godeferd2015structure}.
Consequently, we should strive considering global geometries to develop more realistic models of planetary bodies.

Apart from the interplay with bulk turbulence, the boundary layers could also become turbulent in the presence of strong enough forcings \citep[e.g.][]{noir2009experimental,calkins2010axisymmetric,sous2013friction}, which could modify the boundary-driven geostrophic flows. 
The geostrophic shear attached to the critical latitudes should also be further characterised. 
For instance, a naive extrapolation of the aforementioned scaling laws would predict an amplitude for the geostrophic shear velocity of $\sim 10^{-1}$ m.s$^{-1}$ for the lunar precession (figure \ref{fig:extrapolation}),  which is an order of magnitude larger than the expected differential velocity between the lunar core and mantle \citep{williams2001lunar}.
Therefore, the observed scaling laws cannot be valid in the asymptotic regime of very low Ekman numbers, as previously reported in preliminary experiments \citep{morize2010experimental}. 
The intense geostrophic differential rotation at the critical latitudes could also become unstable \citep{sauret2014tide}, for instance due to shear instabilities \citep[e.g.][]{busse1968shear,schaeffer2005quasigeostrophic}, which may lead to space-filling turbulence and mixing. 
Moreover, the shell geometry should be further explored for more accurate planetary applications.  
Additional viscous effects are indeed expected due to the presence of an inner core \citep[e.g. the reflection of inertial waves, see in][]{lin2020libration}, such that exploring shell geometries should be further continued. 
Moreover, we have only validated formula (\ref{eq:coshell}) for a few libration-driven zonal flows, but it could apply to other forcings (e.g. precession) in shell geometries or possibly other geometries. For instance, ignoring the need for joining corner regions \cite[as in][]{wedemeyer1966viscous}, mean zonal flow can be calculated in no-slip half-spheroids \cite[as e.g. in][]{noir2012experimental} by summing the contribution of the plane boundary layer \cite[i.e. half the mean zonal flow in the cylinder, given by][]{wang1970cylindrical} and the contribution of the curved boundary (i.e. half the one of the full spheroid). 

Finally, the core-mantle boundary of most planets exhibits roughness \citep{narteau2001small,le2006dissipation}, but scant attention has been given to the flow dynamics in the presence of small-scale topography \citep[e.g.][]{burmann2018effects}. 
However, our asymptotic theory could be used to get further physical insights into topographic effects for planetary applications. 
A small-scale azimuthal roughness could be mimicked here using the multipolar tidal-like forcing (\ref{cond_limite567}) with $\omega=0$ and $\boldsymbol{\Omega}_c=\boldsymbol{0}$ (such that $\boldsymbol{U}= \widehat{\boldsymbol{z}}_R \times \boldsymbol{r}$). 
The mathematical problem is tractable  in the short azimuthal wavelength approximation (i.e. $m \gg 1$), and we obtain the mean zonal flow $f(s) = s^{2(m-2)}/4 \to 0$ when $m \to \infty$ \citep[the mean zonal flow driven by weak librations of a rotating sphere also vanishes in the limit $\omega \gg 1$, see in][] {sauret2013libration}.
Therefore, it appears that a small-scale azimuthal roughness is unlikely to sustain significant mean zonal flows in planetary interiors via this mechanism. 
Investigating this problem deserves further numerical work, as well as exploring the flows driven by other small-scale topographies.

\backsection[Acknowledgements]{J.V. and D.C. acknowledge Dr Benjamin Favier for sharing his mapping to model non-homoeoidal shells in \textsc{Nek5000}.
D.C. acknowledges Dr Lo\"{i}c Huder for his expert support in \textsc{Python}.}

\backsection[Funding]{This work received funding from the European Research Council (ERC) under the European Union's Horizon 2020 research and innovation programme via the \textsc{theia} project (grant agreement no. 847433). 
ISTerre is part of Labex OSUG@2020 (ANR10 LABX56).
The \textsc{xshells} code received funding from the European Union's Horizon 2020 research and innovation programme under the ChEESE project, grant agreement no. 823844.
This work was granted access to the HPC resources of TGCC and CINES under allocation A0080407382 attributed by GENCI (Grand Equipement National de Calcul Intensif).}

\backsection[Declaration of interests]{The authors report no conflict of interest.}

\backsection[Data availability statement]{The open-source codes \textsc{xshells} and \textsc{Nek5000} are available at \url{https://nschaeff.bitbucket.io/xshells} and \url{https://nek5000.mcs.anl.gov/}, respectively. 
The analytical calculations have been checked using Maple software.}

\backsection[Author ORCID]{D.C., \href{https://orcid.org/0000-0002-3579-8281}{0000-0002-3579-8281};
J.V., \href{https://orcid.org/0000-0002-3654-6633}{0000-0002-3654-6633};
N.S., \href{https://orcid.org/0000-0001-5206-3394}{0000-0001-5206-3394};
A.S., \href{https://orcid.org/0000-0001-7874-5983}{0000-0001-7874-5983}}

\backsection[Author contributions]{The theory was initiated by A.S. during his PhD. D.C then corrected his calculations during the first COVID-19 lockdown, and extended them to account for other forcings and geometries. 
J.V. conducted the DNS using \textsc{Nek5000}.
D.C. and N.S. performed the DNS of libration-driven flows with \textsc{xshells}, whereas the DNS of precession were conducted by N.S. and A.B. 
The discussion was led by J.V. and D.C., who both drafted the manuscript. N.S and A.S proof-checked the article, and the authors gave final approval for submission.}

\appendix

\section{Central regularity conditions with finite differences}
\label{sec:jvreg}
We detail here how the central conditions is implemented with finite differences in \textsc{xshells}. We expand the velocity field $\boldsymbol{V}$ onto the set of spherical harmonics $Y_l^m$ using the poloidal-toroidal decomposition in spherical coordinates $(r, \theta, \phi)$
\begin{subequations}
\label{eq:poltorappendix}
\begin{equation}
    \boldsymbol{V} = \sum_{l\geq 1} \sum_{|m| \leq l} \boldsymbol{V}_l^m, \quad
    \boldsymbol{V}_l^m = \boldsymbol{\nabla} \times \boldsymbol{\nabla} \times (P_l^m (r) Y_l^m \, \boldsymbol{r}) + \boldsymbol{\nabla} \times (T_l^m (r) Y_l^m \, \boldsymbol{r}),
    \tag{\theequation \emph{a,b}}
\end{equation}
\end{subequations}
where $[P_l^m(r),T_l^m(r)]$ are the poloidal-toroidal radial scalars.
These scalars must satisfy regularity conditions at the centre for the vector field to be regular and infinitely differentiable.
To do this, the two scalars and the Cartesian components of the velocity fields must behave like monomials in the Cartesian coordinates $(x,y,z)$. 
This is ensured by expanding $[P_l^m(r),T_l^m(r)]$ in the regular form \citep[e.g.][]{dudley1989time}
\begin{equation}
    \left [ P_l^m(r), T_l^m(r) \right ] = \sum_{j\geq 0} [A_j, B_j] \, r^{2j+l},
    \label{eq:poltorPOL}
\end{equation}
where $[A_j, B_j]$ are unknown coefficients. 
At the centre $r=0$, the $(l,m)$ components for the velocity (\ref{eq:poltorappendix}) then reduce to 
\begin{equation}
    \boldsymbol{V}_l^m = \begin{cases}
    2 A_0 \left (Y_1^m, \partial_\theta Y_1^m, (1/\sin \theta) \, \partial_\phi Y_1^m \right)^\top & \text{for} \quad l=1, \\
    (0,0,0)^\top &\text{for} \quad l\neq 1,
    \end{cases}
    \label{eq:central1}
\end{equation}
with $A_0 = \left. \partial_r P_1^m \right|_{r=0}$. 
Within \textsc{xshells}, the poloidal-decomposition is implemented using vector spherical harmonics that depend on the radial scalar $l(l+1) P_l^m/r$, the spheroidal scalar $S_l^m =(1/r) \, \partial_r (r P_l^m)$ and the toroidal scalar $T_l^m$. 
Therefore, we obtain from (\ref{eq:central1}) the following regularity conditions for a finite difference scheme
\begin{equation}
    P_l^m (r=0) = 0, \quad T_l^m  (r=0) = 0, \quad S_l^m (r=0) = \begin{cases}
    2 \left . \partial_r P_1^m \right|_{r=0} & \text{for} \quad l = 1, \\
    0  & \text{for} \quad l \neq 1.
    \end{cases}
    \label{eq:central_cnd}
\end{equation}
This allows a non-zero velocity at the centre, corresponding to a flow going through it.

These regularity conditions apply to the velocity, but also to the vorticity $\boldsymbol{W}_l^m$, which is related to the velocity
\begin{eqnarray}
    \boldsymbol{W}_l^m &=& \boldsymbol{\nabla} \times \boldsymbol{\nabla} \times \boldsymbol{\nabla} \times (P_l^m (r) Y_l^m \, \boldsymbol{r}) + \boldsymbol{\nabla} \times \boldsymbol{\nabla} \times (T_l^m (r) Y_l^m \, \boldsymbol{r}) \\
    &=& \boldsymbol{\nabla} \times ( -\Delta P_l^m (r) Y_l^m \, \boldsymbol{r}) + \boldsymbol{\nabla} \times \boldsymbol{\nabla} \times (T_l^m (r) Y_l^m \, \boldsymbol{r})
\end{eqnarray}
Applying the same reasoning as above leads to condition (\ref{eq:central_cnd}) with $P_l^m=0$ replaced by $T_l^m=0$ and $T_l^m=0$ replaced by $\Delta P_l^m=0$.
For $l=1$, this also leads to $\partial_{rr} P_1^m|_{r=0} = 0$.
This vanishing second-order derivative of $P_1^m$ ensures that the error for the 2-point finite difference approximation of $S_1^m(r=0)$ is of order 2
\begin{equation}
S_1^m(r=0) = 2\partial_r P_1^m|_{r=0} = 2 P_1^m(\epsilon)/\epsilon + O(\epsilon^2).
\end{equation}
The above conditions are actually simpler to implement with finite differences than with some spectral descriptions in radius \citep[e.g. see the discussion in][]{livermore2005comparison}.

In addition, to avoid a stringent restriction on the time-step size, the spherical harmonic expansion is truncated near the centre at lower degrees
\begin{equation}
l_{tr}(r) = l_{\max} \, \sqrt{\frac{r}{\alpha \max(r)}} + 1,
\label{eq:ltrunc}
\end{equation}
where $l_{\max}$ is the maximum spherical harmonic degree in the DNS, and $\alpha=0.05$ is found to be an appropriate parameter to avoid spurious numerical errors near $r=0$ while allowing large enough time steps. 
In practice, our resolution for all the DNS ensured that the truncation quickly jumps to $l \geq 6$ at the second radial point (not shown), ensuring a sufficient numerical resolution.

Finally, the above numerical implementation is accurate enough to determine the values of the mean flow rotation rate $f$ near the centre.
Indeed, defining 
\begin{equation}
    \langle \overline{V}_\phi \rangle / s = \sum \limits_{l\geq 1} \overline{\boldsymbol{V}}_l^0 \boldsymbol{\cdot} \widehat{\boldsymbol{\phi}} /s,
\end{equation}
with the cylindrical radius $s = r \sin \theta$, we obtain the value of the rotation rate in the equatorial plane $\theta=\pi/2$ (i.e. $z_0=0$) from the toroidal component $T_1^0(r)$
\begin{equation}
    \langle \overline{V}_\phi \rangle /s = - \partial_r T_1^0|_{r=0} \left. \partial_\theta Y_1^0 \right |_{\theta=\pi/2}
    \label{eq:fTOR}
\end{equation}
Expression (\ref{eq:fTOR}) shows that, as expected, the rotation rate is regular on the axis of rotation \citep[and more generally in coreless geometries, e.g. see][]{lewis1990physical}, with a possible non-zero value $\propto \partial_r T_1^0|_{r=0}$ due to the $l=1$ spherical harmonic in full spheres. 

The error for the 2-point finite difference approximation of $f(r=0) \propto \partial_r T_1^0(r=0)$ is guaranteed to be of order 1
\begin{equation}
f(r=0) \propto \partial_r T_1^0|_{r=0} = T_1^0(\epsilon)/\epsilon + O(\epsilon).
\end{equation}
We took special care to refine the grid near $r=0$ to ensure the reported values for $f(r=0)$ are meaningful, although they are less accurate than the values away from 0.

\section{Details on the theoretical calculations} \label{sec:ans} \label{sec:appBH}

In this appendix, we first provide some useful formulas in the spheroidal coordinates used in this work (section \ref{sec:ans1}). We then provide details on the calculation of the first-order flows in the bulk (sections \ref{sec:ans2}-\ref{sec:ans23}) and in the boundary layer (section \ref{sec:ans3}).

\subsection{Miscellaneous formula in our spheroidal coordinates} \label{sec:ans1}
The ratio of the spheroid axes is $r_{pol}/r_{eq}=\mathcal{T}^\prime_{(Q_1)}/\mathcal{T}_{(Q_1)}$. For oblate and prolate spheroidal coordinates, it gives respectively $r_{pol}/r_{eq}=\tanh Q_1$ and $r_{pol}/r_{eq}=\coth Q_1$. 
Note that we only have $\widehat{\boldsymbol{n}}=\widehat{\boldsymbol{q}}_1 $ at the boundary $q_1=Q_1$. 
Using these coordinates, the leading-order bulk flow is then $\Omega_0 \, \widehat{\boldsymbol{z}}_R \times \boldsymbol{r}=\Omega_0 \, s\, \widehat{\boldsymbol{\phi}} $, where $s=(x^2+y^2)^{1/2}=a \mathcal{T}_{(q_1)} \sin q_2$ is the cylindrical radius. 
Finally, the iso-surface for $q_1$ becomes more and more spherical when $q_1$ becomes large, with a spherical radius $r$ given by $a \exp(q_1)\approx 2 r $.

Note also that, for every boundary layer-flow $\boldsymbol{v}$ of components $\boldsymbol{v}=(0,{v}_{2},{v}_{\phi})^{\top}$ in our spheroidal coordinates, we have
\begin{eqnarray}
\label{relationdure52}
\widehat{\boldsymbol{n}} \boldsymbol{\cdot} \boldsymbol{\nabla} \times \int_{0}^{\infty}\, \widehat{\boldsymbol{n}} \times \boldsymbol{v} \, \mathrm{d} \zeta &  = &   \frac{1}{a \sin q_2} \left[ \frac{1}{\tilde{h}} \frac{\partial}{\partial q_2} \left( \sin q_2 \int_{0}^{\infty} {v}_{2} \, \mathrm{d}\zeta \right) \right. \nonumber \\
& &\qquad \qquad \qquad \qquad   \left. + \frac{1}{\mathcal{T}_{(q_1)}} \frac{\partial}{\partial \phi} \left( \int_{0}^{\infty} {v}_{\phi}\mathrm{d}\zeta \right)  \right] ,
\end{eqnarray} 
where the last term vanishes when calculating the mean zonal component of $\boldsymbol{v}$.

\subsection{Generic solution for the forced interior flow ${}_P\boldsymbol{U}_0^1$}
\label{sec:ans2}
Using the Cartesian coordinates of our frame of reference, the uniform-vorticity flow satisfying the no-penetration condition is given by \citep[e.g.][]{noir2013precession}
\begin{equation} 
\boldsymbol{U}_0^1 = 
 \, \begin{pmatrix}
f_0^2 \, \mathcal{Q}_y\, z_R -\mathcal{Q}_z \, y_R \\
- f_0^2\, \mathcal{Q}_x \, z_R+ \mathcal{Q}_z x_R \\
f_0^2\, (r_{pol}/r_{eq})^2 \, [\mathcal{Q}_x \, y_R - \mathcal{Q}_y \, x_R]  
\end{pmatrix}
\label{eq:mld0cwm}
\end{equation}
with $f_0^2=2/(1+r_{pol}^2/r_{eq}^2)$, and where the (constant) rotation vector $\boldsymbol{\mathcal{Q}}=(\boldsymbol{\nabla} \times \boldsymbol{U}_0^1)/2$ has the Cartesian components $\boldsymbol{\mathcal{Q}}= \mathcal{Q}_x \, \widehat{\boldsymbol{x}}_R + \mathcal{Q}_y \, \widehat{\boldsymbol{y}}_R + \mathcal{Q}_z \, \widehat{\boldsymbol{z}}_R$ given by
\begin{subequations}
\label{eq:besoin}
\begin{align}
\mathcal{Q}_x &= \widehat{\mathcal{Q}}_x \cos (\omega t) + \widehat{\mathcal{Q}}_x^\star \sin (\omega t ), \\
\mathcal{Q}_y &= (\widehat{\mathcal{Q}}_y \sin \omega t + \widehat{\mathcal{Q}}_y^\star \cos \omega t), \\
\mathcal{Q}_z &= -\widehat{\mathcal{Q}}_z \cos (\omega t).
\end{align} 
\end{subequations}
This generic form encompasses all the particular cases considered in this article, and is compliant with the ansatz used below to integrate equations (\ref{eq_CL01DF}). 

For instance, the flow $\boldsymbol{U}_0^1$ driven by longitudinal librations in the mantle frame of reference is obtained with $\boldsymbol{\mathcal{Q}}=-\cos (\omega t) \, \widehat{\boldsymbol{z}}_R$. 
For latitudinal librations, the viscous flow in the mantle frame of reference is obtained from \citet{vantieghem2015latitudinal}. 
We have corrected a few typos in their expressions (3.29)-(3.31), which leads to
\begin{subequations}
\begin{eqnarray}
\widehat{\mathcal{Q}}_x &=&\frac{(\omega^2-f^2)(f_0^2-\omega^2)-K^2(f_0^2+\omega^2)}{(\omega^2-f^2)^2+2K^2(f^2+\omega^2)+K^4}, \\
\widehat{\mathcal{Q}}_x^\star &=&K \omega \frac{\omega^2+f^2-2f_0^2+K^2}{(\omega^2-f^2)^2+2K^2(f^2+\omega^2)+K^4}, \\
\widehat{\mathcal{Q}}_y &=& \omega \frac{\beta_{bc}(\omega^2-f^2)-K^2(1+f^2/f_0^2)}{(\omega^2-f^2)^2+2K^2(f^2+\omega^2)+K^4}, \\
\widehat{\mathcal{Q}}_y^\star &=& K \frac{\omega^2+f^2-2\omega^2 f^2/f_0^2+K^2}{(\omega^2-f^2)^2+2K^2(f^2+\omega^2)+K^4},
\end{eqnarray}
\end{subequations}
with $\beta_{bc}=(f^2-f_0^2)/f_0^2$, $K = \Lambda E^{1/2}$ where $\Lambda \geq 0$ is the viscous damping factor, and with the eigenfrequency $f$ of the spin-over mode, that is $f=f_0^2$ for the spheroid as in \citep{vantieghem2015latitudinal}. 
Yet, note the erroneous presence in their formula (3.30) of $\beta_{bc}$ in $\widehat{\mathcal{Q}}_y^\star$, as well as that of $\widehat{\boldsymbol z}_M$ instead of $\widehat{\boldsymbol z}_I$ in their expression (A5). 
In the inviscid limit $E=0$, we have $\widehat{\mathcal{Q}}_x^\star=\widehat{\mathcal{Q}}_y^\star=0$ and, for a spheroid,
\begin{subequations}
\begin{equation}
\widehat{\mathcal{Q}}_x = (f-\omega^2 )/(\omega^2-f^2), \quad
\widehat{\mathcal{Q}}_y = \omega \beta /(\omega^2-f^2),
\tag{\theequation \emph{a,b}}
\end{equation}
\end{subequations}
where $\beta=(r_{eq}^2-r_{pol}^2)/(r_{eq}^2+r_{pol}^2)$, giving a diverging flow at $\omega=f$ for any spheroids (except the sphere $\beta= 0$, where the solution reduces to $\widehat{\mathcal{Q}}_x =-1 $ and $\widehat{\mathcal{Q}}_y=0$). 
Finally, the precession-driven flow $\boldsymbol{U}_0^1$ can be written under this form in the mean rotating frame (with $\omega=0$), also called precessing frame, or in the mantle frame \citep[][]{noir2013precession}.

\subsection{Calculation of ${}_H\boldsymbol{U}_0^1$ for longitudinal librations}
\label{sec:ans22}
In their analytical study, \citet{greenspan1963time} considered longitudinal librations at $\omega \ll 1$ in axisymmetric arbitrary containers of revolution around the $\widehat{\boldsymbol{z}}_R$, and obtained ${}_H\boldsymbol{U}_0^1$ in the form of a quasi-geostrophic flow. 
Their assumption $\omega \ll 1$ implies a steady boundary layer at leading order, whereas the boundary layer can be time dependent when $\omega \gg E^{1/2}$ (as considered in the main text), in particular when $\omega \geq \mathcal{O}(1)$. 
Since the two regimes overlap for $E^{1/2} \ll \omega \ll 1$, we aim to understand the transition between these two situations. 
To do so, we determine below the quasi-geostrophic component ${}_g\boldsymbol{U}_0^1$ of ${}_H\boldsymbol{U}_0^1$ for arbitrary values of $\omega$, extending the study of \citet{greenspan1963time} to time-dependent boundary-layer flows. The slight differences between ${}_H\boldsymbol{U}_0^1$ and  ${}_g\boldsymbol{U}_0^1$ will then be briefly studied by performing the exact calculation in the particular case $\omega>2$ (i.e. without any critical latitude).

Using cylindrical coordinates $(s,\phi,z)$ in the mean rotating frame, we consider a fluid within an arbitrary axisymmetric container  $\tilde{g}_1(s) \leq z \leq \tilde{g}_2(s)$, with $\tilde{g}_1(s) \leq 0$ and $\tilde{g}_2(s) \geq 0$. The zonal (i.e. $m=0$) component  ${}_{zo}\boldsymbol{U}_0^1$ of ${}_H\boldsymbol{U}_0^1$ can be written as
\begin{equation}
{}_{zo}\boldsymbol{U}_0^1 = - \boldsymbol{\nabla} \times( \Psi_0^1(s,z,t) \, \widehat{\boldsymbol{\phi}} ) + \mathcal{V}_0^1(s,z,t) \, \widehat{\boldsymbol{\phi}} =[\partial_z \Psi_0^1, \mathcal{V}_0^1,-\partial_s(s \, \Psi_0^1)/s  ]^{\top} ,
\end{equation}
which satisfies divergenceless condition (\ref{eq:legi0}b). 
Then, the curl of equation (\ref{eq:legi0}a), that is
\begin{equation} 
\label{eq:legi}
(\partial_t+\Omega_0 \partial_{{\phi}}) \, \boldsymbol{\nabla} \times ( {}_H\boldsymbol{U}_0^1) -  2\, \Omega_s^0\, (\widehat{\boldsymbol{z}}_R \boldsymbol{\cdot} \boldsymbol{\nabla} ) {}_H\boldsymbol{U}_0^1 = \boldsymbol{0},
\end{equation}
can be written in this case
\begin{subequations}
 \label{eq:bessel}
\begin{equation}
  \partial_t \mathcal{V}_0^1  = - 2 \,  \partial_z \Psi_0^1   , \qquad 
\partial_t (\tilde{\mathcal{L}} \Psi_0^1)  = 2 \,  \partial_z \mathcal{V}_0^1,
\tag{\theequation \emph{a,b}}  
\end{equation}
\end{subequations}
with $\tilde{\mathcal{L}}\Psi_0^1=(\nabla^2-s^{-2}) \Psi_0^1=\partial_s[s^{-1} \partial_s(s \Psi_0^1)]+\partial_{zz}^2 \Psi_0^1$. 
To obtain the quasi-geostrophic component ${}_g\boldsymbol{U}_0^1$ of ${}_H\boldsymbol{U}_0^1$, the velocity components perpendicular to the rotation axis are assumed to be $z-$invariant  \citep{labbe2015magnetostrophic}. 
Then, equation (\ref{eq:bessel}) naturally reduces to the Taylor-Proudman constraint $\partial_z ({}_g\boldsymbol{U}_0^1)=0 $. 
Using the axisymmetric decompositions
\begin{subequations}
\begin{align}
{}_g\boldsymbol{U}_0^1 + E^{1/2} \boldsymbol{U}_1^1 &= - E^{1/2} \, \boldsymbol{\nabla} \times( \chi_I \, \widehat{\boldsymbol{\phi}} ) + \mathcal{V}_I \, \widehat{\boldsymbol{\phi}}, \\
{}_g\boldsymbol{u}_0^1 + E^{1/2} \boldsymbol{u}_1^1 &= - E^{1/2} \, \boldsymbol{\nabla} \times( \chi_B \, \widehat{\boldsymbol{\phi}} ) + \mathcal{V}_B \, \widehat{\boldsymbol{\phi}},
\end{align}
\end{subequations}
where we have anticipated that the meridional stream functions are of the order $E^{1/2}$ \citep[as in][]{greenspan1963time}, we obtain $\partial_{\tau} \mathcal{V}_I+2 \, \partial_z \chi_I =0$ and then
\begin{equation}
\label{eq:appel}
\chi_I=- (z/2) \, \partial_{\tau} \mathcal{V}_I  + \chi_I^0,
\end{equation}
as previously obtained by \citet{greenspan1963time}. 
As we follow closely the approach and the notations of \citet{greenspan1963time}, we do not remind below all the intermediate steps. 
In the boundary layer, equations (5.7)-(5.10) of \citet{greenspan1963time} are modified into \citep[see also equation (3.2) in][]{sauret2013libration}
\begin{subequations}
\label{eq:packar}
\begin{equation}
\left( \partial^2_{\zeta \zeta}  - \partial_t \right) \mathcal{V}_B-2 \, \widehat{\boldsymbol{n}} \boldsymbol{\cdot} \widehat{\boldsymbol{z}}_R \  \partial_{\zeta} \chi_B = 0, \quad 
\left( \partial^2_{\zeta \zeta}  - \partial_t \right) \partial^2_{\zeta \zeta}  \chi_B  +2 \, \widehat{\boldsymbol{n}} \boldsymbol{\cdot} \widehat{\boldsymbol{z}}_R \ \partial_{\zeta} \mathcal{V}_B  = 0,
\tag{\theequation \emph{a,b}}
\end{equation}
\end{subequations}
which can be integrated by considering the ansatz $\mathrm{e}^{ \mathrm{i} \omega t}$ for $[\chi_I,\chi_B,\mathcal{V}_I,\mathcal{V}_B]$, leading to 
\begin{eqnarray} \label{eq:tosh}
\mathcal{V}_B&=&\frac{ s\, (1-f_0)}{4}  \left( \mathrm{e}^{-\lambda_+ \zeta}+ \mathrm{e}^{-\lambda_-^* \zeta} \right) \mathrm{e}^{\mathrm{i}\, \omega t}  + \text{c.c.}, \end{eqnarray}
with $\mathcal{V}_I= \Im_m \left( f_0 \, s \,  \mathrm{e}^{\, \mathrm{i} \,\omega t}  \right) $, and $\text{c.c.}$ the complex conjugate. 
Note that equation (\ref{eq:tosh}) can naturally be exactly recovered with equation (\ref{eq:dcfou3}) by considering the BC $\boldsymbol{u}_0^1+\boldsymbol{U}_0^1  =  \boldsymbol{V}_{\Sigma}^1 $ with a non-zero $_g\boldsymbol{U}_0^1$.
From equation (\ref{eq:packar}) we get
\begin{subeqnarray}
 \label{eq:dellu}
\chi_B(\zeta=0)|_{z=\tilde{g}_i}&=& \mathrm{sgn}(\tilde{g}_i) \, \frac{ s(1-f_0) \, \tilde{\Lambda}_i}{4} \, \mathrm{e}^{\mathrm{i}\, \omega t}  + \text{c.c.} , \\
\tilde{\Lambda}_i &=& \frac{1}{2 \, | \widehat{\boldsymbol{n}}_i \boldsymbol{\cdot} \widehat{\boldsymbol{z}}_R |} \left[\lambda_+ + \lambda_-^*-\mathrm{i}\, \omega \left(\frac{1}{\lambda_+ }+\frac{1}{\lambda_-^* } \right) \right],
\end{subeqnarray}
with $|\widehat{\boldsymbol{n}}_i \boldsymbol{\cdot} \widehat{\boldsymbol{z}}_R | =[1+({\mathrm{d}_s \tilde{g}_i})^2]^{-1/2}$, and where the terms $\lambda_+$ and $\lambda_-^*$ involved in $\tilde{\Lambda}_i$ have to be calculated with the associated $\widehat{\boldsymbol{n}}_i \boldsymbol{\cdot} \widehat{\boldsymbol{z}}_R$ (noting ${\mathrm{d}_s \tilde{g}_i}$  the derivative of the one-variable functions $\tilde{g}_1(s) \leq z \leq \tilde{g}_2(s)$ describing the container geometry). 
Then, the BC $\chi_B+\chi_I=0$ can be written using equations (\ref{eq:appel}) and (\ref{eq:dellu}) as
\begin{subequations}
\label{eq:dfcbw}
\begin{align} 
 \partial_{\tau} \mathcal{V}_I &=  2 \, [\chi_B(\zeta=0)|_{z=\tilde{g}_2}-\chi_B(\zeta=0)|_{z=\tilde{g}_1}][\tilde{g}_2-\tilde{g}_1]^{-1},  \\
\chi_I^0 &=  [\tilde{g}_1\, \chi_B(\zeta=0)|_{z=\tilde{g}_2} - \tilde{g}_2\,  \chi_B(\zeta=0)|_{z=\tilde{g}_1} ][\tilde{g}_2-\tilde{g}_1]^{-1} ,
\end{align}
\end{subequations}
which provides $\chi_I^0= \Im_m \left( A_0 \,  \mathrm{e}^{\, \mathrm{i} \,\omega t}  \right) $ and
\begin{subequations}
\label{eq:phili}
 \begin{equation}  
 f_0=(1+\mathrm{i} \, \omega \, \mathcal{S}^{-1} \, E^{-1/2})^{-1} ,  \qquad 
\mathcal{S}= (\tilde{\Lambda}_1+\tilde{\Lambda}_2)/(\tilde{g}_2- \tilde{g}_1) .
       \tag{\theequation \emph{a,b}}
 \end{equation}
\end{subequations} 
Equation (\ref{eq:phili}) gives $\mathcal{V}_I$, but also $\chi_I$ using equation (\ref{eq:appel}), providing both the geostrophic part of $\boldsymbol{U}_0^1$ and the associated component of $\boldsymbol{U}_1^1$. 
\begin{figure}
    \centering
    \begin{tabular}{cc}
    \includegraphics[width=0.49\textwidth]{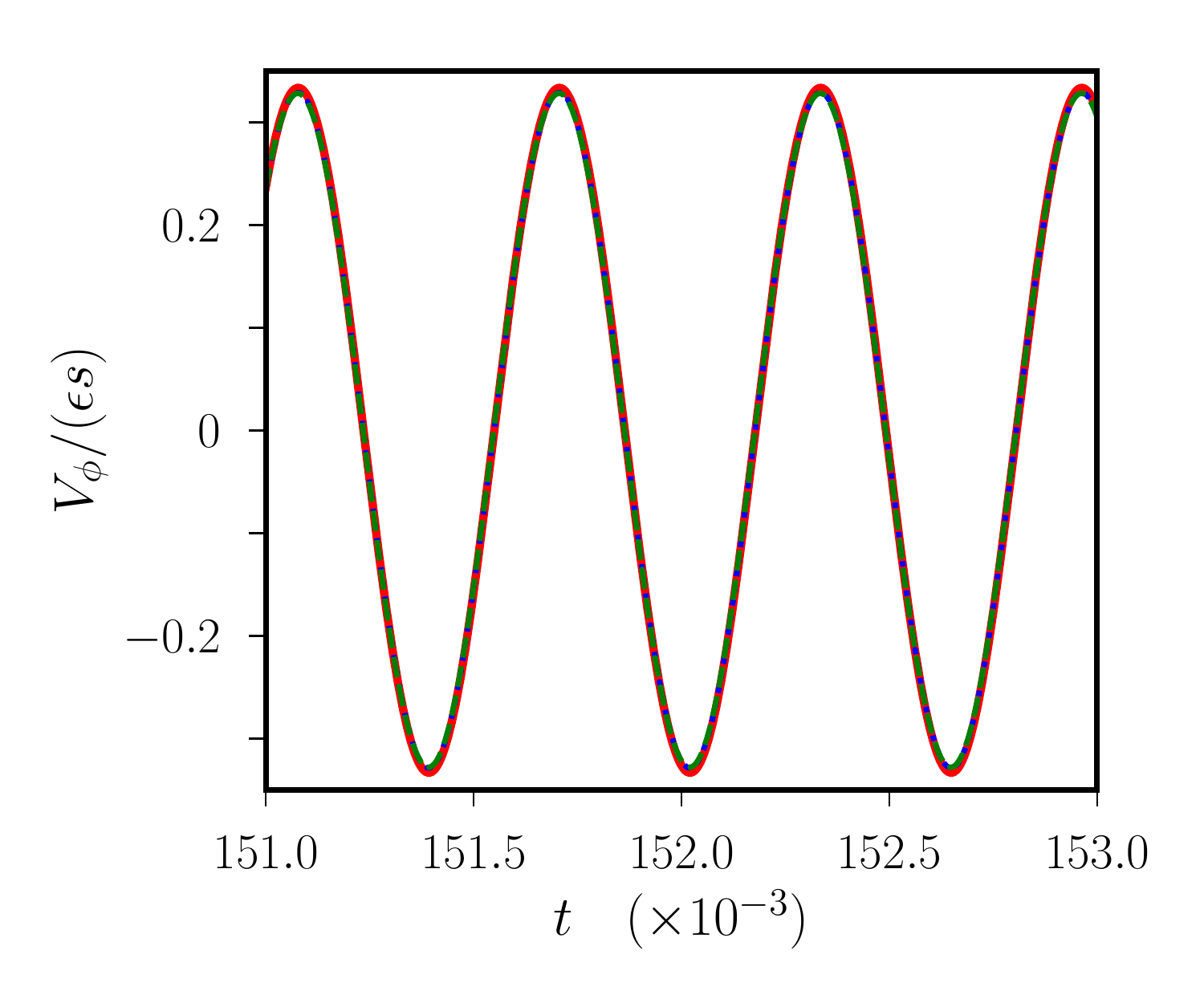} & 
    \includegraphics[width=0.49\textwidth]{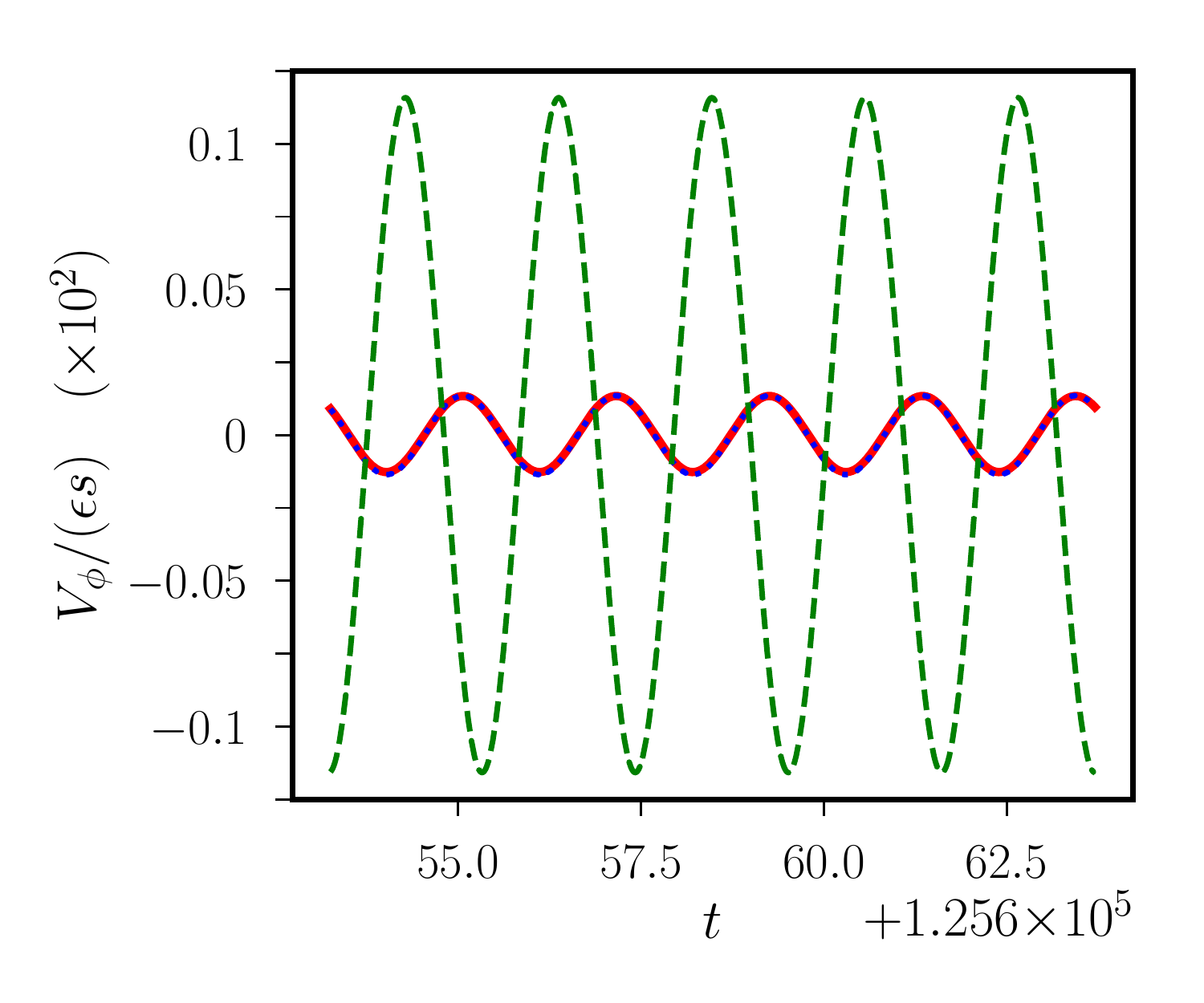} \\
    (a) $\omega=10^{-2}$ & (b) $\omega=3$\\
    \end{tabular}
    \caption{Normalised (instantaneous) rotation rate $V_{\phi}/(\epsilon s)$ as a function of time for longitudinal librations of a sphere ($E=10^{-6}$ and $\epsilon=10^{-4}$), at position $(s=0.9,q_2=\pi/2)$ in the equatorial plane. DNS are given by the solid red lines and the theory $V_{\phi}/(\epsilon s)=\Im_m \left( f_0 \,  \mathrm{e}^{\, \mathrm{i} \,\omega t}  \right)$, where $f_0$ is given by equation (\ref{eq:phili}a), is shown as dotted blue and dashed green lines, when calculating $\mathcal{S}$ with equation (\ref{eq:phili}b) or with its approximation for $\omega=0$, respectively.}
    \label{fig:timedepcirc}
\end{figure}
The asymptotic regime studied by \citet{greenspan1963time} is recovered by using $\omega=0$ in $\mathcal{S}$ (i.e. $\omega \ll 1$), that is when the boundary layer can be assumed to be steady (giving e.g. $\mathcal{S} \approx (1-s^2)^{-3/4}$  within the sphere in this regime). 
In figure \ref{fig:timedepcirc}(a), we confirm that this approximation is in excellent agreement with DNS for $\omega \ll 1$. 
By contrast, figure \ref{fig:timedepcirc}(b) shows that only the more complete theory developed here is in excellent agreement with DNS for $\omega \geq \mathcal{O}(1)$, as expected. We now come back to our initial question of the decrease of the bulk flow $\mathcal{V}_I$ in the limit $\omega \gg E^{1/2}$, for the opposite regimes $\omega \ll 1$ and $\omega \gg 1$. Since $\mathcal{S}$ is independent of $\omega$ in the particular regime $\omega \ll 1$, equation (\ref{eq:phili}) shows that $\mathcal{V}_I$ decreases toward $0$ as $\mathcal{V}_I \sim E^{1/2}/\omega$. 
By contrast, $\mathcal{S} \sim \sqrt{\omega}$ for $\omega \gg 1$, showing that  $\mathcal{V}_I$ decreases toward $0$ as $\mathcal{V}_I \sim \sqrt{E/\omega}$ in this regime.

\begin{figure}
    \centering
    \begin{tabular}{cc}
    \includegraphics[width=0.49\textwidth]{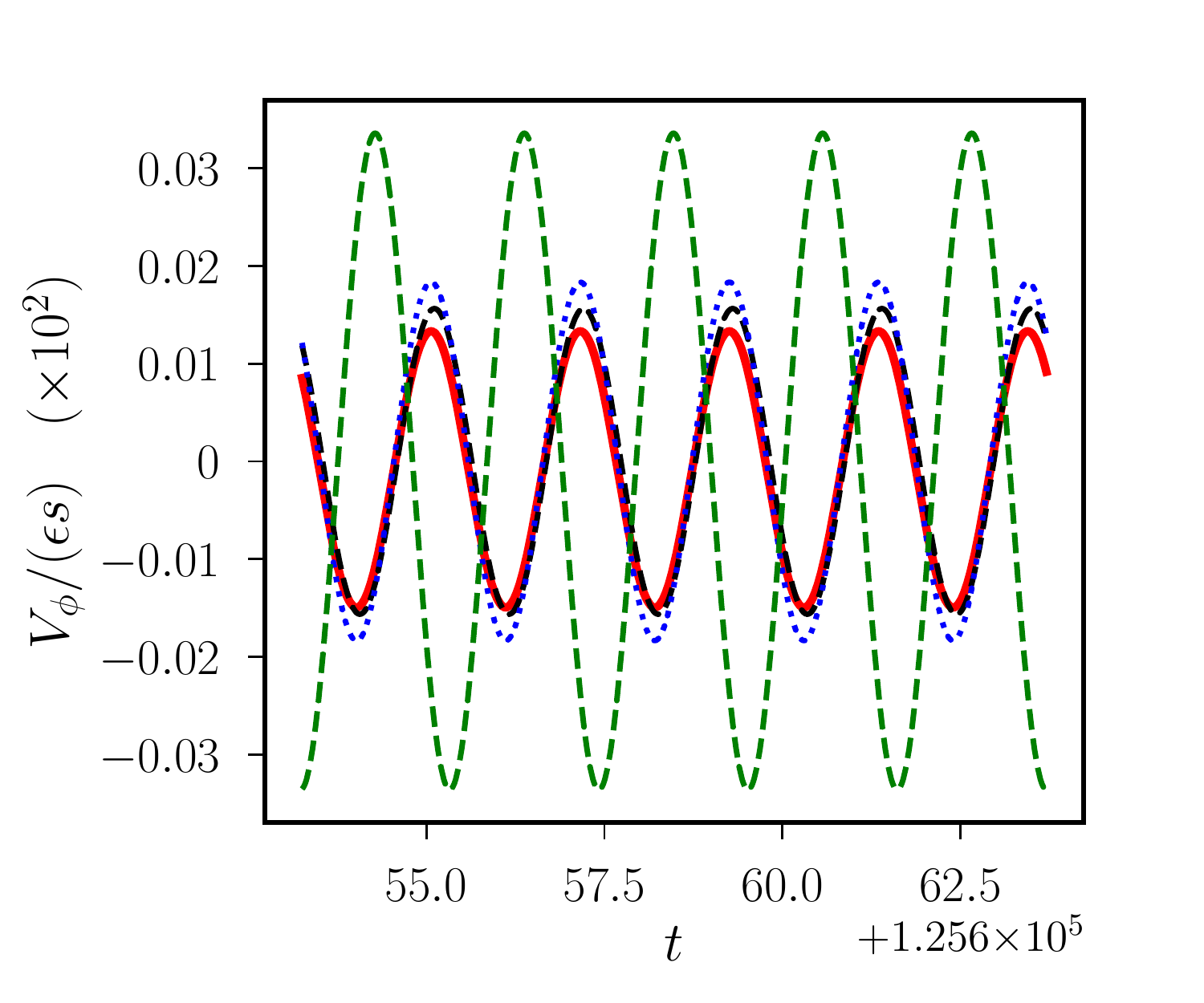} & 
    \includegraphics[width=0.35\textwidth,trim=2cm 0cm 0 0cm, clip]{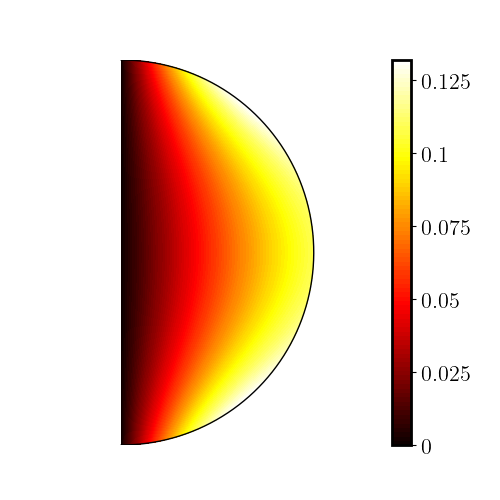} \\
    (a)  & (b) \\
    \end{tabular}
    \caption{(a) Same as in Figure \ref{fig:timedepcirc}(b) but at $s=0.1$ and showing the solution obtained from equation (\ref{eq:pd5e23}) in black dashed line. (b) Snapshot $\widehat{v}_{\phi} = - 2 \partial_z \widehat{\Psi}_0^1/(\mathrm{i} \omega) $ in a meridional section of the theoretical azimuthal flow obtained from equation (\ref{eq:pd5e23}). We obtain an excellent agreement with the analogous snapshot from DNS (not shown).}
    \label{fig:deouf}
\end{figure}

To confirm further that ${}_g\boldsymbol{U}_0^1$ is a good approximation of ${}_H\boldsymbol{U}_0^1$, we now aim at obtaining ${}_H\boldsymbol{U}_0^1$ directly from equation (\ref{eq:bessel}) without the quasi-geostrophic assumption. 
To this end, we have adapted the calculation of \cite{wang1970cylindrical} performed in cylinders. 
With the ansatz $(\mathcal{V}_0^1 ,\Psi_0^1)=(\widehat{\mathcal{V}}_0^1,\widehat{\Psi_0^1}) \mathrm{e}^{\mathrm{i} \omega t} +c.c.$ and noting $\tilde{\omega}=1-4/\omega^2$, equation (\ref{eq:bessel}) gives
\begin{eqnarray}
\partial_s [ \partial_s (s \widehat{\Psi}_0^1) /s ] +\tilde{\omega} \, \partial^2_{zz} \widehat{\Psi}_0^1 =0 ,
\end{eqnarray}
whose general solution is (imposing regularity at $s=0$)
\begin{eqnarray}
\widehat{\Psi}_0^1 &=& \sum_{k}  \mathrm{J}_1(\tilde{C}_k s) \left[ \tilde{A}_k \mathrm{e}^{\tilde{C}_k z/\tilde{\omega}^{1/2}} +\tilde{B}_k \mathrm{e}^{-\tilde{C}_k z/\tilde{\omega}^{1/2}}   \right] ,
\label{eq:pd5e230}
\end{eqnarray}
with $\mathrm{J}_1$ the Bessel function of the first kind and with the constants $[\tilde{A}_k,\tilde{B}_k,\tilde{C}_k]$. In the quasi-static regime $\omega \to \infty$, the leading order naturally recovers the linear $z$-dependency of quasi-geostrophic solution (\ref{eq:appel}). In symmetric containers with $\tilde{g}_1= -\tilde{g}_2$, we have $\widehat{\Psi}_0^1(z=0) =0$ by symmetry, such that $\tilde{A}_k=-\tilde{B}_k$. Equation (\ref{eq:pd5e230}) then reduces to
\begin{eqnarray}
\widehat{\Psi}_0^1 &=& \sum_{k} \tilde{A}_k\, \mathrm{J}_1(\tilde{C}_k s)   \sinh{\frac{\tilde{C}_k z}{\tilde{\omega}^{1/2}}} .
\label{eq:pd5e23}
\end{eqnarray}
The (complex-valued) constants $[\tilde{A}_k,\tilde{C}_k]$ are then fixed by the BC $\Psi_0^1+E^{1/2} \chi_B=0$ at $\zeta=0$, where $\chi_B$ is given by equation (\ref{eq:dellu}). 
Contrary to the cylinder, the properties of Fourier-Bessel series cannot be used to obtain the constants \citep{wang1970cylindrical}. 
Using $f_0=\widehat{v}_{\phi}/s$ and $\widehat{v}_{\phi} = - 2 \partial_z \widehat{\Psi}_0^1/(\mathrm{i} \omega) $ in $\chi_B(\zeta=0)$, the BC reads
\begin{eqnarray}
 \frac{ \tilde{\Lambda}_2 }{4}  s+ \sum_{k}  \tilde{a}_k  \mathrm{J}_1(\tilde{C}_k s) \left[ \sinh{\frac{\tilde{C}_k z}{\tilde{\omega}^{1/2}}}- \frac{ \mathrm{i}\,  \tilde{\Lambda}_2 \, \tilde{C}_k}{2 \omega\, \tilde{\omega}^{1/2}}  \cosh{\frac{\tilde{C}_k z}{\tilde{\omega}^{1/2}}} \right]  =0 , \label{eq:dlm45yyhuu}
\end{eqnarray}
with $\tilde{a}_k=\tilde{A}_k/E^{1/2} $. When $\omega<2$, many terms are required in equation (\ref{eq:dlm45yyhuu}) to accurately fulfil this BC (due to the divergence at the critical latitude). In this case, bulk inertial modes can also be excited \citep[e.g.][]{aldridge1969axisymmetric,zhang2017theory}, and they can then constitute a better basis to describe $\widehat{\Psi}_0^1$. By contrast, we find that only few terms are necessary for $\omega>2$, and there is no excitation of inertial mode. Considering for instance $\omega=3$ in the sphere ($s=\sin \theta$, $z=\cos \theta$) and the first three terms of $\widehat{\Psi}_0^1$, we impose the BC at six equally spaced values of the colatitude $\theta \in ]0, \pi/2[$. The numerical integration of this nonlinear system of six equations provides then the $6$ constants, with $\tilde{a}_1 \approx 0.26 + 0.65  \mathrm{i}$ and $\tilde{C}_1 \approx 0.63 - 0.21 \mathrm{i}$ for the leading-order term ($\tilde{a}_2$ and $\tilde{a}_3$ are respectively $\sim 10$ and $\sim 10^3$ times smaller, showing convergence of the series). One can then check a posteriori that BC (\ref{eq:dlm45yyhuu}) is verified on the whole range $\theta \in [0, \pi/2]$ with a maximum error $<2 \times 10^{-6}$. 
Using these constants, we can then calculate the three components of velocity and compare with DNS. This solution naturally recovers the excellent agreement shown in figure \ref{fig:timedepcirc}(b), and even agrees slightly better with DNS as shown in figure \ref{fig:deouf}(a) for $s=0.1$. 
As can be expected, this difference is due to the weak departure of the flow from quasi-geostrophy, as shown in figure \ref{fig:deouf}(b) that is in perfect agreement with our DNS. 
To conclude this section, note that the mean zonal component of $({}_H\boldsymbol{U}_0^1 \boldsymbol{\cdot} \boldsymbol{\nabla}) \, {}_H\boldsymbol{U}_0^1$ is non-zero, and ${}_H\boldsymbol{U}_0^1$ should thus a priori modify the bulk mean zonal flow when not considering the no spin-up regime $\omega \gg E^{1/2}$.

\subsection{Inertial modes excited by longitudinal librations and mean zonal flows}
\label{sec:ans23}
Longitudinal librations can excite inertial modes through the Ekman pumping $\boldsymbol{u}_1^1$ generated by the oscillating Ekman layer \cite[see sections 2.12 and 2.14 in][]{greenspanbook}. 
The amplitude of such forced inertial modes
is of the order $\mathcal{O}(\epsilon E^{0})$ when the forcing frequency $\omega \gg E^{1/2}$ matches the frequency of an inertial mode \citep{greenspanbook,aldridge1969axisymmetric,zhang2017theory}.
This inertial mode excitation is not an inviscid resonance \citep[as encountered e.g. with latitudinal librations, see in][]{vantieghem2015latitudinal},
and thus its amplitude remains finite in the limit $E \to 0$ (but vanishes when $E=0$). 
For a sphere in longitudinal libration at the inertial mode frequency $\omega=\omega_{12}=\sqrt{12/7} \approx 1.309 $, \citet{zhang2017theory} obtained the following flow in spherical coordinates $(r,\theta,\phi)$ in the container frame (rotating at $\Omega_c=1+\epsilon \sin \omega t $ here)
\begin{subequations}
\label{eq:gjvid}
\allowdisplaybreaks
\begin{eqnarray}
{}_P\boldsymbol{U}_0^1 &=& \Re_e \left( \mathrm{i} s \,  \mathrm{e}^{\mathrm{i} \omega_{12} t} \, \widehat{\boldsymbol{\phi}} \right),  \\
{}_H\boldsymbol{U}_0^1 &=&\Re_e \left( \frac{15 \tilde{\mathcal{A}}_{12}}{4} \left[ \frac{\mathrm{i} \sqrt{3}}{2 \sqrt{7} } \left( (r^3-r)(1+3 \cos 2 \theta) \, \widehat{\boldsymbol{r}}+\sin 2 \theta (3r-5r^3) \, \widehat{\boldsymbol{\theta}} \right) \right. \right.  \nonumber \\ & & \left. \left.  + \sin \theta \, (2r^3+r^3 \cos 2 \theta -r)  \, \widehat{\boldsymbol{\phi}} \right]  \mathrm{e}^{\mathrm{i} \omega_{12} t}  \right),  \\
{}_P\boldsymbol{u}_0^1 &=& \Re_e \left(\sin \theta \left[ - \frac{\mathrm{i} \widehat{\boldsymbol{\phi}} - \widehat{\boldsymbol{\theta}} }{2} \, \mathrm{e}^{\lambda_+ \zeta}- \frac{\mathrm{i} \widehat{\boldsymbol{\phi}} + \widehat{\boldsymbol{\theta}} }{2} \,  \mathrm{e}^{\lambda_-^* \zeta} \right]  \mathrm{e}^{\mathrm{i} \omega_{12} t} \right),  \\
{}_H\boldsymbol{u}_0^1 &=& \Re_e \left( \frac{15 \tilde{\mathcal{A}}_{12}}{4} \left[ \frac{\mathrm{i}}{2} \left( 1+\cos 2 \theta - \omega_{12} \cos \theta \right) \sin \theta (\mathrm{i} \widehat{\boldsymbol{\phi}} - \widehat{\boldsymbol{\theta}}) \,  \mathrm{e}^{\lambda_+ \zeta} \right. \right. \nonumber \\
& & \left. \left.  + \frac{\mathrm{i}}{2} \left( 1+\cos 2 \theta + \omega_{12} \cos \theta \right) \sin \theta \,  (\mathrm{i} \widehat{\boldsymbol{\phi}} + \widehat{\boldsymbol{\theta}}) \,  \mathrm{e}^{\lambda_-^* \zeta}  \right] \mathrm{e}^{\mathrm{i} \omega_{12} t} \right),
\end{eqnarray}
\end{subequations}
with $\tilde{\mathcal{A}}_{12}=0.034156-\mathrm{i}0.13641 $, and noting $\boldsymbol{u}_0^1={}_P\boldsymbol{u}_0^1+{}_H\boldsymbol{u}_0^1 $ where ${}_P\boldsymbol{u}_0^1 $ (resp. ${}_H\boldsymbol{u}_0^1 $) is the boundary-layer flow associated with ${}_P\boldsymbol{U}_0^1$ (resp.  ${}_H\boldsymbol{U}_0^1$). 
In the container frame considered in \cite{zhang2017theory}, the flow is mainly an apparent one, that is the oscillating solid-body rotation ${}_P\boldsymbol{U}_0^1$ directly related to the frame motion (there is no spin up since $\omega \gg E^{1/2}$). By contrast with the findings of  \cite{zhang2017theory}, the bulk flow reduces to the inertial mode in the mean rotating frame (where ${}_P\boldsymbol{U}_0^1=\boldsymbol{0}$), and is then strongly dependent on $\omega$ \cite[similar to the findings of][who found a way to measure this effect]{aldridge1969axisymmetric}. 
Note however that ${}_P\boldsymbol{u}_0^1 \neq \boldsymbol{0}$ in the mean rotating frame (due to the oscillating boundary velocity), and this flow generates the mean zonal flow in the absence of any other bulk flows (e.g. when no inertial mode is excited with $\omega >2$).

\begin{figure}
	\centering
	\begin{tabular}{cc}
		\includegraphics[width=0.49\textwidth]{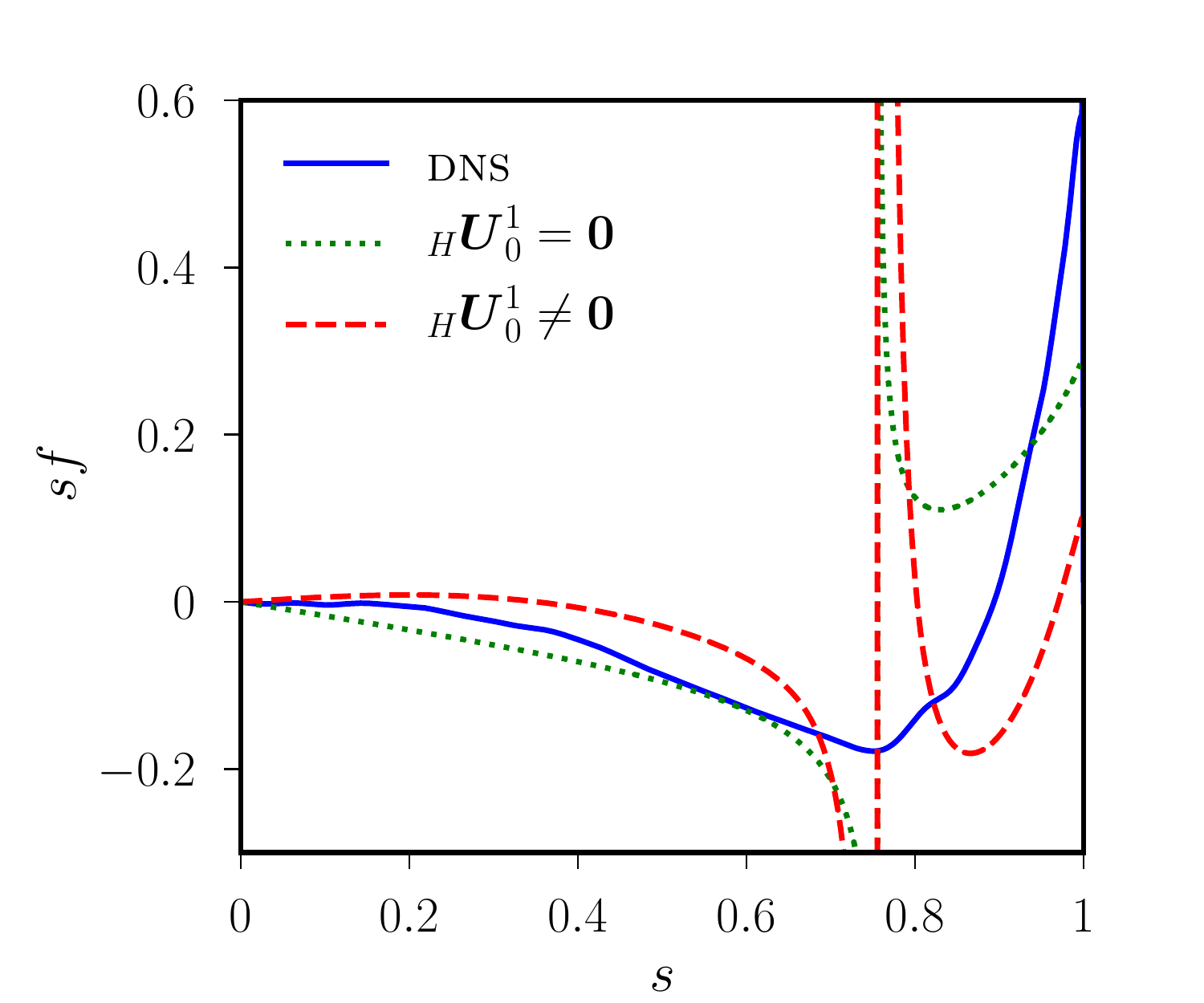} &
        \includegraphics[width=0.49\textwidth]{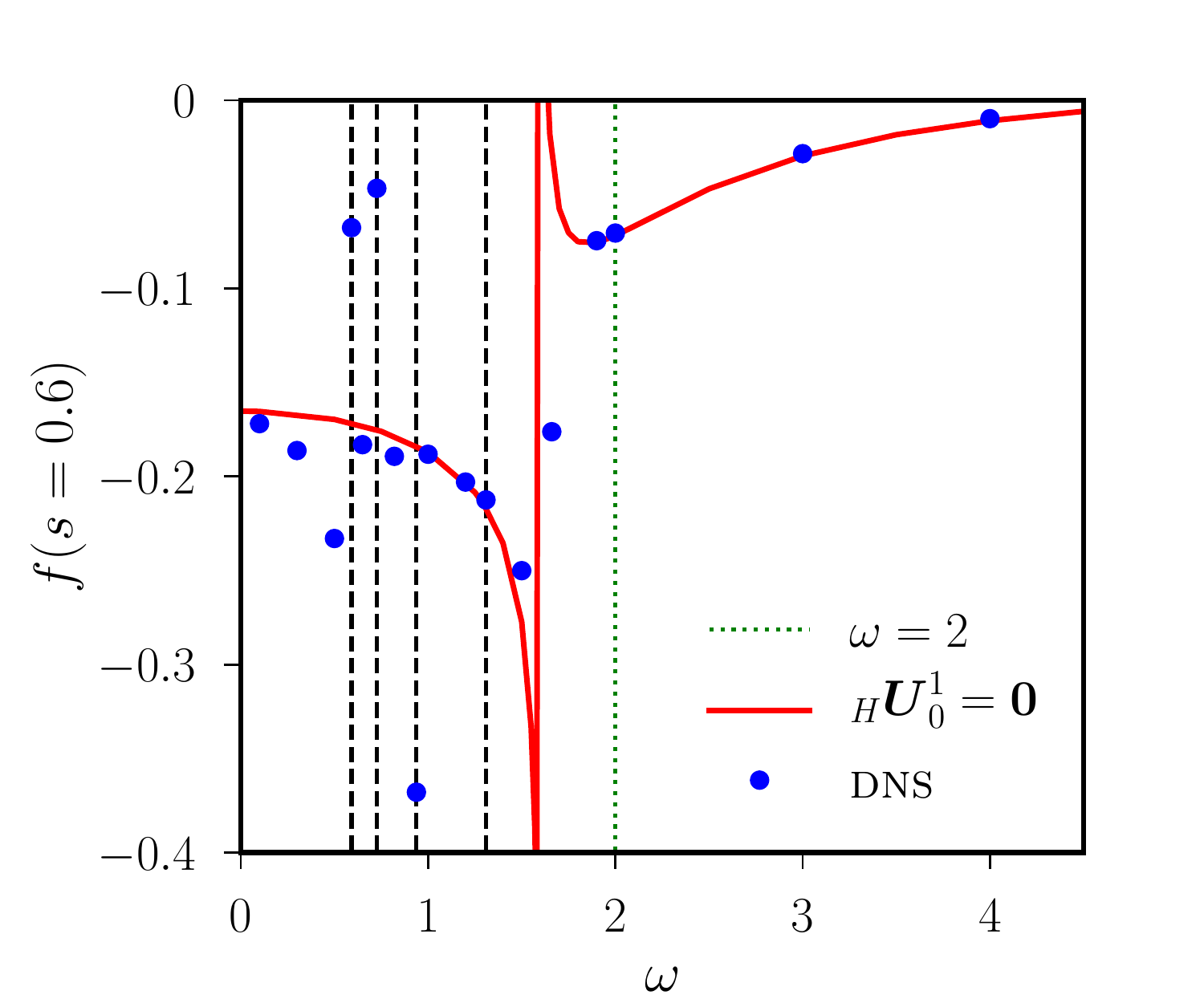} \\
		(a) & (b) \\
	\end{tabular}
	\caption{Mean zonal flow driven by longitudinal librations of a sphere, from theory and DNS ($E=10^{-7}$, $\epsilon=10^{-4}$).  (a) Geostrophic velocity $sf$, as a function of the cylindrical radius $s$, at the inertial mode eigenfrequency $\omega=\sqrt{12/7}$ \citep{zhang2017theory}. DNS (solid line) does not agree well with the theory, whether ${}_H\boldsymbol{U}_0^1$ is taken into account (dashed line) using equation (\ref{eq:gjvid}) or not (dotted line). (b) Rotation rate $f$ at $s=0.6$, as a function of the libration frequency $\omega$, given by the theory (solid line) assuming ${}_H\boldsymbol{U}_0^1=\boldsymbol{0}$.
	Dashed lines show the four least damped inertial modes in figure 3 of \citet{aldridge1969axisymmetric}, including $\omega=\sqrt{12/7}$.}
	\label{fig:ZF_IM}
\end{figure}

Beyond the mean zonal flow obtained by assuming ${}_H\boldsymbol{U}_0^1=\boldsymbol{0}$ (i.e. generated by ${}_P\boldsymbol{u}_0^1$ only), as mainly considered in this work, one can use equation (\ref{eq:gjvid}) to calculate how this mean flow is modified by the bulk inertial mode (i.e. by the non-zero ${}_H\boldsymbol{U}_0^1$ and ${}_H\boldsymbol{u}_0^1$). To do so, we first note that the mean zonal component of $ ({}_H\boldsymbol{U}_0^1\boldsymbol{\cdot}\boldsymbol{\nabla}) \, {}_H\boldsymbol{U}_0^1+2 \boldsymbol{\Omega}_c^1 \times {}_H\boldsymbol{U}_0^1$ is zero. One can thus proceed exactly as in sections \ref{sec:2bulk}-\ref{sec:2mean} to integrate equations (\ref{eq:lf421})-(\ref{plafplouf32}). 
The result is shown in figure \ref{fig:ZF_IM}(a), where it is compared with DNS and with the mean zonal flow generated by ${}_P\boldsymbol{u}_0^1$ only. 
Note that the presence of the critical latitude is difficult to spot on the DNS mean zonal flow (by contrast with $\omega=0.1$ or $\omega=1$, see in figure \ref{fig:oscillation}b).
Moreover, the disagreement suggests that considering a single forced inertial mode ${}_H\boldsymbol{U}_0^1$ in equations (\ref{eq:lf421})-(\ref{plafplouf32}) is not sufficient to explain the generation of the observed jets. 
Contributions neglected in the theory are thus expected to be significant, such as the presence of several inertial modes in the bulk, or nonlinear interactions within the internal shear layers \citep[as in][]{lin2020libration}.

To perform a more systematic comparison, we show in figure \ref{fig:ZF_IM}(b) the mean zonal flow rotation rate (at $s=0.6$) given by the theory (assuming ${}_H\boldsymbol{U}_0^1=\boldsymbol{0}$) and by DNS. 
One can observe the expected divergence at $\omega=2 \cos \sin^{-1} 0.6=1.6$, due to the presence of the critical latitudes. 
Note also the overall rather good agreement, except near inertial mode frequencies excited by the forcing (the apparent good agreement at the eigenfrequency $\omega=\sqrt{12/7}$ is coincidental and related to the choice $s=0.6$, as shown in figure \ref{fig:ZF_IM}a).

\subsection{First-order boundary-layer flow}
\label{sec:ans3}
Integrating the coupled ordinary differential equations (\ref{eq:mfx1}), we obtain
\begin{subeqnarray}\label{syst_DCF2}
 \boldsymbol{Y}_k = 
 \begin{pmatrix}
  A_1 \mathrm{e}^{\alpha_{k+} \, \zeta}+A_2 \mathrm{e}^{-\alpha_{k+} \, \zeta}+A_3 \mathrm{e}^{\alpha_{k-} \, \zeta}+A_4 \mathrm{e}^{-\alpha_{k-} \, \zeta}  \\
  - \mathrm{i} \left[ A_1 \mathrm{e}^{\alpha_{k+} \, \zeta}+  A_2 \mathrm{e}^{-\alpha_{k+} \, \zeta} \right]+ \mathrm{i} \left[ A_3 \mathrm{e}^{\alpha_{k-} \, \zeta}+  A_4 \mathrm{e}^{-\alpha_{k-} \, \zeta}  \right]
 \end{pmatrix}
\end{subeqnarray}
with $\alpha_{k \pm}=(1 + \mathrm{i} s_{\pm} )\sqrt{|\gamma_1 \pm \gamma_k|}$, and $ s_{\pm} =\mathrm{sgn}(\gamma_1 \pm \gamma_k)$. The boundary layer velocity vanishes when $\zeta \rightarrow \infty$, imposing $A_1=A_3=0$. We thus finally obtain 
\begin{eqnarray}\label{syst_DCF3Z}
 \boldsymbol{Y} = \sum_{k}^{}  \,
\begin{pmatrix}
\left[ A_k \mathrm{e}^{-\alpha_{k+} \, \zeta}+B_k \mathrm{e}^{-\alpha_{k-} \, \zeta} \right] \mathrm{e}^{\mathrm{i}(\omega_k t+m_k {\phi})} \\[2mm] 
 - \mathrm{i} \left[A_k \mathrm{e}^{-\alpha_{k+} \, \zeta}- B_k \mathrm{e}^{-\alpha_{k-} \, \zeta} \right] \mathrm{e}^{\mathrm{i}(\omega_k t+m_k {\phi})}
\end{pmatrix} ,
\end{eqnarray}
where $A_k$ and $B_k$ are directly obtained using BC (\ref{eq_CL01DF}b). 
The velocity $\boldsymbol{V}_\Sigma^1$ imposed at the spheroidal boundary depends on the problem at hand. 
Here, we consider that $\boldsymbol{V}_\Sigma^1$ can generically be written as 
\begin{subequations}
\label{eq:lmffnnw}
\begin{equation}
\boldsymbol{V}_\Sigma^1=\boldsymbol{V}_{m}+\boldsymbol{V}_{st}+\boldsymbol{V}_{un}, \quad \boldsymbol{V}_{m}=  \tilde{Q}_z\, \cos (\omega t)\,  s^{q-1} \cos(m {\phi}) \, \widehat{\boldsymbol{z}}_R \times \boldsymbol{r},
\tag{\theequation \emph{a,b}}
\end{equation}
\end{subequations}
where $\boldsymbol{V}_{m}$ is the multipolar and oscillating extension of the BC used by \citet{suess1971}, who considered the particular case $\omega=0$ and $m=2$. In equation (\ref{eq:lmffnnw}), $\boldsymbol{V}_{un}$ is a uniform-vorticity flow, of the form (\ref{eq:mld0cwm}) but with the rotation vector $\tilde{\boldsymbol{q}}=(\tilde{q}_x,\tilde{q}_y,\tilde{q}_z)$, and the velocity $\boldsymbol{V}_{st}=\boldsymbol{V}_{sb}-\widehat{\boldsymbol{n}} \boldsymbol{\cdot} \boldsymbol{V}_{sb}$ is given by the tangential components on $\Sigma$ of the solid-body rotation $\boldsymbol{V}_{sb}= \tilde{Q}_x\, \cos (\omega t)\,  \widehat{\boldsymbol{x}}_R \times \boldsymbol{r}$. 
Since $ \boldsymbol{U}_0^1-\boldsymbol{V}_{un}$ is a uniform-vorticity flow, we assume that $\tilde{\boldsymbol{q}}$ is of the form of $\boldsymbol{\mathcal{Q}}$, that is
\begin{equation}
\frac{\boldsymbol{\nabla} \times ( \boldsymbol{U}_0^1-\boldsymbol{V}_{un})}{2} =
\boldsymbol{\mathcal{Q}}-\tilde{\boldsymbol{q}}
=
\begin{pmatrix}
 Q_x \cos \omega t + Q_x^\star \sin \omega t  \\
 Q_y \sin \omega t+Q_y^\star \cos \omega t \\
-(\widehat{\mathcal{Q}}_z+\widehat{\tilde{q}}_z) \cos \omega t 
\end{pmatrix} .
\label{eq:lkdic}
\end{equation}
The last component in (\ref{eq:lkdic}) allows us to reproduce the particular case $m=0$ of expression (\ref{eq:lmffnnw}b). 
Since non-zero $\widehat{\mathcal{Q}}_z+\widehat{\tilde{q}}_z$ will only be considered when $m=0$, we simplify the equations by replacing $\tilde{Q}_z$ by $Q_z$ in expression (\ref{eq:lmffnnw}b), with $Q_z=\tilde{Q}_z+\widehat{\tilde{q}}_z-\widehat{\mathcal{Q}}_z$, and by putting $\widehat{\mathcal{Q}}_z+\widehat{\tilde{q}}_z=0 $ in equation (\ref{eq:lkdic}). 
Then, replacing the $m=1$ dependency of $\boldsymbol{V}_{un}+\boldsymbol{V}_{st}$ by the formal $m$ dependency imposed by the ansatz of $\boldsymbol{Y}$, equation (\ref{eq_CL01DF}b) gives
\begin{eqnarray} \label{eq:dingue2}
{\boldsymbol{u}_0^1}&=& 
\boldsymbol{V}_{m}+\boldsymbol{V}_{st}+\boldsymbol{V}_{un}-{\boldsymbol{U}_0^1}  \nonumber \\
&=&  \frac{1}{4} \,
\begin{pmatrix}
 0\\[2mm]
  A_-^{(1)} \mathcal{C}_1^1-A_-^{(-1)} \mathcal{C}_{-1}^{-1}+   A_+^{(1)} \mathcal{C}_1^{-1}-A_+^{(-1)} \mathcal{C}_{-1}^{1} \\[2mm]
 B_-^{(1)}   \mathcal{C}_1^1+B_-^{(-1)} \mathcal{C}_{-1}^{-1}+   B_+^{(1)} \mathcal{C}_1^{-1}+B_+^{(-1)} \mathcal{C}_{-1}^{1} \\
\end{pmatrix}
\label{eq:mccjk}
\end{eqnarray}
at the boundary $q_1=Q_1$, with $\mathcal{C}_k^l=\exp[ \mathrm{i} (k m {\phi}+ l \omega t)]$ and 
\begin{subequations}
\begin{eqnarray}
A_{\pm}^{(j)}&=& \frac{\mathrm{i}\, \tilde{Q}_x\, a  \sinh 2 Q_1}{2\tilde{h}} + a \tilde{h} \tanh(2 Q_1) [\mp j (Q_x^\star \mp Q_y^\star)- \mathrm{i}\,(Q_x \pm Q_y)], \\
B_{\pm}^{(j)} &=& - \tilde{Q}_x a \mathcal{T}^\prime_{(Q_1)}  \cos q_2 + Q_z s^q   \nonumber\\
& & +  a \tanh(2 Q_1)  \mathcal{T}_{(Q_1)} \cos q_2 [(Q_x \pm Q_y) \mp \mathrm{i}\, j\, (Q_x^\star \mp Q_y^\star)] .
\end{eqnarray}
\end{subequations}
Using BC (\ref{eq:mccjk}), equation (\ref{syst_DCF3Z}) gives $\boldsymbol{Y}$, and $\boldsymbol{u}_0^1$ reads
\begin{subequations}
\label{eq:dcfou3}
\allowdisplaybreaks
\begin{align}
\boldsymbol{u}_0^1 \boldsymbol{\cdot} \widehat{\boldsymbol{q}}_1 &=0, \\
\boldsymbol{u}_0^1 \boldsymbol{\cdot} \widehat{\boldsymbol{q}}_2 &=\frac{ 1}{8}\,\left[(\mathcal{A}_+^{(1)} \, \mathrm{e}^{-\lambda_+\,\zeta}+\mathcal{A}_-^{(1)} \, \mathrm{e}^{-\lambda_-^*\,\zeta}) \, \mathcal{C}_1^1-(\mathcal{A}_+^{(-1)} \, \mathrm{e}^{-\lambda_+^*\,\zeta}+\mathcal{A}_-^{(-1)}\, \mathrm{e}^{-\lambda_-\,\zeta}) \, \mathcal{C}_{-1}^{-1} \right. \nonumber \\
& + \left. (\mathcal{B}_+^{(1)} \, \mathrm{e}^{-\kappa_+\,\zeta}+\mathcal{B}_-^{(1)} \, \mathrm{e}^{-\kappa_-^*\,\zeta}) \, \mathcal{C}_1^{-1} -(\mathcal{B}_+^{(-1)} \, \mathrm{e}^{-\kappa_+^*\,\zeta}+\mathcal{B}_-^{(-1)} \, \mathrm{e}^{-\kappa_-\,\zeta}) \, \mathcal{C}_{-1}^{1} \right],  \\
\boldsymbol{u}_0^1 \boldsymbol{\cdot} \widehat{\boldsymbol{\phi}} &= - \frac{\mathrm{i}}{8}\, \left[(\mathcal{A}_+^{(1)} \, \mathrm{e}^{-\lambda_+\,\zeta}-\mathcal{A}_-^{(1)} \, \mathrm{e}^{-\lambda_-^*\,\zeta}) \,  \mathcal{C}_1^1+(\mathcal{A}_+^{(-1)} \, \mathrm{e}^{-\lambda_+^*\,\zeta}-\mathcal{A}_-^{(-1)} \, \mathrm{e}^{-\lambda_-\,\zeta}) \, \mathcal{C}_{-1}^{-1} \right. \nonumber \\ 
& +\left. (\mathcal{B}_+^{(1)} \, \mathrm{e}^{-\kappa_+\,\zeta}-\mathcal{B}_-^{(1)} \,  \mathrm{e}^{-\kappa_-^*\,\zeta}) \, \mathcal{C}_1^{-1} +(\mathcal{B}_+^{(-1)} \, \mathrm{e}^{-\kappa_+^*\,\zeta}-\mathcal{B}_-^{(-1)} \, \mathrm{e}^{-\kappa_-\,\zeta}) \, \mathcal{C}_{-1}^{1} \right],
\end{align}
\end{subequations}
with
\begin{subequations}
\begin{eqnarray}
\mathcal{A}_{\pm}^{(j)}&=&A_-^{(j)} \pm \mathrm{i} B_-^{(j)} \nonumber \\ &=&  a \tanh(2 Q_1) [-\mathrm{i} (Q_x-Q_y) - j(Q_x^\star+Q_y^\star) ](  \tilde{h} \mp  \mathcal{T}_{(Q_1)} \cos q_2 )  \nonumber \\ & & + \mathrm{i} \tilde{Q}_x\, a [\sinh(2 Q_1)/(2 \tilde{h}) \mp  \mathcal{T}^\prime_{(Q_1)} \cos q_2 ] \pm \mathrm{i} \, Q_z s^q \\
\mathcal{B}_{\pm}^{(j)}&=&A_+^{(j)} \pm \mathrm{i} B_+^{(j)} \nonumber \\ &=&  a \tanh(2 Q_1)[-\mathrm{i} (Q_x+Q_y) + j (Q_x^\star-Q_y^\star)](  \tilde{h} \mp  \mathcal{T}_{(Q_1)} \cos q_2 )  \nonumber \\ & & + \mathrm{i} \tilde{Q}_x\, a [\sinh(2 Q_1)/(2 \tilde{h}) \mp  \mathcal{T}^\prime_{(Q_1)} \cos q_2 ]   \pm \mathrm{i} \, Q_z s^q,
\end{eqnarray}
\end{subequations}
and
\begin{subequations} \label{eq:dclmp2}
\begin{equation}
\lambda_{\pm} = [1 + \mathrm{i} \, \mathrm{sgn}(\gamma_1 \pm \gamma_+ )]\sqrt{|\gamma_1 \pm \gamma_+ |}, \quad
\kappa_{\pm} = [1 + \mathrm{i} \, \mathrm{sgn}(\gamma_1 \pm \gamma_- )]\sqrt{|\gamma_1 \pm \gamma_- |},
\tag{\theequation \emph{a,b}}
\end{equation}
\end{subequations}
where we have introduced $\gamma_{\pm}=(m \Omega_0 \pm \omega)/2$. Note also the following identities $\lambda_{\pm}= \mathrm{i} \, \mathrm{sgn}(\gamma_1 \pm \gamma_+ ) \lambda_{\pm}^*$, $  \lambda_{\pm}/ \lambda_{\pm}^*=-  \lambda_{\pm}^*/\lambda_{\pm} $, and $
\lambda_{\pm}/ {\lambda_{\pm}^*}^3=\lambda_{\pm}^*/ \lambda_{\pm}^3$ \citep[see also in][]{busse2010}, with similar identities for $\kappa_{\pm}$. In order to ease the cumbersome calculation of the mean zonal flow, it is also useful to already note that
\begin{subequations}
\begin{equation}
\frac{\partial \lambda_{\pm}}{\partial q_2} = - \mathrm{sgn}( \gamma_1 \pm \gamma_+)  \frac{\gamma_1 {\mathcal{T}^\prime_{(q_1)}}^2 \tan q_2}{\tilde{h}^2 \lambda_{\pm}^*}, \quad
\frac{\partial \kappa_{\pm}}{\partial q_2} =  - \mathrm{sgn}(\gamma_1 \pm  \gamma_- )  \frac{\gamma_1 {\mathcal{T}^\prime_{(q_1)}}^2 \tan q_2}{\tilde{h}^2 \kappa_{\pm}^*},
\tag{\theequation \emph{a,b}}
\end{equation}
\end{subequations}
which also gives the derivative of the complex conjugates by swapping $\lambda_{\pm}$ and $\lambda_{\pm}^*$ (as well as $\kappa_{\pm}$ and $\kappa_{\pm}^*$).

The boundary-layer flow $\boldsymbol{u}_0^1$ is actually generated by the differential velocity $\boldsymbol{u}_0^1= \boldsymbol{V}_{\Sigma}^1 -\boldsymbol{U}_0^1 $ at the boundary, and can thus be the same in various frames of reference. 
Considering for instance longitudinal librations ($m=0$), we have $\tilde{Q}_z=1$ and $\widehat{\mathcal{Q}}_z=0$  in the mean rotating frame (i.e. an oscillating boundary with a zero basic flow), which gives $Q_z=1$. 
By comparison, we have $\widehat{\mathcal{Q}}_z=1$ and $\tilde{Q}_z=0$ in the wall frame  (oscillating basic flow with a zero boundary velocity), which also gives $Q_z=1$ and thus $\boldsymbol{u}_0^1$ is formally the same in both frames of reference.

As illustrating examples, we give below the expression of $\boldsymbol{u}_0^1$ in few particular simpler cases. Considering for instance the latitudinal librations of a sphere in the mean rotating frame of reference, $\boldsymbol{V}_\Sigma^1=\boldsymbol{V}_{un} = {\epsilon}\, \cos(\omega t)\, \widehat{\boldsymbol{x}}_R \times \boldsymbol{r}$ can be imposed using $Q_x=-1$ and $m=1$, discarding the other possible contributions to $\boldsymbol{V}_\Sigma^1$ (i.e. $Q_y=Q_y^\star =Q_x^\star=Q_z=\tilde{Q}_x=0 $). Then, in the spherical geometry limit $q_1 \to \infty$, equation (\ref{eq:dcfou3}) reduces to
\begin{subequations}
\label{sol:ordre1_gleliblat}
\allowdisplaybreaks
\begin{eqnarray}
{u_0^1}_{r} &=&0 , \\
{u_0^1}_{\theta} &=&\frac{1}{8}\,\left[(\mathcal{A}_{+}^{(1)}  e^{-\lambda_+\zeta}+\mathcal{A}_{-}^{(1)}  e^{-\lambda_-^*\zeta}) e^{ \mathrm{i} (\phi+\omega t)}-(\mathcal{A}_{+}^{(1)}  e^{-\lambda_+^*\zeta}+\mathcal{A}_{-}^{(1)}  e^{-\lambda_-\zeta}) e^{- \mathrm{i} (\phi+\omega t)}\right. \nonumber \\
& +&  \left. (\mathcal{A}_{+}^{(1)}  e^{-\kappa_+\zeta}+\mathcal{A}_{-}^{(1)}  e^{-\kappa_-^*\zeta}) e^{ \mathrm{i}  (\phi-\omega t)}-(\mathcal{A}_{+}^{(1)}  e^{-\kappa_+^*\zeta}+\mathcal{A}_{-}^{(1)}  e^{-\kappa_-\zeta}) e^{- \mathrm{i}  (\phi-\omega t)} \right] \\
{u_0^1}_{\phi} &=& - \frac{\mathrm{i}}{8} \left[(\mathcal{A}_{+}^{(1)}  e^{-\lambda_+\zeta}-\mathcal{A}_{-}^{(1)}  e^{-\lambda_-^*\zeta}) e^{ \mathrm{i} (\phi+\omega t)}+(\mathcal{A}_{+}^{(1)}  e^{-\lambda_+^*\zeta}-\mathcal{A}_{-}^{(1)}  e^{-\lambda_-\zeta}) e^{- \mathrm{i} (\phi+\omega t)}\right. \nonumber \\ & +& \left. (\mathcal{A}_{+}^{(1)}  e^{-\kappa_+\zeta}-\mathcal{A}_{-}^{(1)}   e^{-\kappa_-^*\zeta}) e^{ \mathrm{i}  (\phi-\omega t)}+(\mathcal{A}_{+}^{(1)}  e^{-\kappa_+^*\zeta}-\mathcal{A}_{-}^{(1)}  e^{-\kappa_-\zeta}) e^{- \mathrm{i} (\phi-\omega t)} \right] \end{eqnarray}
\end{subequations}
where the coordinates $(q_1,q_2,\phi)$ are naturally mapped on the usual spherical coordinates $(r,\theta,\phi)$ used here, with $\mathcal{A}_{\pm}^{(1)} = \mathrm{i} \, r\, ( 1 \mp  \, \cos \theta )$, and where equation (\ref{eq:dclmp2}) can be simplified using $\gamma_{\pm}= \pm \omega/2$ and $\gamma_1=\cos \theta$.

Considering now the multipolar tidal-like forcing of a spheroid, $\boldsymbol{V}_\Sigma^1 = {\epsilon}\, \cos(\omega t)\, \widehat{\boldsymbol{z}}_R \times \boldsymbol{r}$ can be imposed using $Q_z=1$, discarding the other possible contributions to $\boldsymbol{V}_\Sigma^1$ (i.e. $Q_x=Q_y=Q_y^\star =Q_x^\star=\tilde{Q}_x=0 $). Then, equation (\ref{eq:dcfou3}) reduces to 
\begin{subeqnarray}\label{894fbww}
{u_0^1}_{q_2} &=&\frac{\mathrm{i} \, s^q}{8}\,\left[(\mathrm{e}^{-\lambda_+\,\zeta}-\mathrm{e}^{-\lambda_-^*\,\zeta}) \,\mathrm{e}^{\, \mathrm{i} \,(m \phi+\omega t)}+(\mathrm{e}^{-\lambda_-\,\zeta}-\mathrm{e}^{-\lambda_+^*\,\zeta}) \,\mathrm{e}^{-\, \mathrm{i} \,(m \phi+\omega t)}\right. \nonumber \\
& & + \left. (\mathrm{e}^{-\kappa_+\,\zeta}-\mathrm{e}^{-\kappa_-^*\,\zeta}) \,\mathrm{e}^{\, \mathrm{i} \, (m \phi-\omega t)}+(\mathrm{e}^{-\kappa_-\,\zeta}-\mathrm{e}^{-\kappa_+^*\,\zeta}) \,\mathrm{e}^{-\, \mathrm{i} \, (m \phi-\omega t)} \right] , \\
{u_0^1}_{ \phi} &=& \frac{ s^q}{8}\, \left[(\mathrm{e}^{-\lambda_+\,\zeta}+\mathrm{e}^{-\lambda_-^*\,\zeta}) \,\mathrm{e}^{\, \mathrm{i} \,(m \phi+\omega t)}+(\mathrm{e}^{-\lambda_-\,\zeta}+\mathrm{e}^{-\lambda_+^*\,\zeta}) \,\mathrm{e}^{-\, \mathrm{i} \,(m \phi+\omega t)}\right. \nonumber \\ & & +\left. (\mathrm{e}^{-\kappa_+\,\zeta}+ \mathrm{e}^{-\kappa_-^*\,\zeta}) \,\mathrm{e}^{\, \mathrm{i} \, (m \phi-\omega t)}+(\mathrm{e}^{-\kappa_-\,\zeta}+\mathrm{e}^{-\kappa_+^*\,\zeta}) \,\mathrm{e}^{-\, \mathrm{i} \,(m \phi-\omega t)} \right] . 
\end{subeqnarray}
Equation (\ref{894fbww}) can even be further simplified in the two following particular cases of interest, that are (i) longitudinal librations, where $m=0$ and thus $\lambda_{\pm}=\kappa_{\mp}$, and (ii) the steady tidal-like forcing, where $\omega=0$ and thus $\lambda_{\pm}=\kappa_{\pm}$.

\bibliographystyle{jfm}
\bibliography{biblio}
\end{document}